\documentclass[twocolumn,aps,prd,superscriptaddress,nofootinbib,preprintnumbers,showpacs]{revtex4-1}

\usepackage{bm,amsmath,amsfonts,amssymb,graphicx,xspace,placeins,multirow,diagbox}

\usepackage{hyperref}
\usepackage{breakurl}
\usepackage{array}
\usepackage{slashed}
\usepackage{xspace}
\usepackage[italic]{hepnames}
\usepackage{booktabs}
\usepackage{hhline}
\usepackage{multirow}
\usepackage{header3}
\usepackage{ulem}
\usepackage[]{lineno}
\usepackage{placeins}

\let\oldequation\equation
\let\oldendequation\endequation

\renewenvironment{equation}
  {\linenomathNonumbers\oldequation}
  {\oldendequation\endlinenomath}

\allowdisplaybreaks
\newcommand{\nn}{\nonumber}

\newcommand{\Bbar}{\,\overline{\!B}{}}
\newcommand{\Dbar}{\,\overline{\!D}{}}

\newcommand{\Vub}{V_{ub}}

\newcommand{\Eg}{\ensuremath{E_{\gamma}}\xspace}

\newcommand{\continuum}{\ensuremath{\APelectron \Pelectron \to \Pquark \APquark}\xspace}

\newcommand{\bfRes}{\ensuremath{\mathcal{B}(B^+ \to \mu^+ \, \nu_\mu) = \left( 5.3 \pm 2.0  \pm 0.9 \right) \times 10^{-7}}\xspace}

\newcommand{\bmunu}{\ensuremath{B^+ \to \mu^+ \, \nu_{\mu}}\xspace}
\newcommand{\bmuN}{\ensuremath{B^+ \to \mu^+ \, N}\xspace}
\newcommand{\bmunugam}{\ensuremath{B^+ \to \mu^+ \, \nu_{\mu} \, \gamma}\xspace}
\newcommand{\btaunu}{\ensuremath{B^+ \to \tau^+ \, \nu_{\tau}}\xspace}
\newcommand{\bellnu}{\ensuremath{B^+ \to \ell^+ \, \nu_{\ell}}\xspace}

\newcommand{\GF}{\ensuremath{G_{F}}}

\newcommand{\bulnu}{\ensuremath{b \to u \, \ell\, \nu_{\ell}}\xspace}
\newcommand{\bclnu}{\ensuremath{b \to c \, \ell\, \nu_{\ell}}\xspace}
\newcommand{\bpilnu}{\ensuremath{B \to \pi \, \ell^+\,\nu_{\ell}}\xspace}
\newcommand{\brholnu}{\ensuremath{B \to \rho \, \ell^+\,\nu_{\ell}}\xspace}
\newcommand{\bomegalnu}{\ensuremath{B \to \omega \, \ell^+\,\nu_{\ell}}\xspace}
\newcommand{\betalnu}{\ensuremath{B \to \eta \, \ell^+\,\nu_{\ell}}\xspace}
\newcommand{\betaplnu}{\ensuremath{B \to \eta' \, \ell^+\,\nu_{\ell}}\xspace}
\newcommand{\bdlnu}{\ensuremath{B \to D \, \ell^+\,\nu_{\ell}}\xspace}
\newcommand{\bdslnu}{\ensuremath{B \to D^* \, \ell^+\,\nu_{\ell}}\xspace}

\makeatletter
\g@addto@macro\bfseries{\boldmath}
\makeatother

\begin{document}

\title{Search for \bmunu and \bmuN with inclusive tagging}

\noaffiliation
\affiliation{University of the Basque Country UPV/EHU, 48080 Bilbao}
\affiliation{Beihang University, Beijing 100191}
\affiliation{University of Bonn, 53115 Bonn}
\affiliation{Brookhaven National Laboratory, Upton, New York 11973}
\affiliation{Budker Institute of Nuclear Physics SB RAS, Novosibirsk 630090}
\affiliation{Faculty of Mathematics and Physics, Charles University, 121 16 Prague}
\affiliation{Chonnam National University, Kwangju 660-701}
\affiliation{University of Cincinnati, Cincinnati, Ohio 45221}
\affiliation{Deutsches Elektronen--Synchrotron, 22607 Hamburg}
\affiliation{University of Florida, Gainesville, Florida 32611}
\affiliation{Key Laboratory of Nuclear Physics and Ion-beam Application (MOE) and Institute of Modern Physics, Fudan University, Shanghai 200443}
\affiliation{Justus-Liebig-Universit\"at Gie\ss{}en, 35392 Gie\ss{}en}
\affiliation{Gifu University, Gifu 501-1193}
\affiliation{II. Physikalisches Institut, Georg-August-Universit\"at G\"ottingen, 37073 G\"ottingen}
\affiliation{SOKENDAI (The Graduate University for Advanced Studies), Hayama 240-0193}
\affiliation{Hanyang University, Seoul 133-791}
\affiliation{University of Hawaii, Honolulu, Hawaii 96822}
\affiliation{High Energy Accelerator Research Organization (KEK), Tsukuba 305-0801}
\affiliation{J-PARC Branch, KEK Theory Center, High Energy Accelerator Research Organization (KEK), Tsukuba 305-0801}
\affiliation{Forschungszentrum J\"{u}lich, 52425 J\"{u}lich}
\affiliation{IKERBASQUE, Basque Foundation for Science, 48013 Bilbao}
\affiliation{Indian Institute of Science Education and Research Mohali, SAS Nagar, 140306}
\affiliation{Indian Institute of Technology Guwahati, Assam 781039}
\affiliation{Indian Institute of Technology Hyderabad, Telangana 502285}
\affiliation{Indian Institute of Technology Madras, Chennai 600036}
\affiliation{Institute of High Energy Physics, Chinese Academy of Sciences, Beijing 100049}
\affiliation{Institute of High Energy Physics, Vienna 1050}
\affiliation{Institute for High Energy Physics, Protvino 142281}
\affiliation{INFN - Sezione di Napoli, 80126 Napoli}
\affiliation{INFN - Sezione di Torino, 10125 Torino}
\affiliation{J. Stefan Institute, 1000 Ljubljana}
\affiliation{Institut f\"ur Experimentelle Teilchenphysik, Karlsruher Institut f\"ur Technologie, 76131 Karlsruhe}
\affiliation{Kennesaw State University, Kennesaw, Georgia 30144}
\affiliation{Department of Physics, Faculty of Science, King Abdulaziz University, Jeddah 21589}
\affiliation{Kitasato University, Sagamihara 252-0373}
\affiliation{Korea Institute of Science and Technology Information, Daejeon 305-806}
\affiliation{Korea University, Seoul 136-713}
\affiliation{Kyoto University, Kyoto 606-8502}
\affiliation{Kyungpook National University, Daegu 702-701}
\affiliation{\'Ecole Polytechnique F\'ed\'erale de Lausanne (EPFL), Lausanne 1015}
\affiliation{P.N. Lebedev Physical Institute of the Russian Academy of Sciences, Moscow 119991}
\affiliation{Liaoning Normal University, Dalian 116029}
\affiliation{Faculty of Mathematics and Physics, University of Ljubljana, 1000 Ljubljana}
\affiliation{Ludwig Maximilians University, 80539 Munich}
\affiliation{Luther College, Decorah, Iowa 52101}
\affiliation{Malaviya National Institute of Technology Jaipur, Jaipur 302017}
\affiliation{University of Maribor, 2000 Maribor}
\affiliation{Max-Planck-Institut f\"ur Physik, 80805 M\"unchen}
\affiliation{School of Physics, University of Melbourne, Victoria 3010}
\affiliation{University of Mississippi, University, Mississippi 38677}
\affiliation{University of Miyazaki, Miyazaki 889-2192}
\affiliation{Moscow Physical Engineering Institute, Moscow 115409}
\affiliation{Moscow Institute of Physics and Technology, Moscow Region 141700}
\affiliation{Graduate School of Science, Nagoya University, Nagoya 464-8602}
\affiliation{Kobayashi-Maskawa Institute, Nagoya University, Nagoya 464-8602}
\affiliation{Universit\`{a} di Napoli Federico II, 80055 Napoli}
\affiliation{Nara Women's University, Nara 630-8506}
\affiliation{National Central University, Chung-li 32054}
\affiliation{National United University, Miao Li 36003}
\affiliation{Department of Physics, National Taiwan University, Taipei 10617}
\affiliation{H. Niewodniczanski Institute of Nuclear Physics, Krakow 31-342}
\affiliation{Nippon Dental University, Niigata 951-8580}
\affiliation{Niigata University, Niigata 950-2181}
\affiliation{Novosibirsk State University, Novosibirsk 630090}
\affiliation{Osaka City University, Osaka 558-8585}
\affiliation{Pacific Northwest National Laboratory, Richland, Washington 99352}
\affiliation{Panjab University, Chandigarh 160014}
\affiliation{University of Pittsburgh, Pittsburgh, Pennsylvania 15260}
\affiliation{Punjab Agricultural University, Ludhiana 141004}
\affiliation{Theoretical Research Division, Nishina Center, RIKEN, Saitama 351-0198}
\affiliation{University of Science and Technology of China, Hefei 230026}
\affiliation{Seoul National University, Seoul 151-742}
\affiliation{Showa Pharmaceutical University, Tokyo 194-8543}
\affiliation{Soongsil University, Seoul 156-743}
\affiliation{University of South Carolina, Columbia, South Carolina 29208}
\affiliation{Sungkyunkwan University, Suwon 440-746}
\affiliation{School of Physics, University of Sydney, New South Wales 2006}
\affiliation{Department of Physics, Faculty of Science, University of Tabuk, Tabuk 71451}
\affiliation{Tata Institute of Fundamental Research, Mumbai 400005}
\affiliation{Department of Physics, Technische Universit\"at M\"unchen, 85748 Garching}
\affiliation{Toho University, Funabashi 274-8510}
\affiliation{Department of Physics, Tohoku University, Sendai 980-8578}
\affiliation{Earthquake Research Institute, University of Tokyo, Tokyo 113-0032}
\affiliation{Department of Physics, University of Tokyo, Tokyo 113-0033}
\affiliation{Tokyo Institute of Technology, Tokyo 152-8550}
\affiliation{Tokyo Metropolitan University, Tokyo 192-0397}
\affiliation{Virginia Polytechnic Institute and State University, Blacksburg, Virginia 24061}
\affiliation{Wayne State University, Detroit, Michigan 48202}
\affiliation{Yamagata University, Yamagata 990-8560}
\affiliation{Yonsei University, Seoul 120-749}
\author{M.~T.~Prim}\affiliation{Institut f\"ur Experimentelle Teilchenphysik, Karlsruher Institut f\"ur Technologie, 76131 Karlsruhe} 
\author{F.U.~Bernlochner}\affiliation{University of Bonn, 53115 Bonn} 
\author{P.~Goldenzweig}\affiliation{Institut f\"ur Experimentelle Teilchenphysik, Karlsruher Institut f\"ur Technologie, 76131 Karlsruhe} 
\author{M.~Heck}\affiliation{Institut f\"ur Experimentelle Teilchenphysik, Karlsruher Institut f\"ur Technologie, 76131 Karlsruhe} 
\author{I.~Adachi}\affiliation{High Energy Accelerator Research Organization (KEK), Tsukuba 305-0801}\affiliation{SOKENDAI (The Graduate University for Advanced Studies), Hayama 240-0193} 
\author{K.~Adamczyk}\affiliation{H. Niewodniczanski Institute of Nuclear Physics, Krakow 31-342} 
\author{H.~Aihara}\affiliation{Department of Physics, University of Tokyo, Tokyo 113-0033} 
\author{S.~Al~Said}\affiliation{Department of Physics, Faculty of Science, University of Tabuk, Tabuk 71451}\affiliation{Department of Physics, Faculty of Science, King Abdulaziz University, Jeddah 21589} 
\author{D.~M.~Asner}\affiliation{Brookhaven National Laboratory, Upton, New York 11973} 
\author{H.~Atmacan}\affiliation{University of South Carolina, Columbia, South Carolina 29208} 
% ...
\collaboration{The Belle Collaboration}

\begin{abstract}
We report the result for a search for the leptonic decay of $\bmunu$ using the full Belle data set of 711 fb${}^{-1}$ of integrated luminosity at the $\Upsilon(4S)$ resonance. In the Standard Model leptonic $B$-meson decays are helicity and CKM suppressed. To maximize sensitivity an inclusive tagging approach is used to reconstruct the second $B$ meson produced in the collision. The directional information from this second $B$ meson is used to boost the observed $\mu$ into the signal $B$ meson rest-frame, in which the $\mu$ has a monochromatic momentum spectrum. Though its momentum is smeared by the experimental resolution, this technique improves the analysis sensitivity considerably. Analyzing the $\mu$ momentum spectrum in this frame we find \bfRes with a one-sided significance of 2.8 standard deviations over the background-only hypothesis. This translates to a frequentist upper limit of $\mathcal{B}(\bmunu) < 8.6 \times 10^{-7}$ at 90\% CL. The experimental spectrum is then used to search for a massive sterile neutrino, \bmuN, but no evidence is observed for a sterile neutrino with a mass in a range of 0 - 1.5  GeV. The determined \bmunu branching fraction limit is further used to constrain the mass and coupling space of the type II and type III two-Higgs-doublet models.
\end{abstract}

\pacs{12.15.Hh, 12.38.Gc, 13.20.-v, 14.40.Nd, 14.60.St}

\maketitle

\section{introduction}

Precision measurements of leptonic decays of $B$ mesons offer a unique tool to test the validity of the Standard Model of particle physics (SM). Produced by the annihilation of the $\bar{b}$-$u$ quark-pair and the subsequent emission of a virtual $W^+$-boson decaying into a antilepton and neutrino, this process is both Cabibbo-Kobayashi-Maskawa (CKM) and helicity suppressed in the SM. The branching fraction of the \bellnu\ \cite{chargeConjugateImplied} process is given by
\begin{equation}\label{eq:BF_SM}
 \mathcal{B}(\bellnu) =  \frac{ \GF^2 \, m_B \, m_\ell^2}{8 \pi} \left( 1 - \frac{m_\ell^2}{m_B^2} \right)^2 \, f_B^2 \, \left| \Vub \right|^2 \, \tau_B \, ,
\end{equation}
with $\GF$ denoting Fermi's constant, $m_B$ and $m_\ell$ the $B$ meson and lepton masses, respectively, and $\left| \Vub  \right|$ the relevant CKM matrix element of the process. Further, $\tau_B$ denotes the $B$ meson lifetime and the decay constant $f_B$ parametrizes the $b$-$u$ annihilation process,
\begin{equation}
 \langle 0 | A^\mu | B(p) \rangle = i \, p^\mu \, f_B \, ,
\end{equation}
with $A^\mu = \bar b \gamma^\mu \, \gamma^5 \, u$ the corresponding axial-vector current and $p^\mu$ the $B$ meson four-momentum. The value of $f_B$ has to be determined using non-perturbative methods, such as lattice QCD~\cite{Aoki2017} or QCD sum-rule calculations~\cite{Baker:2013mwa,Gelhausen:2013wia}.

\begin{figure}[b]
  \includegraphics[width=0.23\textwidth]{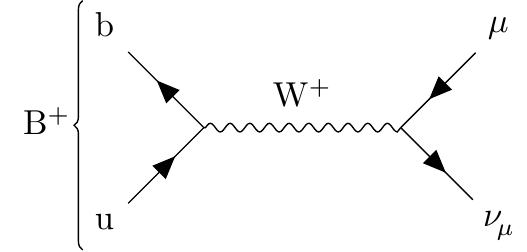} \hfil
  \includegraphics[width=0.23\textwidth]{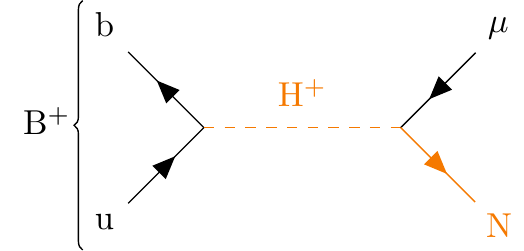} \\
  \vspace{1ex}
  \includegraphics[width=0.23\textwidth]{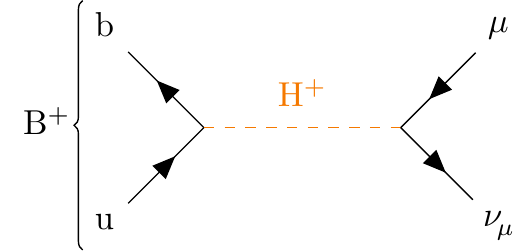} \hfil
  \includegraphics[width=0.23\textwidth]{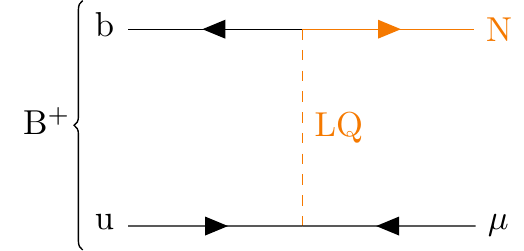} \\
\caption{
   The SM leptonic \bmunu decay process and possible BSM processes with and without a sterile neutrino $N$ in the final state are shown.
 }
\label{fig:btomunu_feynman}
\end{figure}

In this paper an improved search for \bmunu using the full Belle data set is presented. Using the results of $f_B = 184 \pm 4 $ MeV~\cite{Aoki2017} and either inclusive or exclusive world averages for $\left| \Vub  \right|$~\cite{pdg:2018} one finds an expected SM branching fraction of $ \mathcal{B}(\bmunu) = \left( 4.3 \pm 0.8 \right) \times 10^{-7}$ or $ \mathcal{B}(\bmunu) = \left( 3.8 \pm 0.4 \right) \times 10^{-7}$, respectively. This implies an expected total of approximately 300 signal events in the entirety of the Belle data set of 711 fb${}^{-1}$ of integrated luminosity recorded at the $\Upsilon(4S)$ resonance. Thus it is imperative to maximize the overall selection efficiency, which rules out the use of exclusive tagging algorithms, as even advanced machine learning based implementations such as Ref.~\cite{Keck:2018lcd} only achieve efficiencies of a few percent. Events containing a high momentum muon candidate are identified as potential signal events, and the additional charged particles and neutral energy depositions in the rest of the event (ROE) are used to reconstruct the second $B$ meson produced in the collision process. With such an inclusive reconstruction one reduces the background due to non-resonant \continuum  ($\Pquark=\Pup,\Pdown,\Pstrange,\Pcharm$) continuum processes, and, after a dedicated calibration, it is possible to deduce the direction of the signal $B$ meson. This is used to carry out the search in the signal $B$ rest frame, in which the \bmunu decay produces a muon with a monochromatic momentum of $p_\mu^B =  2.64$ GeV. The experimental resolution on the boost vector reconstructed from ROE information broadens this signal signature. The use of this frame, which enhances the expected sensitivity of the search, is the main improvement over the preceding analysis, published in Ref.~\cite{Sibidanov:2017vph}. Further, the modeling of the crucial \bulnu\ semileptonic and continuum backgrounds has been improved with respect to the preceding analysis. 
In Ref.~\cite{Sibidanov:2017vph} a 90\% confidence interval of $[2.9,10.7] \times 10^{-7}$ for the \bmunu branching fraction was determined, while the most stringent 90\% upper limit for this quantity that has been determined is $1 \times 10^{-6}$~\cite{Aubert:2009ar}. 

In the presence of new physics interactions or particles, the CKM and helicity suppression of the \bmunu decay can be lifted: the presence of, for instance, a charged Higgs boson, favored in many supersymmetric extensions of the SM, could strongly enhance the observed \bellnu\ branching fractions. Leptoquarks could have a similar effect. Another interesting exotic particle whose existence can be investigated with this decay are sterile neutrinos. This hypothetical particle acts as a singlet under the fundamental symmetry group of the SM, i.e. they carry no color charge, no weak isospin, nor weak hypercharge quantum numbers. Further, sterile neutrinos do not couple to the gauge bosons of the SM, but their existence could explain, for instance, the dark matter content of the universe~\cite{Boyarsky:2018tvu} or the smallness of the neutrino mass terms~\cite{Akhmedov:1999uz}. The only possibility for a sterile neutrino $N$ to occur in a \bmuN\ final state is due to the existence of a non-SM mediator. Further, the mass of the sterile neutrino has to be $m_N < 5.17 \, \text{GeV} = m_B - m_\mu$ and in the present analysis we are able to probe a mass range of $m_N \in [0,1.5)$ GeV. In Fig.~\ref{fig:btomunu_feynman}  the SM and a selection of beyond the SM (BSM) processes are shown.

The rest of this paper is organized as follows: Section~\ref{sec:data_anastrat} summarizes the used data set, simulated samples and  reconstruction steps. Section~\ref{sec:astart} outlines the inclusive tag reconstruction and calibration of its direction. In addition, the employed background suppression strategies and the used categorization are summarized. In Section~\ref{sec:incl_tag_B_Dpi} the validation of the inclusive tag reconstruction and calibration using $ B^+ \to \bar D^0 \, \pi^+$ decays is described. Section~\ref{sec:stat} introduces the statistical methods used to determine the \bmunu signal yield. In Section~\ref{sec:syst} systematic uncertainties of the measurement are discussed and Section~\ref{sec:bulnu} documents sideband studies to validate the modeling of the crucial \bulnu\ semileptonic and continuum backgrounds. Section~\ref{sec:res} presents the main findings of the paper. Finally, Section~\ref{sec:conc} contains a summary and our conclusions.

\section{Data set and simulated samples}\label{sec:data_anastrat}

\begin{table}[t!]
\caption{
	Used branching fractions for the main background processes are listed.
}
\label{tab:bfs}
\begin{tabular}{lcc}
\hline\hline
 $\mathcal{B}$ & Value $B^+$ & Value $B^0$  \\
 \hline
 \bulnu & \\
 \quad \bpilnu & $\left(7.8 \pm 0.3 \right) \times 10^{-5}$ & $\left(1.5 \pm 0.06 \right) \times 10^{-4}$ \\
 \quad \betalnu & $\left(3.9 \pm 0.5 \right)  \times 10^{-5}$ & - \\
 \quad \betaplnu & $\left(2.3 \pm 0.8 \right) \times 10^{-5}$ & - \\
 \quad  \bomegalnu & $\left(1.2 \pm 0.1 \right)  \times 10^{-4}$ & - \\
 \quad  \brholnu & $\left(1.6 \pm 0.1\right)  \times 10^{-4}$ & $\left(2.9 \pm 0.2\right)  \times 10^{-4}$ \\
 \quad  $B \to X_u \, \ell^+ \, \nu_\ell$ & $\left(2.2 \pm 0.3 \right)  \times 10^{-3}$ & $\left(2.0 \pm 0.3 \right)  \times 10^{-3}$ \\
 \hline
 \bclnu & \\
 \quad $B \to D \, \ell^+ \, \nu_\ell$ & $\left(2.3 \pm 0.1\right) \times 10^{-2} $ & $\left(2.1 \pm 0.1\right)\times 10^{-2} $ \\
 \quad $B \to D^* \, \ell^+ \, \nu_\ell$ & $\left(5.3 \pm 0.1 \right)\times 10^{-2} $ &$\left( 4.9 \pm 0.1 \right)\times 10^{-2} $ \\
 \hline
 \bmunugam & $\left( 1.0 \pm 1.3  \right) \times 10^{-6}$ & - \\
 \hline
 $b \to s/d$ \\
 \quad $B \to K^0_L \, \pi^+$ & $\left( 2.4 \pm 0.1  \right) \times 10^{-5}$  & - \\
 \quad $B \to K^+ \, \pi^0$ & $\left( 1.3 \pm 0.1  \right) \times 10^{-5}$  & - \\
 \quad $B \to \rho^+ \, \pi^-$ & - & $\left( 2.3 \pm 0.2  \right) \times 10^{-5}$  \\
 \hline\hline
\end{tabular}
\end{table}

We analyze the full Belle data set of \mbox{$(772 \pm 10) \times 10^6$} \PB meson pairs, produced at the KEKB accelerator complex~\cite{KEKB} with a center-of-mass energy (c.m.) of $\sqrt{s} = \SI{10.58}{GeV}$ at the $\Upsilon(4S)$ resonance. In addition, we use $\SI{79}{fb^{-1}}$ of collisions recorded $\SI{60}{MeV}$ below the $\Upsilon(4S)$ resonance peak to derive corrections and carry out cross-checks.

The Belle detector is a large-solid-angle magnetic spectrometer that consists of a silicon vertex detector (SVD), a 50-layer central drift chamber (CDC), an array of aerogel threshold  \v{C}erenkov counters (ACC),  a barrel-like arrangement of time-of-flight scintillation counters (TOF), and an electromagnetic calorimeter comprised of CsI(Tl) crystals (ECL) located inside a superconducting solenoid coil that provides a \SI{1.5}{T} magnetic field. An iron flux return located outside of the coil is instrumented to detect $K^0_L$ mesons and to identify muons (KLM). A more detailed description of the detector, its layout and performance can be found in Ref.~\citep{Abashian:2000cg} and in references therein.

Charged tracks are identified as electron or muon candidates by combining the information of multiple subdetectors into a lepton identification likelihood ratio, $\mathcal{L}_\mathrm{LID}$. For electrons the identifying features are the ratio of the energy deposition in the ECL with respect to the reconstructed track momentum, the energy loss in the CDC, the shower shape in the ECL, the quality of the geometrical matching of the track to the shower position in the ECL, and the photon yield in the ACC~\citep{HANAGAKI2002490}. Muon candidates are identified from charged track trajectories extrapolated to the outer detector. The identifying features are the difference between expected and measured penetration depth as well as the transverse deviation of KLM hits from the extrapolated trajectory~\citep{ABASHIAN200269}. Charged tracks are identified as pions or kaons using a likelihood classifier which combines information from the CDC, ACC, and TOF subdetectors. In order to avoid the difficulties understanding the efficiencies of reconstructing $K^0_L$ mesons, they are not explicitly reconstructed in what follows.

Photons are identified as energy depositions in the ECL without an associated track. Only photons with an energy deposition of \mbox{\Eg $ > \SI{100}{MeV}$, $\SI{150}{MeV}$, and $\SI{50}{MeV}$} in the forward endcap, backward endcap and barrel part of the calorimeter, respectively, are considered.

We carry out the entire analysis in the Belle~II analysis software framework~\cite{basf2}: to this end the recorded Belle collision data and simulated Monte Carlo (MC) samples were converted using the software described in Ref.~\cite{b2b2}. MC samples of \PB meson decays and continuum processes are simulated using the \texttt{EvtGen} generator~\citep{EvtGen}. The used sample sizes correspond to approximately ten and six times the Belle collision data for \PB meson and continuum decays, respectively. The interactions of particles traversing the detector are simulated using \texttt{Geant3}~\citep{Geant3}. Electromagnetic final-state radiation (FSR) is simulated using the \texttt{PHOTOS}~\citep{Photos} package. The efficiencies in the MC are corrected using data-driven methods.

Signal \bmunu and \bmuN\ decays are simulated as two-body decay of a scalar initial-state meson to a lepton and a massless antineutrino. The effect of the non-zero sterile neutrino mass is incorporated by adjusting the kinematics of the simulated events.

The most important background processes are semileptonic \bulnu\ decays and continuum processes, which both produce high-momentum muons in a momentum range similar to the \bmunu process. Charmless semileptonic decays are produced as a mixture of specific exclusive modes and non-resonant contributions: Semileptonic \bpilnu\ decays are simulated using the BCL form factor parametrization~\citep{Bourrely:2008za} with central values and uncertainties from the global fit carried out by Ref.~\citep{HFLAV16}. The processes of \brholnu\ and \bomegalnu\ are modeled using the BCL form factor parametrization. We fit the measurements of Refs.~\cite{Sibidanov:2013rkk,Lees:2012mq,delAmoSanchez:2010af} in combination with the light-cone sum rule predictions of Ref.~\cite{Bharucha:2012wy} to determine a set of central values and uncertainties. The subdominant processes of \betalnu and \betaplnu are modeled using the ISGW2 model~\cite{Scora:1995ty}. In addition to these narrow resonances, we produce non-resonant \bulnu\ decays with at least two pions in the final state using the DFN model~\cite{DeFazio:1999ptt}. In this model, the triple differential rate is regarded as a function of the four-momentum transfer squared ($q^2$), the lepton energy ($E_\ell^B$), and the hadronic invariant mass squared ($m_X^2$) at next-to-leading order precision in the strong coupling constant $\alpha_s$. The triple differential rate is convolved with a non-perturbative shape function using an ad-hoc exponential model. The free parameters in this model are the $b$ quark mass in the $1S$ scheme, $m_{b}^{1S} = (4.69 \pm 0.04)\,\mathrm{GeV}$ and a non-perturbative parameter $a = 1.9 \pm 0.5$. The values of these parameters were determined in Ref.~\cite{HFLAV16} from a fit to $\bclnu$  information. The non-perturbative parameter $a$ is related to the average momentum squared of the $b$ quark inside the $B$ meson and controls the second moment of the shape function. It is defined as $a = \frac{3 \overline \Lambda^2}{- \lambda_1} -1$ with the binding energy $\overline \Lambda = m_B - m_b^{1S}$ and the hadronic matrix element expectation value $\lambda_1$. Hadronization of parton-level DFN predictions for the \bulnu\ process is accomplished using the JETSET algorithm~\cite{SJOSTRAND199474} to produce two or more final state mesons. The inclusive and exclusive \bulnu\ predictions are combined using a so-called `hybrid' approach, which is a method originally suggested by Ref.~\cite{hybrid}: to this end we combine both predictions such that the partial branching fractions in the triple differential rate of the inclusive ($ \Delta \mathcal{B}_{ijk}^{\rm incl}$) and combined exclusive  ($ \Delta \mathcal{B}_{ijk}^{\rm excl}$)  predictions reproduce the inclusive values. This is achieved by assigning weights to the inclusive contributions $w_{ijk}$ such that
\begin{equation}
\begin{aligned}
 \Delta \mathcal{B}_{ijk}^{\rm incl} = & \,   \Delta \mathcal{B}_{ijk}^{\rm excl} + w_{ijk} \times  \Delta \mathcal{B}_{ijk}^{\rm incl} \, ,
\end{aligned}
\end{equation}
with $i,j,k$ denoting the corresponding bin in the three dimensions of $q^2$, $E_\ell^B$, and $m_X$:
\begin{equation}
\begin{aligned}
 q^2 & = [0,2.5,5,7.5,10,12.5,15,20,25] \, \text{GeV}^2 \, , \nn \\
 E_\ell^B & = [0,0.5,1,1.25,1.5,1.75,2,2.25,3] \, \text{GeV} \, , \nn \\
 m_X & = [0,1.4,1.6,1.8,2,2.5,3,3.5]  \, \text{GeV}  \, .
\end{aligned}
\end{equation}
To study the model dependence of the DFN shape function and possible effects of next-to-next-to-leading order corrections in $\alpha_s$, we also determine weights using the BLNP model of Ref.~\cite{Lange:2005yw}.

The modeling of simulated continuum background processes is corrected using a data-driven method, which was first proposed in Ref.~\cite{Martschei_2012}: a boosted decision tree (BDT) is trained to distinguish between simulated continuum events and the recorded off-resonance data sample. This allows the BDT to learn differences between both samples, and a correction weight, $w = p / \left( 1 - p \right)$, accounting for differences in both samples can be derived directly from the classifier output $p$. As input for the BDT we use the same variables used in the continuum suppression approach (which is further detailed in Section~\ref{sec:astart}) and, additionally, the signal-side muon momentum in the signal $B$ meson frame.

The semileptonic background from \bclnu\ decays is dominated by \bdlnu\ and \bdslnu\ decays. The \bdlnu\ form factors are modeled using the BGL form factors~\cite{Boyd:1994tt} with central values and uncertainties taken from the fit in Ref.~\cite{Glattauer:2015teq}. For \bdslnu\ we use the BGL implementation proposed by Refs.~\cite{Grinstein:2017nlq,Bigi:2017njr} with central values and uncertainties from the fit of the preliminary measurement of Ref.~\cite{Abdesselam:2017kjf}. The measurement is insensitive to the precise details of the modeling of \bclnu\ involving higher charm resonances.

For the contributions of \bmunugam\ we use the recent experimental bounds of Ref.~\cite{Gelb:2018end}. In this process, structure-dependent corrections, which are suppressed by the electromagnetic coupling constant $\alpha_{\rm em}$, lift the helicity suppression of the \bmunu decay. We simulate this process using the calculation of Ref.~\citep{PhysRevD.61.114510} and only allow daughter photons with $E_\gamma > 300$ MeV, to avoid overlap with the FSR corrections simulated by PHOTOS as corrections to the \bmunu final state. In the following, we treat these two processes separately. 

The small amount of background from rare $b \to s/d$ processes is dominated by $B^+ \to K^0_L \, \pi^+$ decays. Subdominant contributions are given by the decays $B^+ \to K^+ \, \pi^0$ and $B^0 \to \rho^+ \, \pi^-$. We adjust those branching fractions to the latest averages of Ref.~\cite{pdg:2018}.

Table~\ref{tab:bfs} summarizes the branching fractions used for all important background processes.

\section{Analysis strategy, inclusive tag reconstruction and calibration}\label{sec:astart}

\begin{figure}[ht!]
	\includegraphics[width=0.45\textwidth]{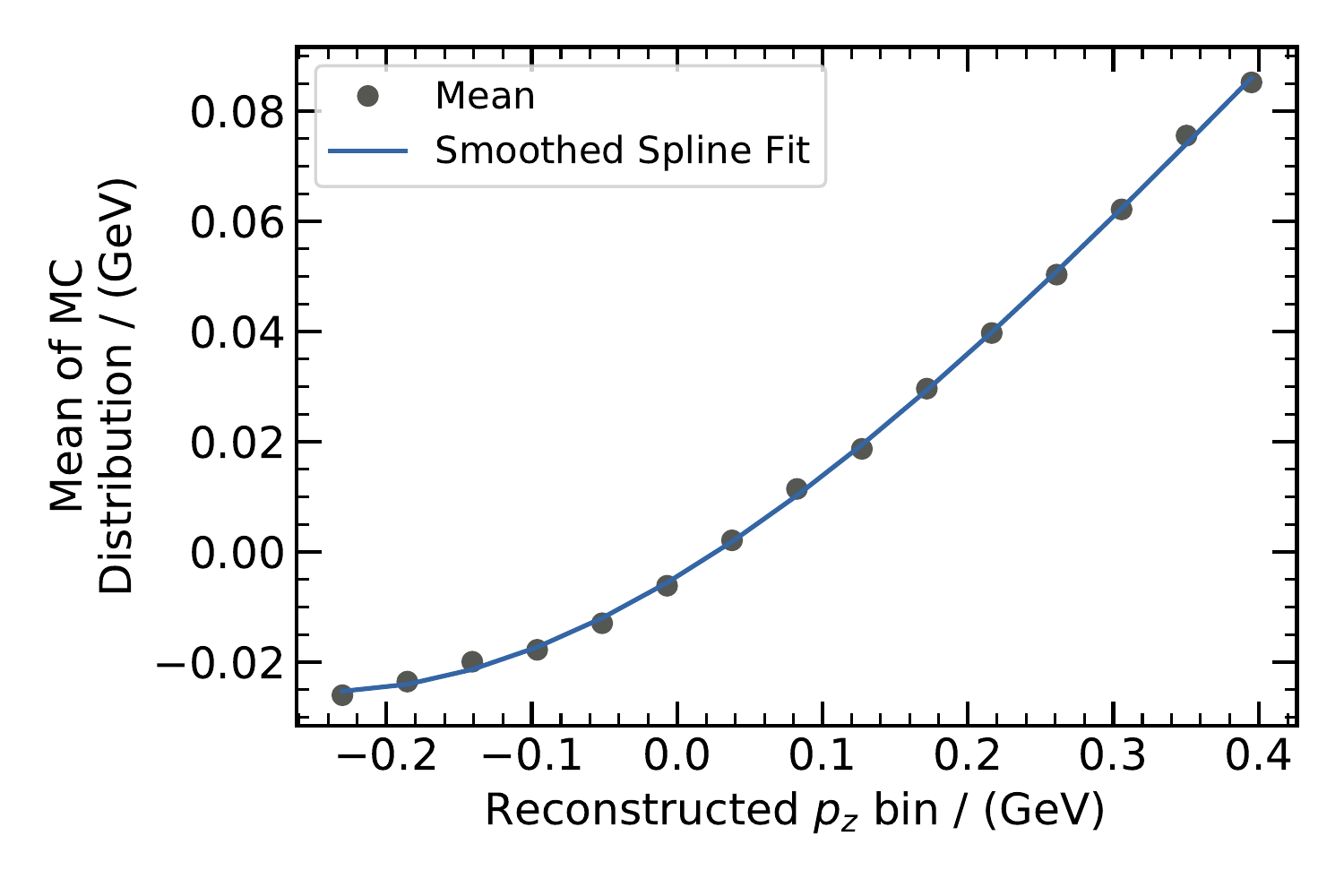}
\caption{
	The functional dependence between the reconstructed and true momentum $z$ component of the inclusively reconstructed tag-side B meson is shown.
}
\label{fig:calibration}
\end{figure}

\begin{figure}[ht!]
  \includegraphics[width=0.45\textwidth]{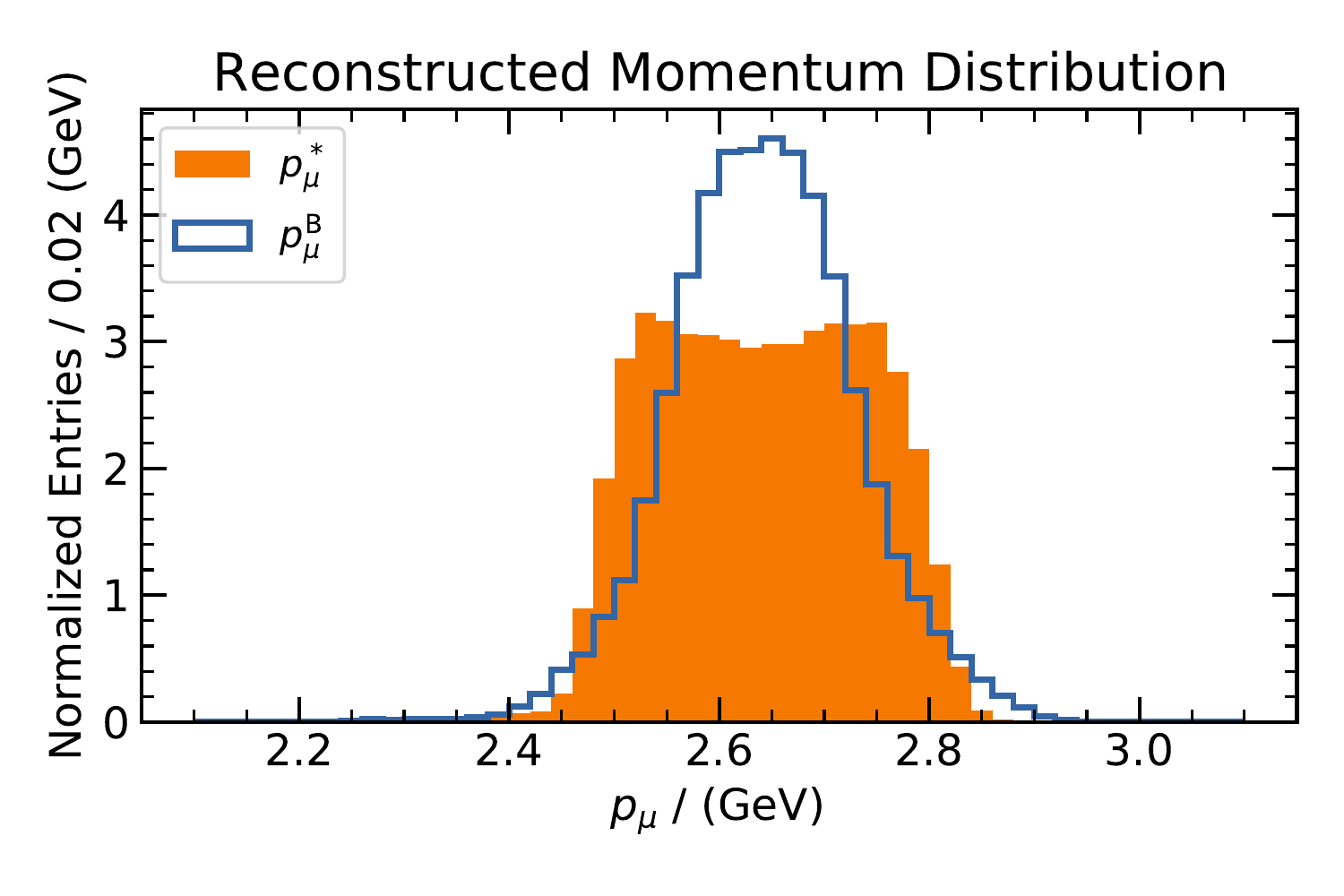}
\caption{
   The signal resolution of \bmunu is compared for signal events reconstructed in the $\APelectron\Pelectron$ c.m. ($p_\mu^*$) and the signal $B$ rest frame ($p_\mu^B$).
 }
\label{fig:sig_res}
\end{figure}

We select $B \bar B$ candidate events by requiring at least three charged particles to be reconstructed and a significant fraction of the c.m. energy to be deposited in the ECL. We first reconstruct the signal side: a muon candidate with a momentum of $p_{\mu}^{\ast} > \SI{2.2}{GeV} $ in the c.m. frame of the colliding $e^+ \, e^-$-pair. The candidate is required to have a distance of closest approach to the nominal interaction point transverse to and along the beam axis of $\text{d} r < \SI{0.5}{cm}$ and $|\text{d} z| < \SI{2}{cm}$, respectively. This initial selection results in a signal-side efficiency of  $\approx 82.2$\%. After this
the remaining charged tracks and neutral depositions are used to reconstruct the ROE to allow us to boost this signal muon candidate into the rest frame of the signal-side $B$ meson. A looser selection on the ROE tracks is imposed, $\text{d} r < \SI{10}{cm}$ and $|\text{d} z| < \SI{20}{cm}$, to also include charged particle candidates which are displaced from the interaction region. All ROE charged particles are treated as pions and no further particle identification is performed. Track candidates with a transverse momentum of $p_T < 275$ MeV do not leave the CDC, but curl back into the detector. To avoid double counting of those tracks, we check if such are compatible with another track. If the track parameters indicate that  this is the case, we veto the lower momentum track. When we combine the momentum information with ROE photon candidates (reconstructed as described in Section~\ref{sec:data_anastrat}) we determine the three-momentum ($  \mathbf{p}_{\rm tag}^{\rm lab} $) and energy ($E_{\rm tag}^{\rm lab}$) of the tag-side $B$ meson in the laboratory frame as
\begin{equation}
\begin{aligned}
  \mathbf{p}_{\rm tag}^{\rm lab} & = \sum_{i}^{\rm tracks} \mathbf{p}_i^{\rm lab} + \sum_{j}^{\rm photons} \mathbf{E}_j^{\rm lab} \, ,  \\
 E_{\rm tag}^{\rm lab} & = \sqrt{  \left( \mathbf{p}_{\rm tag}^{\rm lab} \right)^2 + m_B^2 } \, .
\end{aligned}
\end{equation}
Here $ \mathbf{p}_i^{\rm lab}$ and $\mathbf{E}_j^{\rm lab} $ denote the three-momentum of tracks and photons in the ROE. We proceed by boosting the tag-side four-vector into the c.m. frame of the $e^+ \, e^-$-collision. Due to the two-body nature of the $\Upsilon(4S) \to B \, \bar B$ decay, we have precise knowledge of the magnitude of tag- and signal-side $B$ meson in this frame: $|\mathbf{p}_{B}^{\ast}| = 330$ MeV. We thus correct after the boost the energy component of the tag-side four-vector to be exactly
\begin{equation}
  E_{\rm tag}^{\ast} =  \sqrt{  \left( \mathbf{p}_{B}^{\ast} \right)^2 + m_B^2 } =   \sqrt{  \left( 330 \, \text{MeV} \right)^2 + m_B^2 }  \, ,
\end{equation}
keeping the direction of the three-momentum unchanged.
This improves the resolution with respect to using the boosted absolute three-momentum $\mathbf{p}_{\rm tag}^{\ast}$. Due to the asymmetric beam energies of the colliding $e^+ \, e^-$-pair, all produced $B$ meson decay products are boosted in the positive $z$ direction in the laboratory frame. Thus it is more likely that charged and neutral particles escape the Belle detector acceptance in the forward region and bias the inclusive tag reconstruction. This bias degrades the resolution of the reconstructed $z$ component of the $\mathbf{p}_{\rm tag}^{\ast}$ momentum vector. The resolution is significantly improved by applying a calibration function derived from simulated $e^+ \, e^- \to \Upsilon(4S) \to B \, \bar B$ decays, where one $B$ decays into a $\mu\nu_\mu$-pair. The goal of this function is to map the reconstructed mean momentum $z$ component, $\left(\mathbf{p}_{\rm tag}^{\ast}\right)_z$, to the mean of the simulated true distribution. The functional dependence between the reconstructed and true momentum $z$ component is shown in Fig.~\ref{fig:calibration}. 
In addition, an overall correction factor $\zeta$ is applied to the calibrated three-momentum, chosen such that the difference between the corrected and the simulated three-momentum becomes minimal. The corrected tag-side $z$ and transverse momentum components are then
\begin{equation}
\begin{aligned}
   \left(\mathbf{p}_{\rm tag, cal}^{\ast}\right)_z  & = \zeta \,  f[ \left(\mathbf{p}_{\rm tag}^{\ast}\right)_z ]\, , \\
    \left(\mathbf{p}_{\rm tag, cal}^{\ast}\right)_T &  = \zeta \, \sqrt{ \left( \mathbf{p}_{\rm tag}^{\ast}\right)^2  -  \left(\mathbf{p}_{\rm tag, corr}^{\ast}\right)_z^2  }\, ,
\end{aligned}
\end{equation}
with $f$ the calibration function. The absolute difference between corrected and simulated three-momentum is found to be minimal for $\zeta = 0.58$. Using the calibrated tag-side $B$ meson three-momentum $\mathbf{p}_{\rm tag, cal}^{\ast}$, we boost the signal-side muon candidate into the signal-side $B$ meson rest frame using
\begin{equation}
\mathbf{p}_{\rm sig} = - \mathbf{p}_{\rm tag, cal}^{\ast}\, .
\end{equation}
Figure~\ref{fig:sig_res} compares the muon momentum spectrum for signal \bmunu decays in the $\APelectron\Pelectron$ c.m. frame with the obtained resolution in the $B$ rest frame (further denoted as $p_\mu^B$) using the calibrated momentum vector. Carrying out the boost into the approximated $B$ meson rest frame improves the resolution of the reconstructed muon momentum by $7\,\%$ with respect to the resolution in the c.m. frame.

\begin{figure}[tb]
  \includegraphics[width=0.5\textwidth]{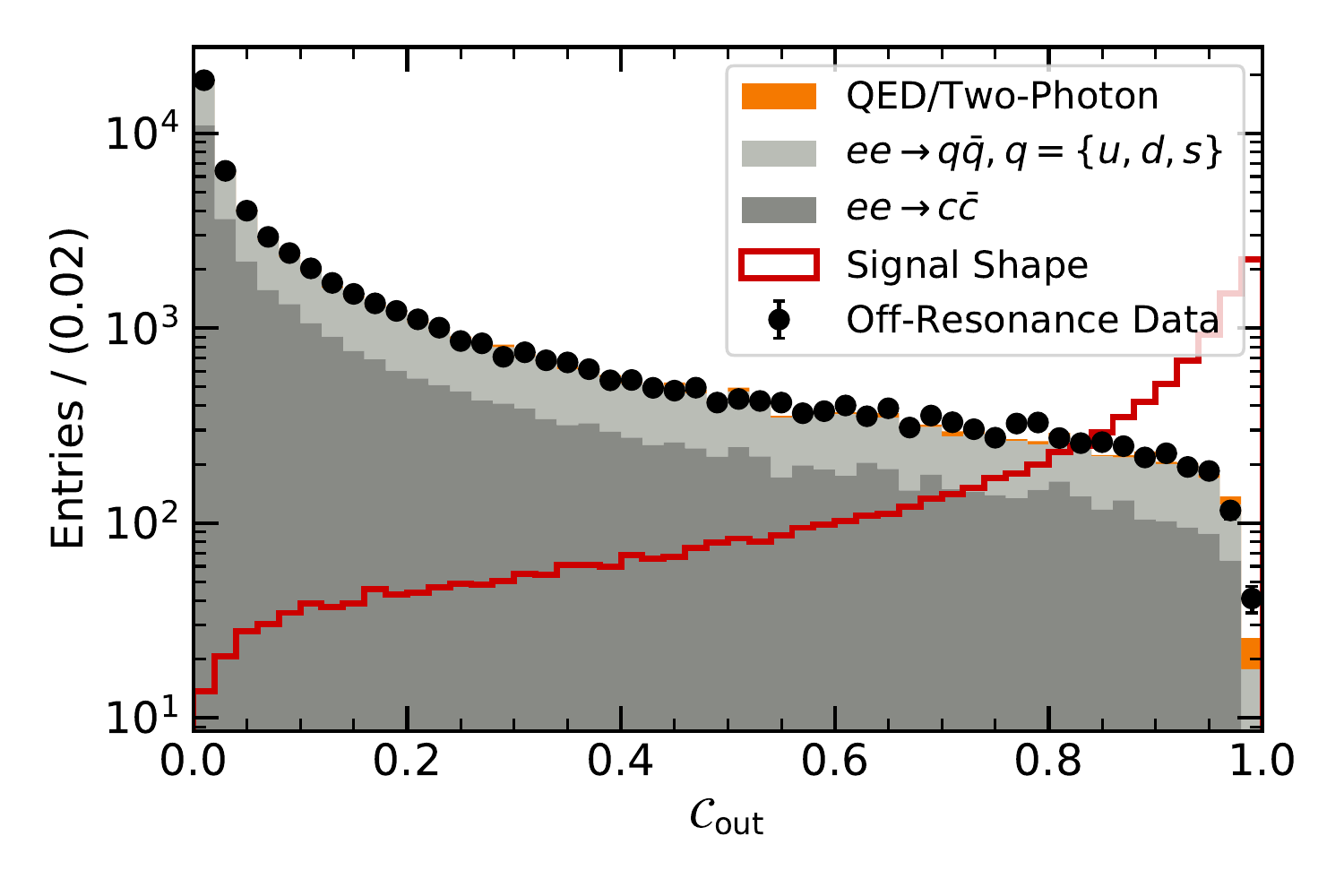} \hfill
  \includegraphics[width=0.5\textwidth]{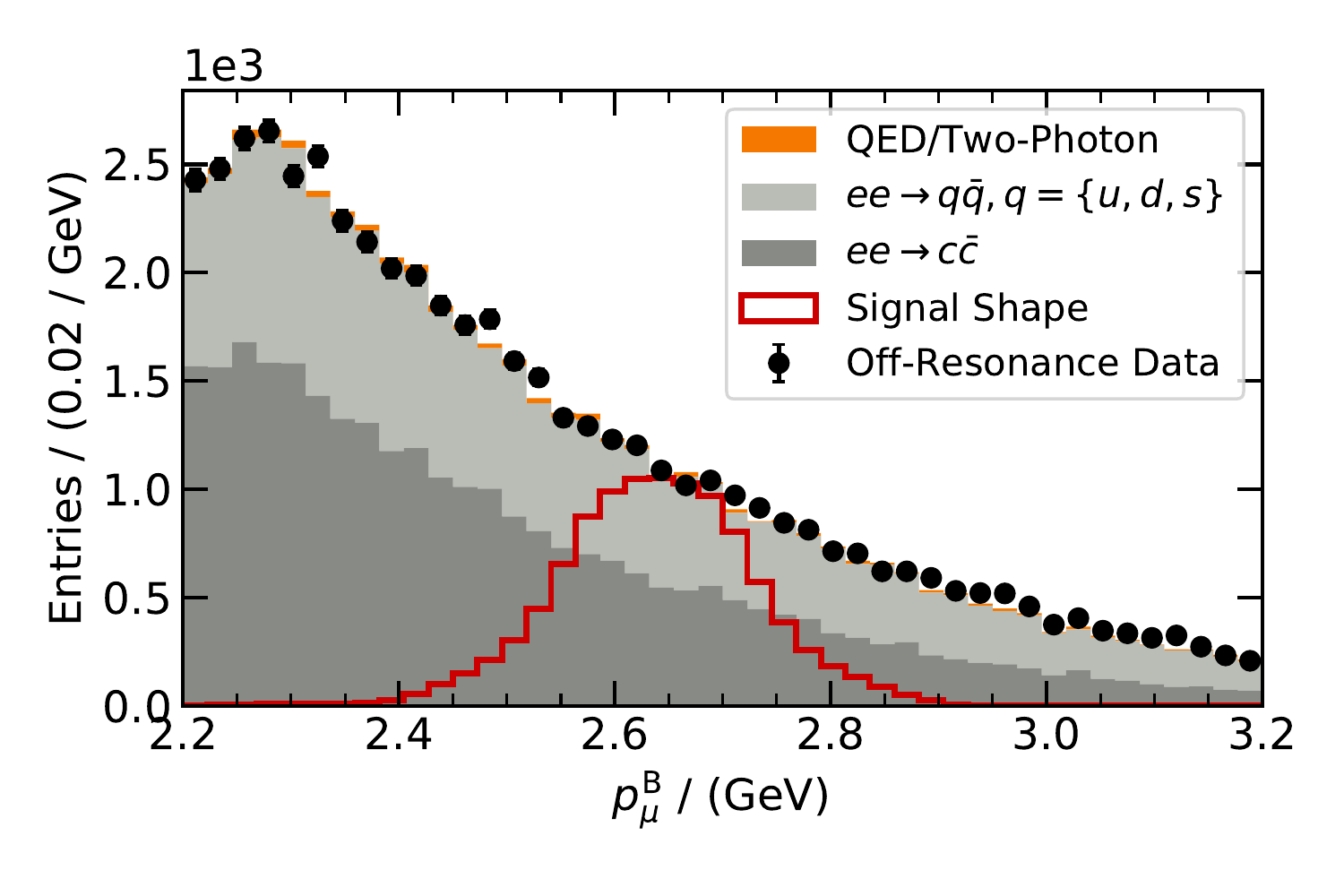}
\caption{
   The classifier output $C_{\rm out}$ and the $p_\mu^B$ distribution of off-resonance data are compared to the simulated continuum background, after applying the correction described in Section~\ref{sec:data_anastrat}.
 }
\label{fig:off_res_classifier}
\end{figure}

To reduce the sizable background from continuum processes, a multivariate classifier using an optimized implementation of gradient-BDTs~\citep{Keck2017} is used and trained to distinguish \bmunu signal decays from continuum processes. The BDT exploits the fact that the event topology for non-resonant $e^+ \, e^-$-collision processes differ significantly from the resonant $e^+ \, e^- \to \Upsilon(4S) \to B \, \bar B$ process. Event shape variables, such as the magnitude of the thrust of final-state particles from both $B$ mesons, the reduced Fox-Wolfram moment $R_2$, the modified Fox-Wolfram moments \citep{SFW} and CLEO Cones \citep{cleocones}, are highly discriminating. To these variables we add as additional inputs to the BDT the number of tracks in the ROE, the number of leptons (electrons or muons) in the ROE, the normalized beam-constrained mass of the tag-side $B$ meson defined as
\begin{equation}
 \widehat m_{bc}^{\rm tag} = \sqrt{ s/4 - \left( \mathbf{p}_{\rm tag, cal}^{\ast} \right)^2 } / \left( \sqrt{s}/2 \right) \, ,
 \end{equation}
and the normalized missing energy defined as
\begin{equation}
\begin{aligned}
  \Delta\widehat E & =  \left( E_{\rm tag, reco}^{\ast} - \sqrt{s}/2 \right) / \left( \sqrt{s}/2 \right) \, ,
\end{aligned}
\end{equation}
with $ E_{\rm tag, reco}^{\ast} $ denoting the energy from boosting the ROE four-vector from the laboratory into the c.m. frame. This list of variables and $p_\mu^B$ are used in the data-driven correction described in Section~\ref{sec:data_anastrat} to correct the simulated continuum events. We apply a loose set of ROE preselection cuts: only events with at least two tracks, fewer than three leptons, $ \widehat m_{bc}^{\rm tag}  > 0.96$, $\Delta \widehat E \in [-0.5,0.1)$, and $R_2 < 0.5$ are further considered. Figure~\ref{fig:off_res_classifier} compares the classifier output $C_{\rm out}$ and $p_\mu^B$ distributions of the predicted simulated and corrected continuum contribution with recorded off-resonance collision events. Both variables show good agreement.

Using this classifier and the cosine of the angle between the calibrated signal $B$ meson in the c.m. system and the muon in the $B$ rest frame ($\cos \Theta_{B\mu}$) we define four mutually exclusive categories. The first two of these are signal enriched categories with $C_{\rm out} \in [0.98,1)$ and split with respect to their $\cos \Theta_{B\mu}$ values. For \bmunu signal decays no preferred direction in $\cos \Theta_{B\mu}$ is expected. For the semileptonic and continuum background events, which pass the selection, the muons are emitted more frequently in the direction of the reconstructed $B$ meson candidate. The second two categories have $C_{\rm out} \in [0.93,0.98)$, and they help separate \bulnu\ and continuum processes from \bmunu signal decays. Table~\ref{tab:cats} summarizes the four categories. The chosen cut values were determined using a grid search and by fits to Asimov data sets (using the fit procedure further described in Section~\ref{sec:stat}).

\begin{table}[b!]
\caption{
	The definition of the four signal categories is shown.
}
\label{tab:cats}
\begin{tabular}{lll|c}
\hline\hline
Category & $C_{\rm out}$ & $ \cos \Theta_{B\mu}$ & Signal Efficiency  \\
 \hline
\quad I & [0.98,1.00) & [-0.13,1.00)  & 6.5{\,}\% \\
\quad II & [0.98,1.00) & [-1.00,-0.13) & 5.9{\,}\% \\
\quad III & [0.93,0.98) & [0.04,1.00) & 7.1{\,}\% \\
\quad IV & [0.93,0.98) & [-1.00,0.04) & 8.3{\,}\% \\
 \hline\hline
\end{tabular}
\end{table}

In Section~\ref{sec:bulnu} the signal-depleted region of $C_{\rm out} \in [0.9,0.93)$ is analyzed and simultaneous fits  in two categories,  $\cos \Theta_{B\mu} < 0$ and $ \cos \Theta_{B\mu} > 0$, are carried out to validate the modeling of the important \bulnu\ background and to extract a value of the inclusive $\mathcal{B}(B \to X_u \, \ell^+ \, \nu)$ branching fraction. The selection efficiencies of \bmunu signal and the background processes are summarized in Table~\ref{tab:cutflow}.

\begin{table}[b!]
\caption{
	The cumulative selection efficiencies of \bmunu signal decays and dominant background processes throughout the selection is listed. See text for details about the various selection steps.
}
\label{tab:cutflow}
\begin{tabular}{lc|cc}
\hline\hline
 Efficiency  & \bmunu & \bulnu & Continuum \\
 \hline
  $B \, \bar B$ \& Muon reco. & 82{\,}\% & 10{\,}\% & 0.9{\,}\%  \\
 ROE Presel. & 55{\,}\% & 1.4{\,}\% & 0.03{\,}\%  \\
 $C_{\rm out}$ cut & 28{\,}\% & 0.2{\,}\% & 0.001{\,}\%  \\
 \hline\hline
\end{tabular}
\end{table}

\section{Inclusive tag validation using $ B^+ \to \overline D^{\,0} \, \pi^+$ decays}\label{sec:incl_tag_B_Dpi}

In order to validate the quality of the inclusive tag reconstruction and rule out possible biases introduced by the calibration method, we study the hadronic two-body decay of $B^+ \to \overline D^{\,0} \, \pi^+$ with $\overline D^{\,0} \to K^+ \, \pi^-$. Due to the absence of any neutrino in this decay, we are able to fully reconstruct the $B^+$ four-vector and boost the prompt $\pi^+$ into the \PBplus rest frame. Alternatively, we use the ROE, as outlined in the previous section, to reconstruct the very same information. Comparing the results from both allows us to determine if the calibration introduces potential biases and also to validate the signal resolution predicted in the simulation. In addition, we use this data set to test the validity of the continuum suppression and the data-driven continuum corrections outlined in Section~\ref{sec:data_anastrat}.

\begin{figure}[t!]
  \includegraphics[width=0.5\textwidth]{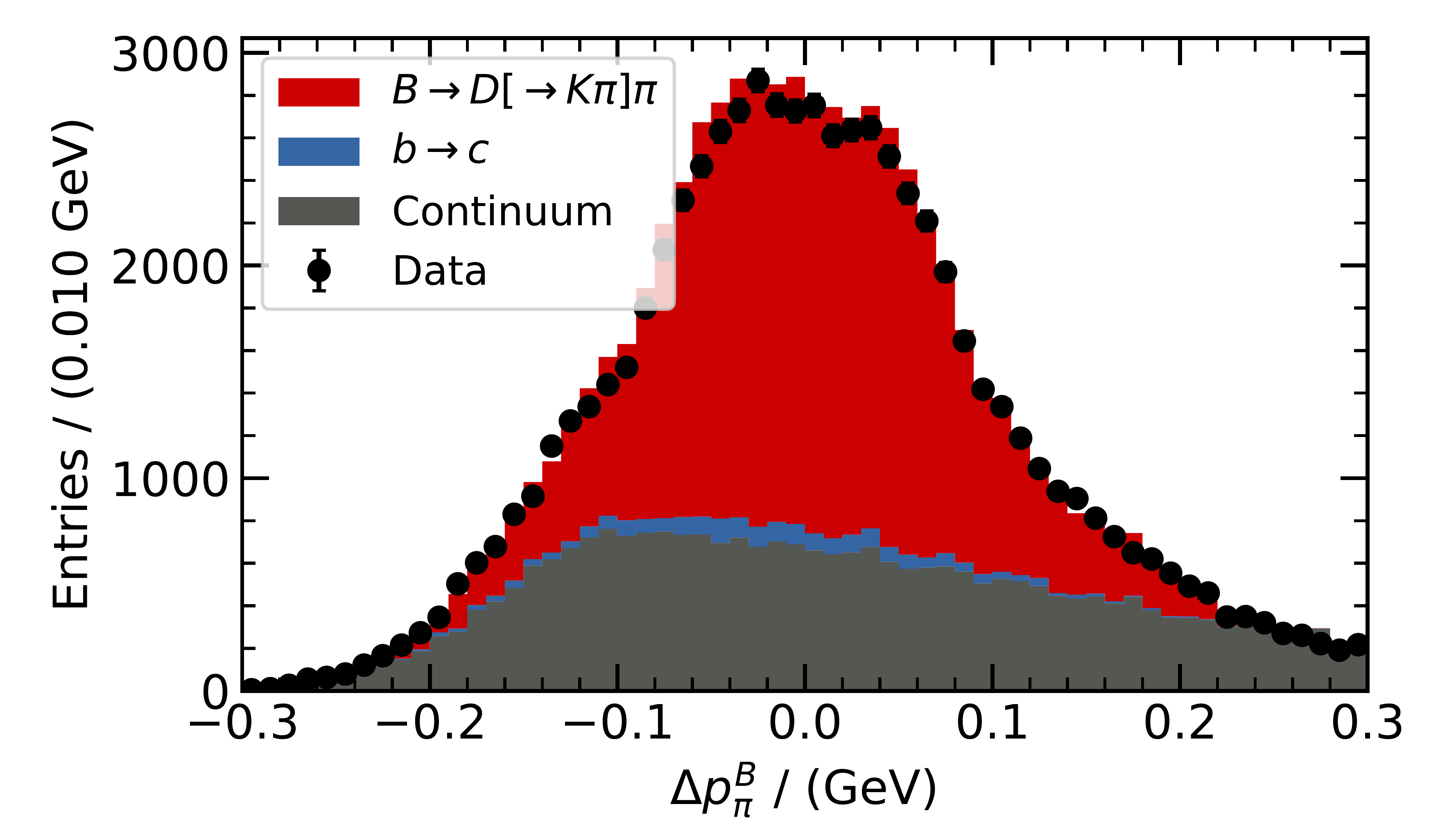}
  \includegraphics[width=0.5\textwidth]{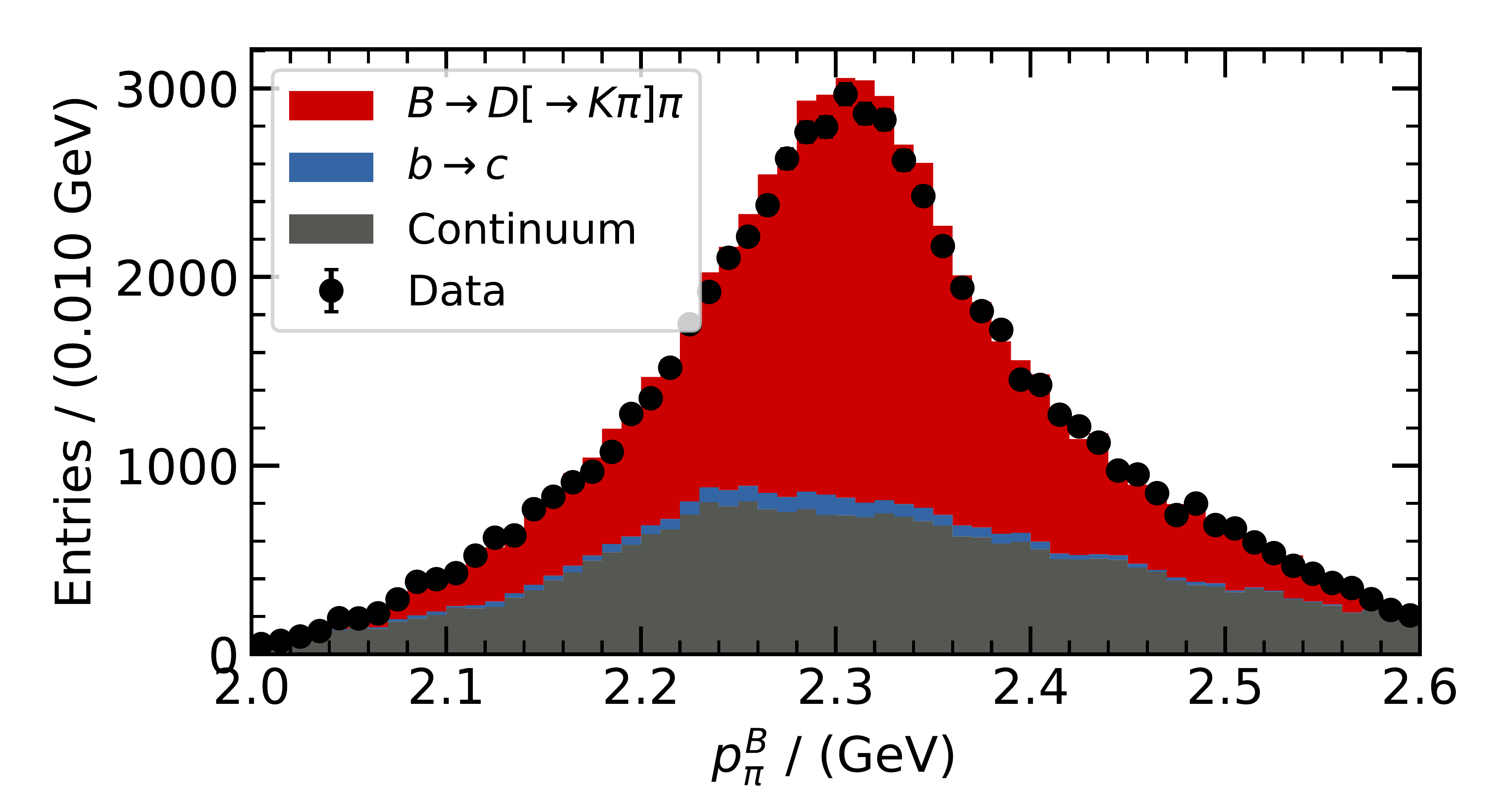}
  \includegraphics[width=0.5\textwidth]{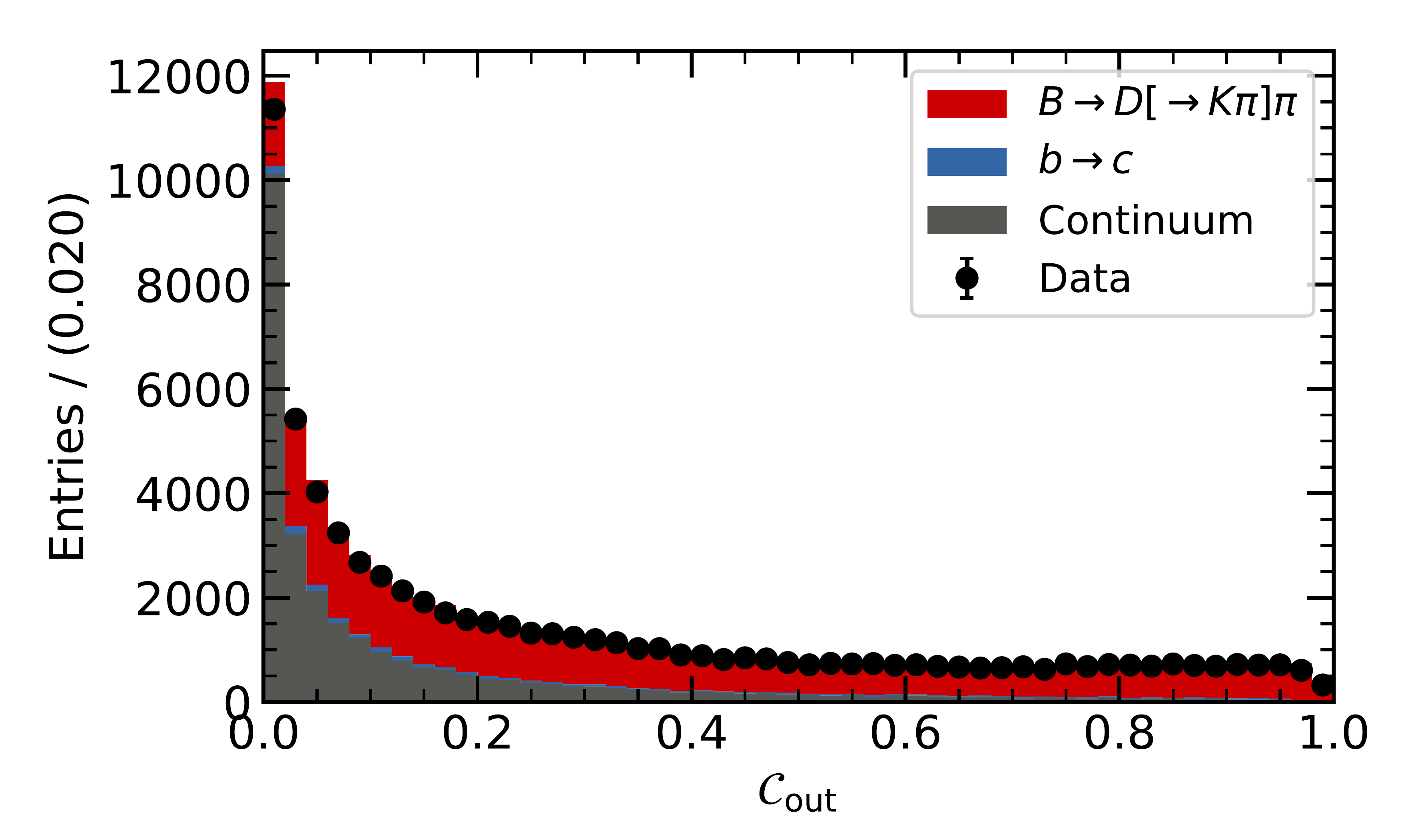}
\caption{
   The $p_\pi^B$ distribution and the residual $\Delta p_\pi^B$ for $ B^+ \to \overline D^{\,0} \, \pi^+$ decays with $\overline D^{\,0} \to K^+ \, \pi^-$ are shown in the reconstructed rest frame of the $B^+$ meson. The $p_\pi^B$ distribution is derived from the inclusive tag reconstruction method described in the text and the residual shows the difference with respect to using the full $B^+$ decay chain to determine the same information. In addition, the continuum classifier of simulated and reconstructed collision events are compared.
 } \label{fig:control_channel}
\end{figure}

We reconstruct the $ B^+ \to \overline D^{\,0} \, \pi^+$ with $\overline D^{\,0} \to K^+ \, \pi^-$ using the same impact parameter requirements used in the \bmunu analysis. For the prompt $\pi^+$ candidate we require a momentum of more than $2.1$ GeV in the c.m.\,frame. For the $\overline D^{\,0}$ decay product candidates a looser requirement is imposed, selecting charged tracks with a three-momentum of at least $0.3$ GeV in the laboratory frame. To identify the kaon and pion candidates, we use the particle identification methods described in Section~\ref{sec:data_anastrat}. To further suppress contributions from background processes we require that the reconstructed $\overline D^{\,0}$ mass is to be within 50 MeV of its expected value. Using the reconstructed four-vector of the $ B^+ \to \overline D^{\,0} \, \pi^+$ candidate we impose additional cuts to enhance the purity of the selected sample by using the beam-constrained mass and energy difference:
\begin{equation}
\begin{aligned}
m_{bc} & = \sqrt{ s/4 - \left( \mathbf{p}_{B^+}^{\ast} \right)^2 } \, > 5.2 \, \text{GeV}  \, , \\
| \Delta E | & =  | E_{B^+}^{\ast} - \sqrt{s}/2 | \quad < 0.2 \, \text{GeV}  \, .
\end{aligned}
\end{equation}
Here $ \mathbf{p}_{B^+}^{\ast} $ and $E_{\rm B^+}^{\ast} $ denote the reconstructed $B^+$ three-momentum and energy in the c.m. frame of the colliding $e^+ \, e^-$-pair, respectively. The inclusive tag is reconstructed in the same way as outlined in the previous section and Fig.~\ref{fig:control_channel} shows the reconstructed prompt $\pi^+$ absolute three-momentum $p_\pi^B$ after using the inclusive tag information to boost into the $B^+$ meson frame of rest. The simulated and reconstructed $ B^+ \to \overline D^{\,0} \, \pi^+$ decays show good agreement. Using the signal side information, we also reconstruct the residual $ \Delta p_\pi^B = p_\pi^B - p_\pi^{B_{\rm sig}} $, with $ p_\pi^{B_{\rm sig}}$ denoting the absolute three-momentum in the $B^+$ rest frame when reconstructed using the signal-side $B^+$ decay chain. The mean and variance of this distribution between simulated and reconstructed sample show good agreement and are compatible within their statistical uncertainties. We obtain a data-driven estimate for the inclusive tag resolution for  $p_\pi^B$ of $0.11$ GeV.

To validate the response of the multivariate classifier used to suppress continuum events, we remove the reconstructed $\overline D^{\,0} $ decay products from the signal side to emulate the \bmunu decay topology. Using the same BDT weights as for \bmunu we then recalculate the classifier output $C_{\rm out}$. Its distribution is shown in Fig.~\ref{fig:control_channel} and simulated and reconstructed events are in good agreement.  In Table~\ref{tab:eff_control} we further compare the selection efficiency denoted as $\epsilon$ between simulated and reconstructed events for the four signal selection categories of the \bmunu analysis. The efficiency is defined as the fraction of reconstructed candidates with $C_{\rm out} > 0.93$ or $0.98$, respectively, with respect to the total number of reconstructed candidates. The efficiency from simulated and reconstructed events are in agreement within their statistical uncertainty and we do not assign additional corrections or uncertainties to the \bmunu\ analysis in the following. 

 \begin{table}[tbh!]
\caption{Selection efficiencies of the category cuts defined in Table~\ref{tab:cats} of simulated and reconstructed data events. The quoted uncertainty is the statistical error.} \label{tab:eff_control}
\begin{tabular}{lccc}
\hline\hline
   Categories  & I-IV &  I+II &  III+IV \\ \hline
  $\epsilon^{\rm Data}$ & $0.030 \pm 0.001$ & $0.0047 \pm 0.0003$ & $0.024 \pm 0.001$ \\
  $\epsilon^{\rm MC}$ & $0.030 \pm 0.001$ & $0.0051 \pm 0.0003$ & $0.025 \pm 0.001$ \\
 \hline\hline
\end{tabular}
\end{table}

\section{Statistical analysis and limit setting procedure}\label{sec:stat}

In order to determine the \bmunu or \bmuN\ signal yield and to constrain all background yields, we perform a simultaneous binned likelihood fit to the $p_\mu^B$ spectra using the four event categories defined in Section~\ref{sec:astart}. The total likelihood function we consider has the form
\begin{equation} \label{eq:likelihood}
 \mathcal{L} = \prod_c \, \mathcal{L}_c \, \times \prod_k \, \mathcal{G}_k \, ,
\end{equation}
with the individual category likelihoods $\mathcal{L}_c$ and nuisance-parameter (NP) constraints $\mathcal{G}_k$. The product in Eq.~\ref{eq:likelihood} runs over all categories $c$ and fit components $k$, respectively. The role of the NP constraints is detailed in Section~\ref{sec:syst}. Each category likelihood  $\mathcal{L}_c$  is defined as the product of individual Poisson distributions $\mathcal{P}$,
\begin{equation}
   \mathcal{L}_c = \prod_i^{\rm bins} \, \mathcal{P}\left( n_i ; \nu_i \right)  \, ,
\end{equation}
with $n_i$ denoting the number of observed data events and $\nu_i$ the total number of expected events in a given bin $i$. We divide the muon momentum spectrum into 22 equal bins of 50 MeV, ranging over $p_\mu^B \in [2.2 ,3.3)$ GeV, and the number of expected events in a given bin, $\nu_i$ is estimated using simulated collision events. It is given by
\begin{equation}\label{eq:nui}
 \nu_i = \sum_k^{\rm processes} \, f_{ik} \, \eta_k  \, ,
\end{equation}
with $\eta_k$ the total number of events from a given process $k$ with the fraction $f_{ik}$ of such events being reconstructed in the bin $i$.

The likelihood Eq.~\ref{eq:likelihood} is numerically maximized to fit the value of four different components, $\eta_k$ using the observed events using the sequential least squares programming method implementation of Ref.~\cite{scipy}.

The four components we determine are:
\begin{enumerate}
 \item Signal \bmunu  Events.
 \item Background \bulnu\ Events; simulated as described in Section~\ref{sec:data_anastrat}.
 \item Background \bclnu\ Events; dominated by $B \to D^{(*)} \ell^+ \, \nu_\ell$ decays and simulated as described in Section~\ref{sec:data_anastrat}
 \item Background Continuum Events; dominated by $e^+ \, e^- \to q \bar q$ and $e^+ \, e^- \to \tau^+ \, \tau^-$ processes.
\end{enumerate}
Two additional background components, \bmunugam\, and other rare $b \to s$ processes, are constrained in the fit to the measurement of Ref.~\cite{Gelb:2018end} and world averages of Ref.~\cite{pdg:2018}.  Both mimic the signal shape and are allowed to vary in the fit within their corresponding experimental uncertainties. Further details on how this is implemented are found in Section~\ref{sec:syst}.

We construct confidence levels for the components using the profile likelihood ratio method. For a given component $\eta_k$ the ratio is
\begin{equation} \label{eq:test_stat}
  \Lambda(\eta_k) =  - 2 \ln \frac{ \mathcal{L}( \eta_k, \boldsymbol{\widehat \eta_{\eta_k}}, \boldsymbol{\widehat \theta_{\eta_k}}  ) }{  \mathcal{L}( \widehat  \eta_k,  \boldsymbol{ \widehat \eta}, \boldsymbol{ \widehat \theta}  )  } \, ,
\end{equation}
where $ \widehat  \eta_k$, $\boldsymbol{\widehat \eta}$, $\boldsymbol{ \widehat \theta} $ are the values of the component of interest, the remaining components, and a vector of nuisance parameters that unconditionally maximize the likelihood function, whereas the remaining components $\boldsymbol{\widehat \eta_{\eta_k}}$ and nuisance parameters $\boldsymbol{\widehat \theta_{\eta_k}} $ maximize the likelihood under the condition that the component of interest is kept fixed at a given value $\eta_k$. In the asymptotic limit, the test statistic Eq.~\ref{eq:test_stat} can be used to construct approximate confidence intervals (CI) through
\begin{equation}
  1 - \text{CI} = \int_{ \Lambda(\eta_k) }^{\infty} \, f_{\chi^2}(x; 1 \, \text{dof}) \, \text{d} x \, ,
\end{equation}
with $f_{\chi^2}(x; 1 \, \text{dof})$ denoting the $\chi^2$ distribution with a single degree of freedom. In the absence of a significant signal, we determine Frequentist and Bayesian limits. For the Frequentist one-sided (positive) limit, we modify our test statistic according to Ref.~\cite{Aad:2012an,Cowan:2010js} to
\begin{equation}\label{eq:imp_test_stat}
\begin{aligned}
  q_0(\eta_k) = \bigg\{ \begin{matrix} \Lambda(\eta_k) &\quad \eta_k \ge 0 \\
           - \Lambda(\eta_k)  & \quad \eta_k < 0 \end{matrix}  \quad ,
\end{aligned}
\end{equation}
to maximize our sensitivity. This test statistic is asymptotically distributed as
\begin{equation}
 f(q_0) = \frac{1}{2}  f_{\chi^2}(-q_0; 1 \, \text{dof}) + \frac{1}{2}  f_{\chi^2}(q_0; 1 \, \text{dof})
\end{equation}
and with an observed value $q_0^{\rm obs}$ we evaluate the (local) probability of an observed signal, $p_0$, as
\begin{equation}\label{eq:loc_prob}
 p_0 = \int_{q_0^{\rm obs}}^\infty \, f(q_0) \, \text{d} q_0 \, .
\end{equation}
For the Bayesian limit, we convert the likelihood Eq.~\ref{eq:likelihood} using a vector of observed event yields in the given bins of all categories  $\boldsymbol{n}$ (denoted as $\mathcal{L} = \mathcal{L}( \boldsymbol{n} | \eta_k)$ in the following) into a probability density function $\mathcal{F}$ of the parameter of interest $\eta_k$ using a flat prior $\pi( \eta_k)$ to exclude unphysical negative branching fractions. This $\mathcal{L} $ is numerically maximized for given values of the parameter of interest $\eta_k$, by floating the other components and nuisance parameters. The probability density function $\mathcal{F}$ is then given by
\begin{equation} \label{eq:Bayesian_PDF}
 \mathcal{F}( \eta_k | \boldsymbol{n}) = \frac{  \mathcal{L}( \boldsymbol{n} | \eta_k) \, \pi(\eta_k)  }{ \int_0^\infty  \mathcal{L}( \boldsymbol{n} | \eta_k) \, \pi(\eta_k) \, \text{d} \eta_k  } \, ,
\end{equation}
with the prior $\pi(\eta_k) = $ constant for $\eta_k \ge 0$ and zero otherwise.

To quote the significance over the background-only hypothesis for the search for \bmunu and \bmuN\  we adapt Eq.~\ref{eq:imp_test_stat} and set $\eta_k = 0$. For the search for a heavy sterile neutrino we do not account for the look-elsewhere effect.

We validate the fit procedure using ensembles of pseudoexperiments generated for different input branching fractions for \bmunu and \bmuN\ decays and observe no biases, undercoverage or overcoverage of CI.

Using a SM branching fraction of \mbox{$\mathcal{B}(\bmunu) = \left(4.3 \pm 0.8 \right) \times 10^{-7}$}, calculated assuming an average value of $\left| \Vub \right| = \left( 3.94 \pm0.36 \right) \times 10^{-3}$~\cite{pdg:2018} we construct Asimov data sets for all four categories. These are used to determine the median expected significance of our analysis. We find a value of $2.4^{+0.8}_{-0.9}$ standard deviations incorporating all systematic uncertainties and $2.6^{+1.0}_{-0.9}$ standard deviations if we only consider statistical uncertainties. The quoted uncertainties on the median expected significance correspond to the 68\% CL intervals. 

\section{Systematic uncertainties}\label{sec:syst}

There are several systematic uncertainties that affect the search for \bmunu and \bmuN. The most important uncertainty stems from the modeling of the dominant semileptonic  \bulnu\ background decays. As we determine the overall normalization of these decays directly from the measured collision events, we only need to evaluate shape uncertainties. The most important here stem from the modeling of the \bpilnu, \brholnu, and \bomegalnu\ form factors, the branching fractions for these processes, \betalnu, \betaplnu\, and inclusive \bulnu\ decays. The uncertainty of the non-resonant \bulnu\ contributions in the hybrid model approach is estimated by changing the underlying model from DFN to BLNP. In addition, the uncertainty on the DFN parameters $m_b^{1S}$ and $a$ are included in the shape uncertainty (see Section~\ref{sec:data_anastrat}). There is no sizable shape uncertainty contribution owing to either muon identification or track reconstruction. The second most important uncertainty for the reported results is from the shape of the continuum template: the off-resonance data sample, which was used to correct the simulated continuum events, introduces additional statistical uncertainties. We evaluate the size of these using a bootstrapping procedure. The \bclnu\ background near the kinematic endpoint for such decays is dominated by $B \to D \, \ell^+ \, \nu_\ell$ and $B \to D^* \, \ell^+ \, \nu_\ell$ decays. We evaluate the uncertainties in the used BGL form factors and their branching fractions for both channels. For the \bmunu signal, and the fixed backgrounds from \bmunugam\ and rare $b \to s$ processes, we also evaluate the impact on the efficiency of the lepton-identification uncertainties, the number of produced $B$ meson pairs in the Belle data set, and the overall tracking efficiency uncertainty. In addition, we propagate the experimental uncertainty on the used \bmunugam\ branching fraction. The rare $b \to s/d$ template is dominated by $B^+ \to K^0_L\, \pi^+$ events (which make up about 32\% of all selected events) and we assign an uncertainty on the measured branching fraction and the two next-most occurring decay channels, $B^+ \to K^+ \, \pi^0$ (5\%) and $B^0 \to \rho^+ \, \pi^-$ (4\%), in the template. The statistical uncertainty on the generated MC samples is also evaluated and taken into account. A full listing of the systematic uncertainties is found in Table~\ref{tab:systematic_uncertainties}.

\begin{table}[tbh!]
\caption{
	The fractional uncertainty on the extract \bmunu branching fraction are shown. For definitions of additive and multiplicative errors please see text. 
}
\label{tab:systematic_uncertainties}
\begin{tabular}{lc}
\hline\hline
  &  \multicolumn{1}{c}{Fractional uncertainty }  \\
  Source of uncertainty & \bmunu \\ 
\hline \\
 {\bf Additive uncertainties} \\
 \quad \bulnu modeling & 11{\,}\%  \\
 \qquad \bpilnu\ FFs & 4.8{\,}\%\\
 \qquad \brholnu\ FFs & 3.4{\,}\%\\
 \qquad \bomegalnu\ FFs & 3.0{\,}\%\\
 \qquad $\mathcal{B}(\bpilnu)$ & 3.4{\,}\%\\
 \qquad $\mathcal{B}(\brholnu)$ & 3.2{\,}\%\\
 \qquad $\mathcal{B}(\bomegalnu)$ & 3.1{\,}\% \\
 \qquad $\mathcal{B}(B \to X_u \, \ell^+ \, \nu)$ & 4.0{\,}\%\\
 \qquad DFN parameters & 4.0{\,}\%\\
 \qquad Hybrid model & 4.2{\,}\%\\
 \qquad MC statistics & 2.6{\,}\%\\
 \quad Continuum modeling & 13{\,}\%\\
 \qquad shape correction & 4.1{\,}\%\\
 \qquad MC statistics & 12.2{\,}\%\\
  \quad \bclnu modeling & 2.5{\,}\% \\
 \quad  \bmunu MC statistics & 1.0{\,}\% \\
 \quad $\mathcal{B}(b \to s)$ processes & 1.0{\,}\%  \\ 
  \quad $\mathcal{B}(\bmunugam)$ & 0.1{\,}\% \\
  {\bf Multiplicative uncertainties}  \\
 \quad $\mathcal{L}_{\rm LID}$ efficiency & 2.0{\,}\%  \\
 \quad $N_{\rm B \bar B}$ & 1.4{\,}\%  \\
 \quad Tracking efficiency & 0.3{\,}\%  \\
 \\
 {\bf Total syst. uncertainty} & {\bf 17{\,}\%} \\
\hline\hline
\end{tabular}
\end{table}

The effect of systematic uncertainties is directly incorporated into the likelihood function. For this we introduce a vector of NPs, $\boldsymbol{\theta}_k$, for each fit template $k$. Each vector element represents one bin of the fitted $p_\mu^B$ spectrum in all four categories. The NPs are constrained in the likelihood Eq.~\ref{eq:likelihood} using multivariate Gaussian distributions $\mathcal{G}_k = \mathcal{G}_k( \boldsymbol{0}; \boldsymbol{\theta}_k, \Sigma_k ) $, with $\Sigma_k$ denoting the systematic covariance matrix for a given template $k$. The systematic covariance is constructed from the sum over all possible uncertainty sources affecting a template $k$, i.e.
\begin{equation}
 \Sigma_k = \sum_{s}^{\text{error sources}} \Sigma_{ks} \, ,
\end{equation}
with $\Sigma_{ks} $ the covariance matrix of error source $s$ which depends on an uncertainty vector $\boldsymbol{\sigma_{ks}}$. The vector elements of $\boldsymbol{\sigma_{ks}}$ represent the absolute error in bins of $p_\mu^B$ of fit template $k$ across the four event categories. We treat uncertainties from the same error source either as fully correlated, or, for MC or other statistical uncertainties as uncorrelated, such that $\Sigma_{ks}  = \boldsymbol{\sigma_{ks}} \otimes \boldsymbol{\sigma_{ks}}$ or $\Sigma_{ks} = \text{Diag}\left( \boldsymbol{\sigma_{ks}}^2 \right)$. The impact of nuisance parameters is included in Eq.~\ref{eq:nui} as follows. First, the fractions $f_{ik}$ for all templates are rewritten as
\begin{equation}
 f_{ik} = \frac{ \eta_{ik}^{\rm MC} }{ \sum_j \eta_{jk}^{\rm MC} } \to  \frac{ \eta_{ik}^{\rm MC} \left( 1 + \theta_{ik} \right) }{ \sum_j \eta_{jk}^{\rm MC} \left( 1 + \theta_{jk} \right)  },
\end{equation}
to take into account shape uncertainties. These uncertainties are listed as `Additive uncertainties' in Table~\ref{tab:systematic_uncertainties}. Here $\theta_{ik}$ represents the NP vector element of bin $i$ and $\eta_{ik}^{\rm MC}$ the expected number of events in the same bin for event type $k$ as estimated from the simulation. Note that this notation absorbs the size of the absolute error into the definition of the NP. Second, we include for the \bmunu signal template and fixed background templates overall efficiency and luminosity related uncertainties: this is achieved by rewriting the relevant fractions as
\begin{equation}
  \eta_k \to \eta_k \left( 1 + \theta_{ks} \right),
\end{equation}
with $\theta_{ks}$ the NP parameterizing the uncertainty in question. The uncertainty sources treated this way include the overall lepton identification and track reconstruction efficiency uncertainty and the uncertainty on the number of $B$ meson pairs produced in the full Belle data set and are labeled as `Multiplicative uncertainties' in Table~\ref{tab:systematic_uncertainties}. For the fixed background templates the corresponding uncertainties from branching fractions are also included this way.

\section{\bulnu\ and Off-resonance control region}\label{sec:bulnu}

\begin{figure}[t!]
  \includegraphics[width=0.45\textwidth]{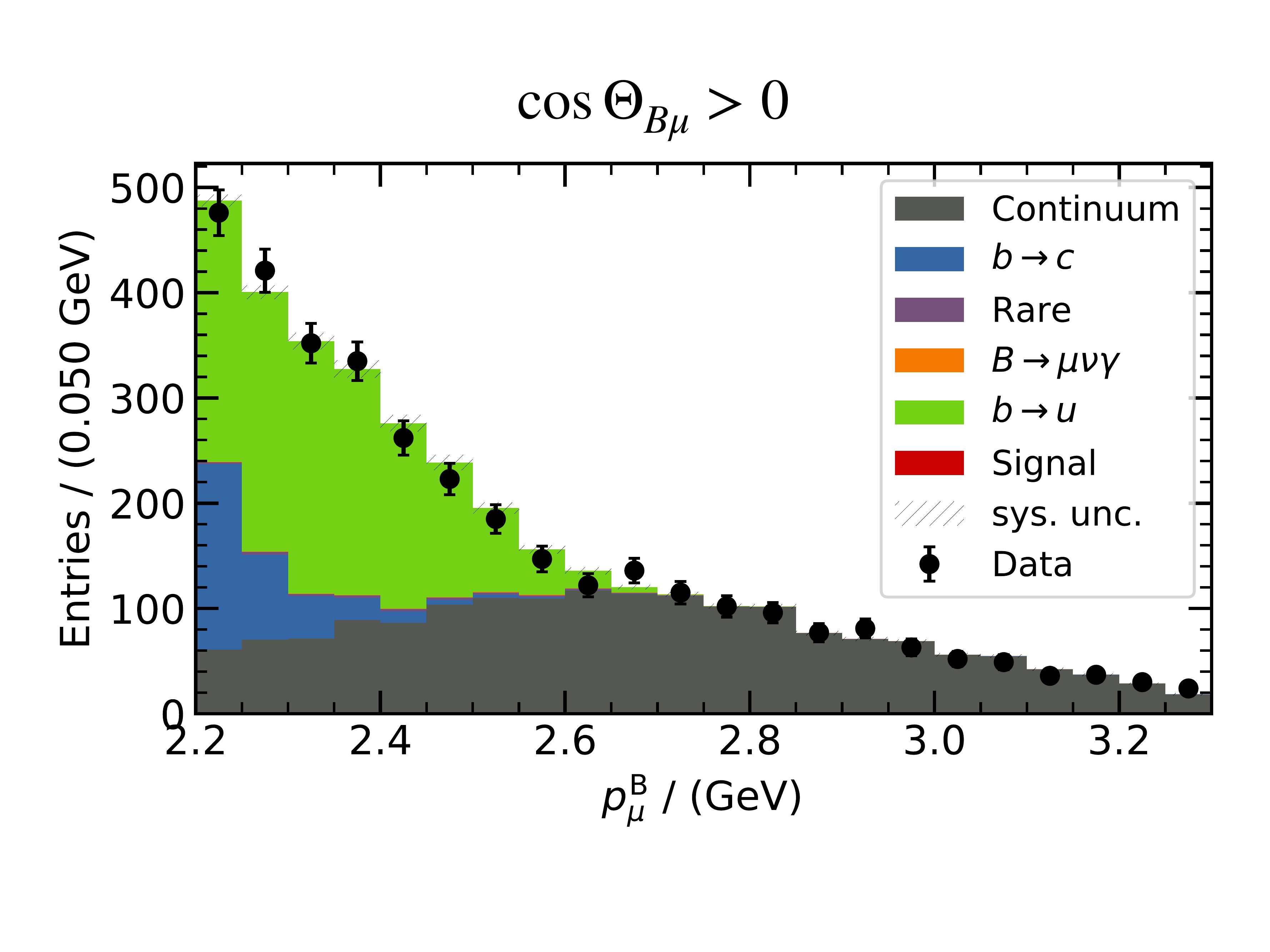} \hfil \\  \vspace{-5ex}
  \includegraphics[width=0.45\textwidth]{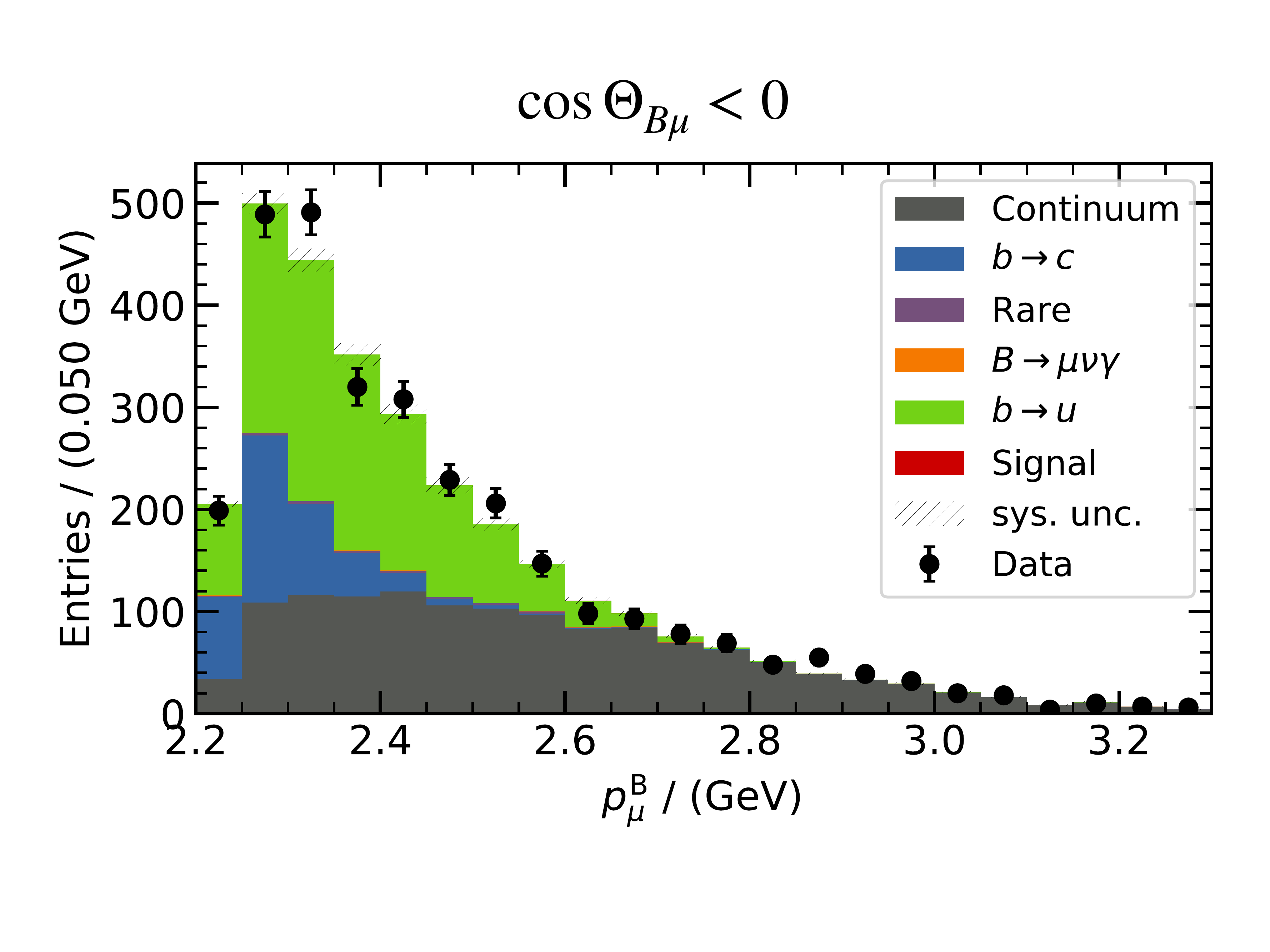} \\
\caption{
   The \bulnu\ control region fit is shown.
 }
\label{fig:bulnu_categories_postfit}
\end{figure}

To test the simulation of the crucial semileptonic \bulnu\ background, we construct a signal-depleted region with moderate continuum contamination. This is achieved by selecting events with continuum suppression classifier values of $C_{\rm out} \in [0.9,0.93)$. In this sample, the region of high muon momentum $p_\mu^B$ is used to test the validity of the continuum description and the region with a muon momentum between 2.2 and 2.6 GeV is dominated by semileptonic \bulnu\ and \bclnu\ decays. To also test the modeling of both backgrounds with respect to the employed signal categorization exploiting the angle between the muon and the signal $B$ meson, we further split the selected events using $\cos \Theta_{B\mu} > 0$ and $\cos \Theta_{B\mu} < 0$. The full likelihood fit procedure including all systematic uncertainties detailed in Sections~\ref{sec:stat} and \ref{sec:syst} is then carried out. Figure~\ref{fig:bulnu_categories_postfit} depicts the fit result: the individual contributions are shown as histograms and the recorded collision events are displayed as data points. The size of the systematic uncertainties is shown on the histograms as a hatched band. In the fit the signal \bmunu yield was fixed to the SM expectation and in both categories we expect about 15 \bmunu events. Both the \bulnu\ and \bclnu, and continuum dominated regions are described well by the fit templates. Assuming that for most bins the statistical uncertainty is approximately Gaussian, we calculate a $\chi^2$ of $30.4$ over 41 degrees of freedom by comparing predicted and observed yields in each bin and by taking into account the full systematic uncertainties. This approximation is justified for most of the $p_\mu^B$ region, but breaks down for the high momentum bins due to low statistics. The value still gives an indication that the fit model is able to describe the observed data well. We also carry out a fit in which the  \bmunu signal template is allowed to float: we determine a value $-37 \pm 61$ events, which is compatible with the SM expectation.

\begin{figure*}[th!]
  \includegraphics[width=0.45\textwidth]{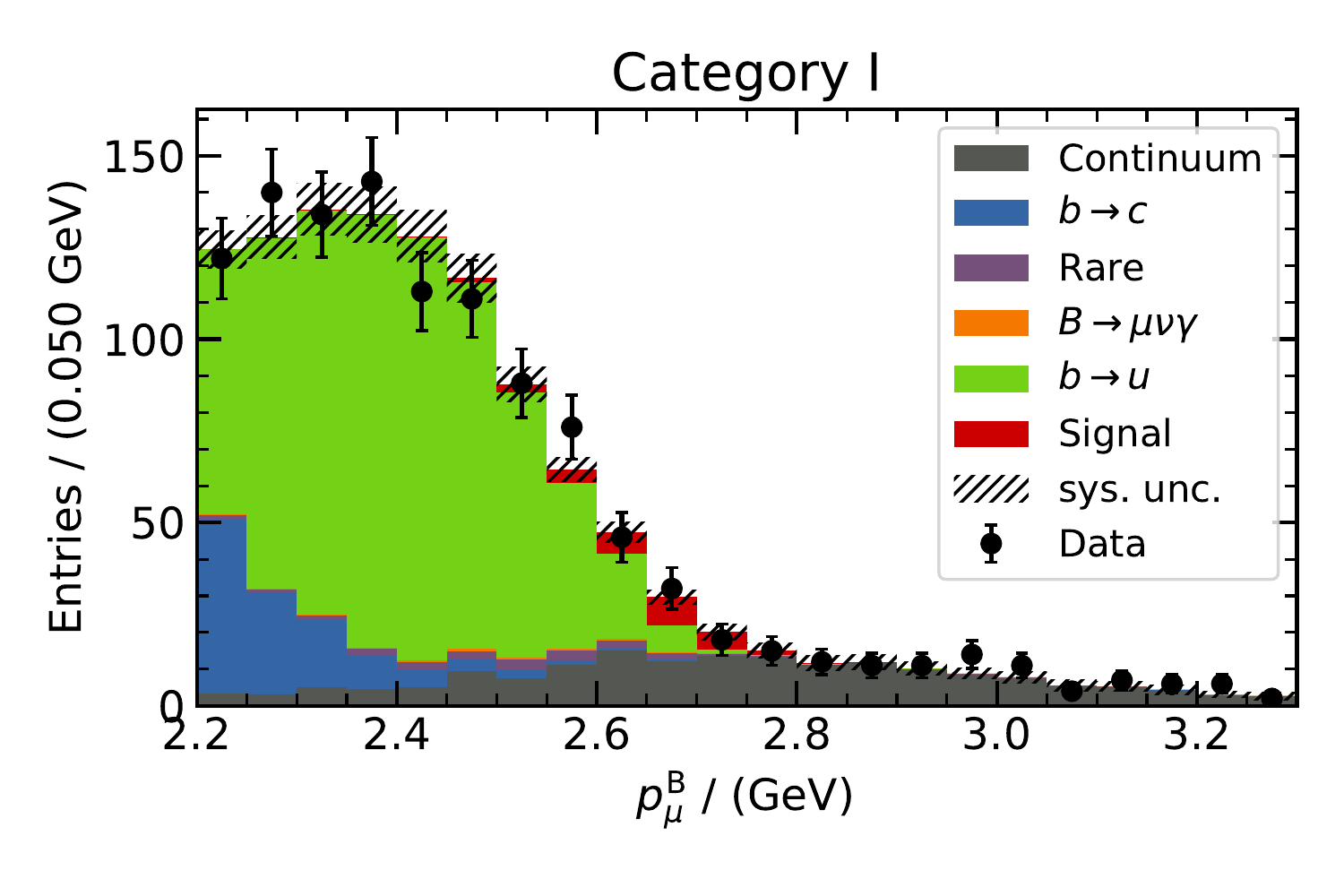} \hfil
  \includegraphics[width=0.45\textwidth]{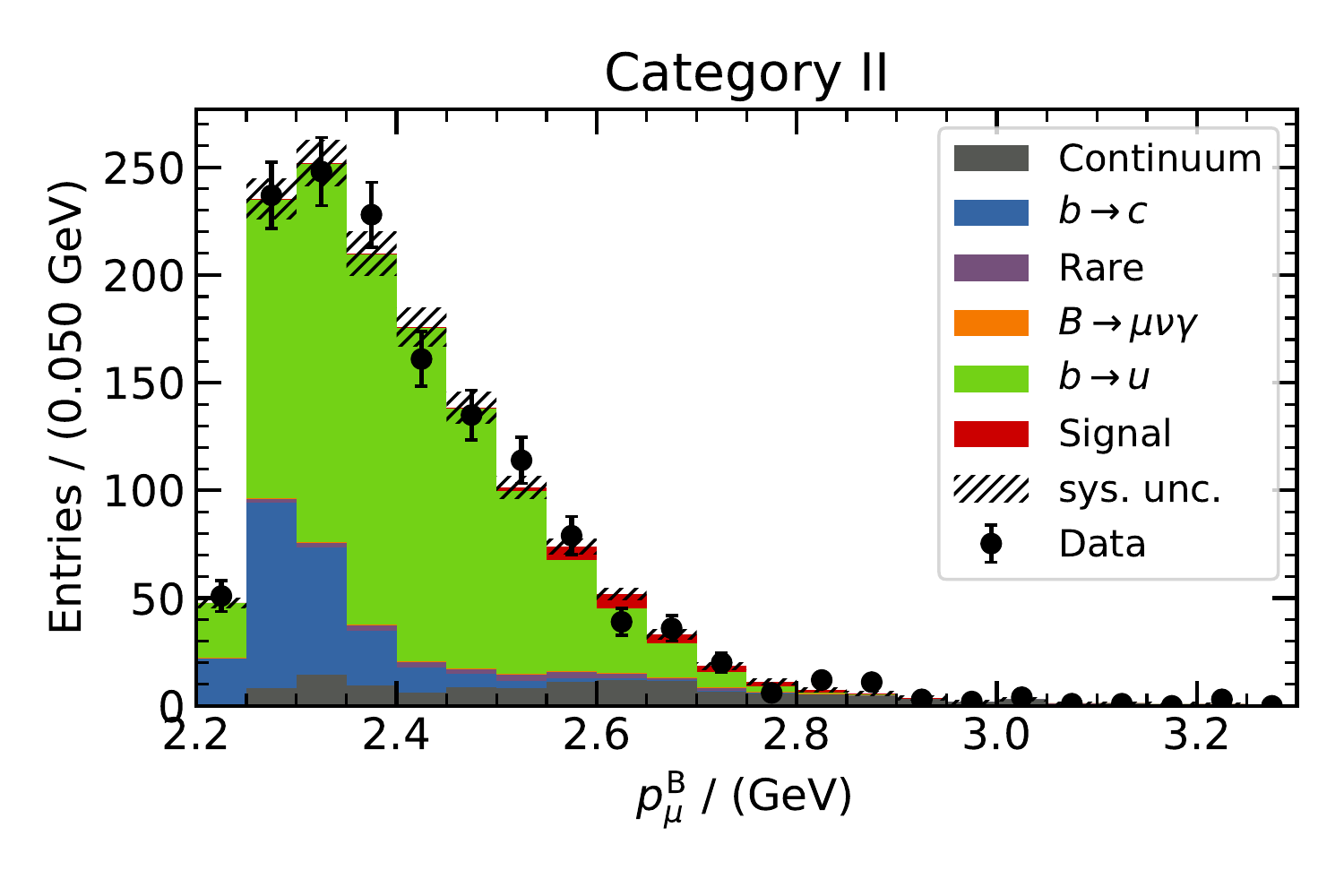} \\
  \vspace{1ex}
  \includegraphics[width=0.45\textwidth]{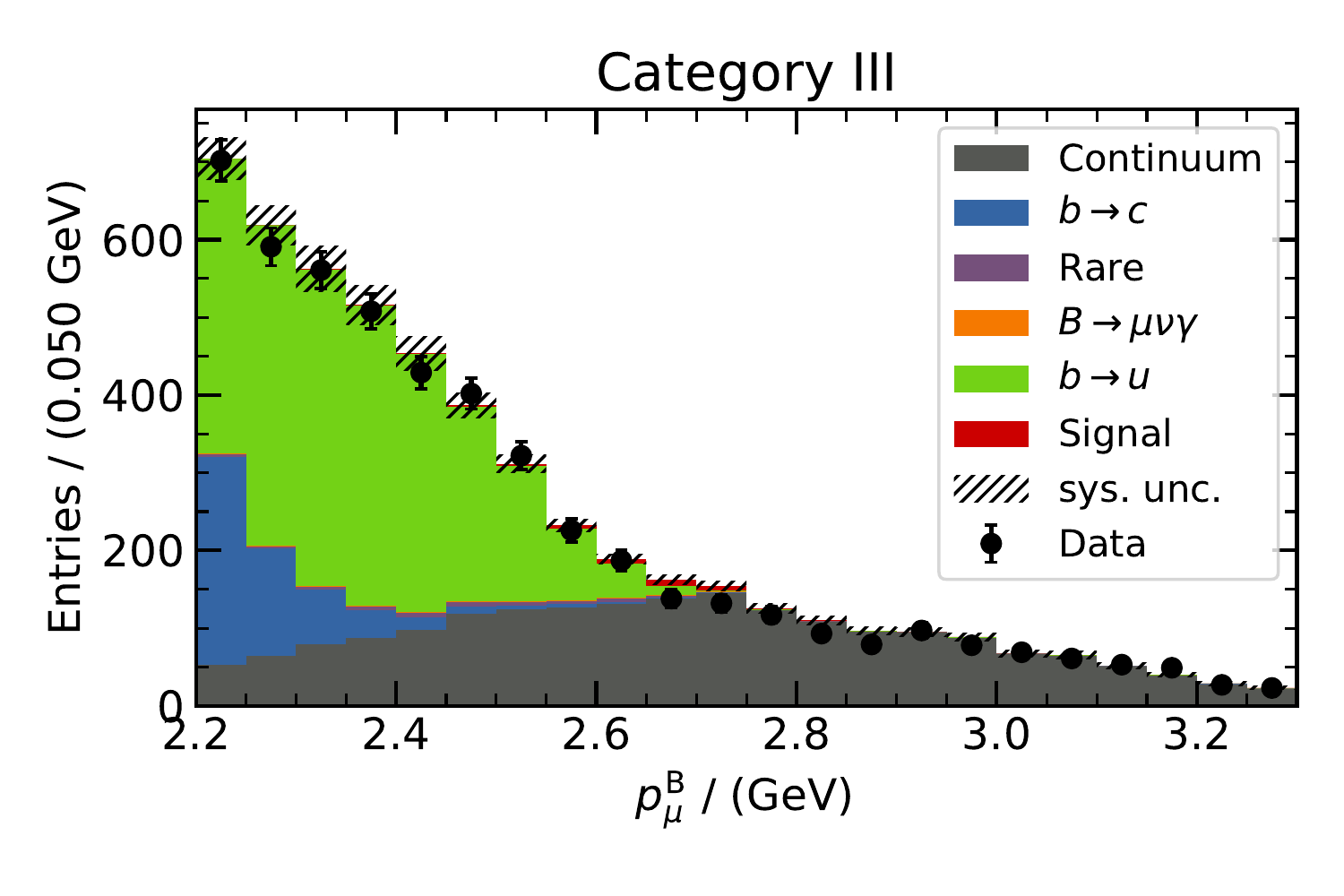} \hfil
  \includegraphics[width=0.45\textwidth]{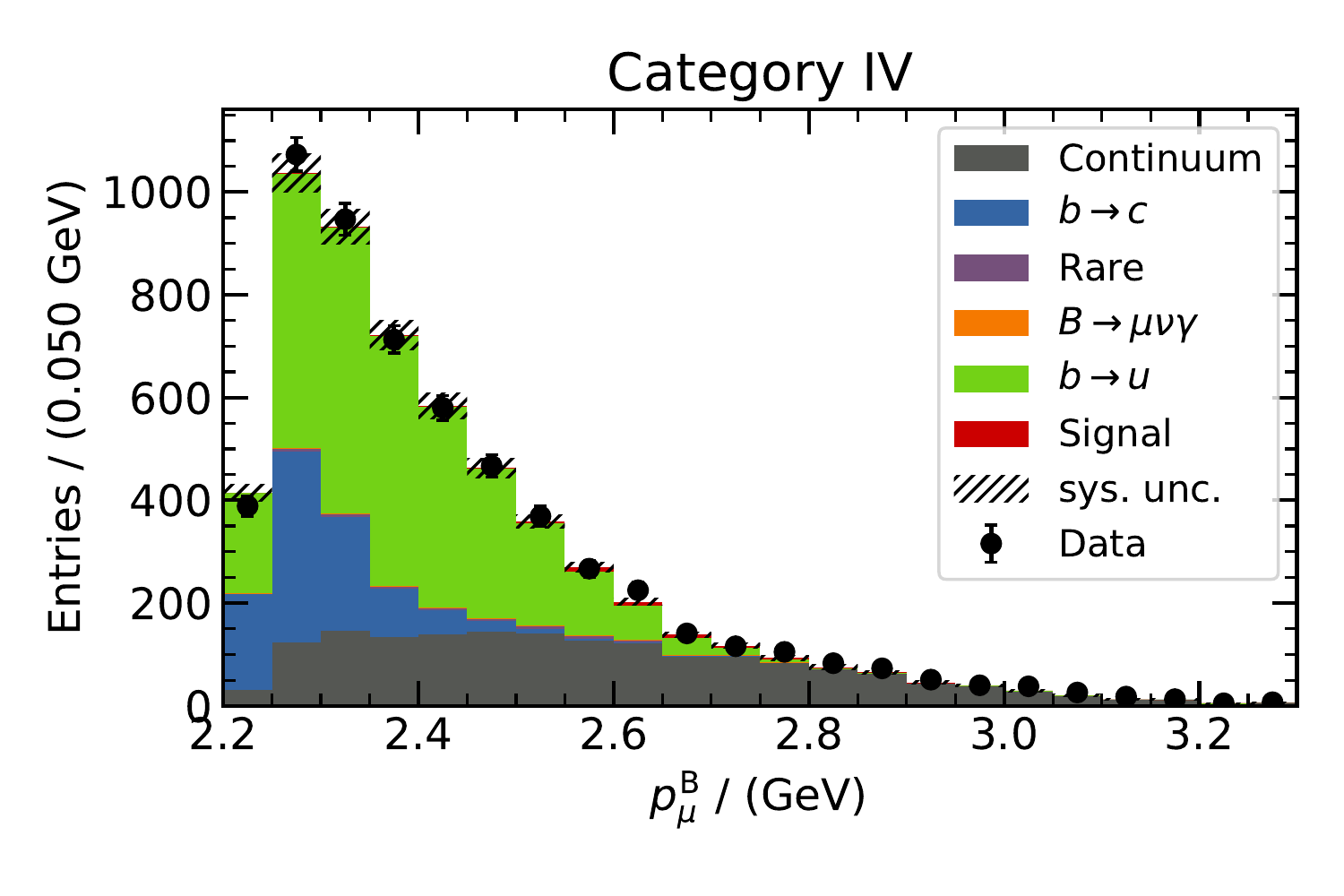} \\
\caption{
   The fitted distribution of $p_\mu^B$ for the four signal categories described in the text. The signal and background templates are shown as histograms and the recorded collision events as data points with uncertainties. The systematic uncertainties on the signal and background templates are shown as a hatched band.
 }
\label{fig:categories_postfit}
\end{figure*}

For the inclusive \bulnu\ branching fraction in which the signal template is kept fixed at its SM expectation, we find
\begin{equation}
 \mathcal{B}(B \to X_u \, \ell^+ \, \nu) = \left(2.04 \pm 0.10 \right) \times 10^{-3} \, ,
\end{equation}
where the uncertainty corresponds to the statistical error. The central value is compatible with the world average of Ref.~\cite{pdg:2018}, \mbox{$\mathcal{B}(B \to X_u \, \ell^+ \, \nu) = \left(2.13 \pm 0.31 \right) \times 10^{-3}$}. Note that Ref.~\cite{pdg:2018} inflated the quoted uncertainty to account for incompatibilities between the measurements used in the average.

We also apply the signal continuum classifier selection of $C_{\rm out} \in [0.93,1)$ on the recorded off-resonance data. With these events we carry out a two-component fit, determining the yields of \bmunu signal and continuum events. This allows us to determine whether the classifier selection could cause a sculpting of the background shape, which in turn would result in a spurious signal. The low number of events passing the selection does not allow further categorization of the events using angular information as only 39 off-resonance events pass the selection. We fit $37 \pm 10$ background events and $2 \pm 7$ signal events.

\section{Results}\label{sec:res}

\begin{figure}[t!]
  \includegraphics[width=0.45\textwidth]{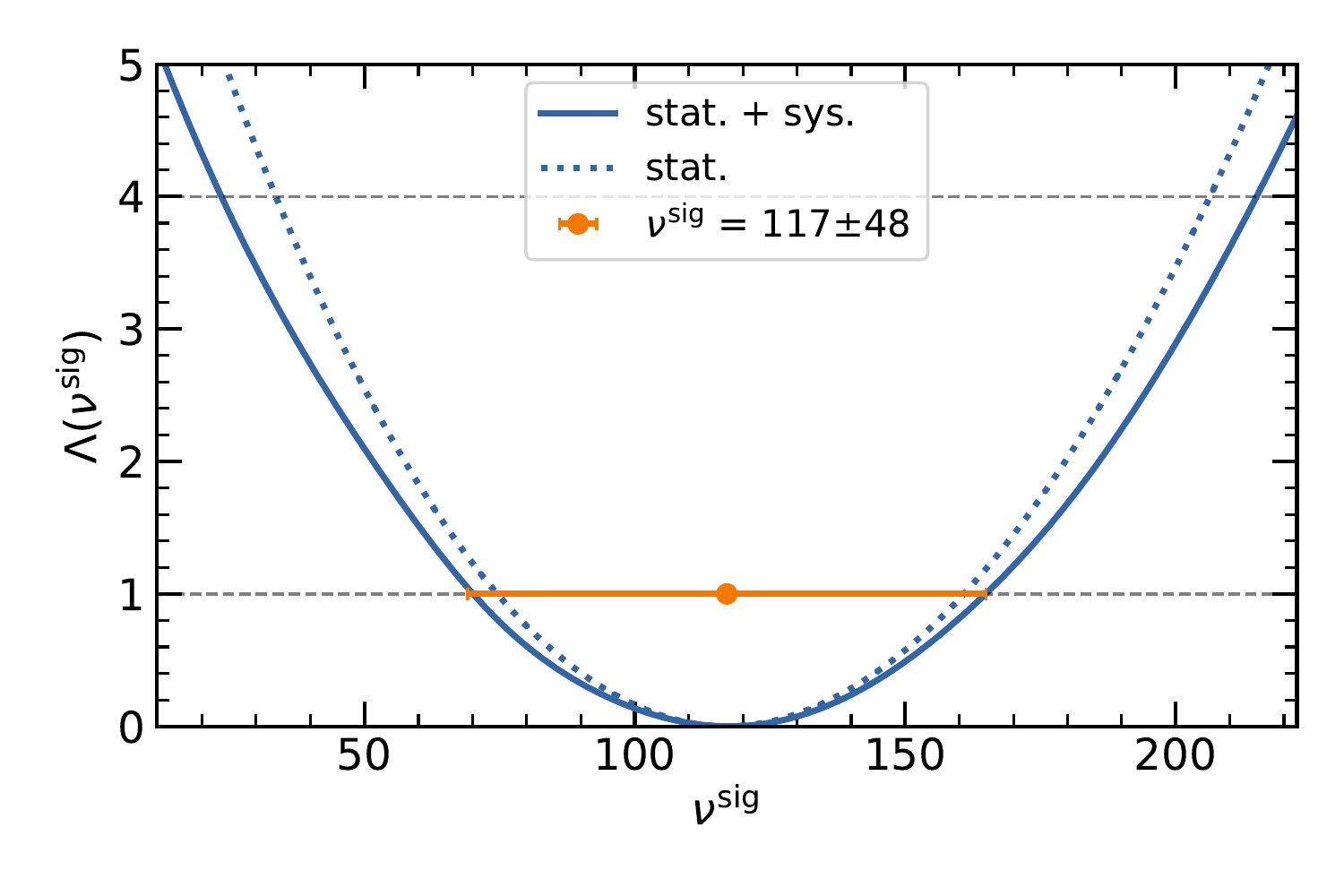}
\caption{
   The likelihood ratio contour $\Lambda(\nu_{\text{sig}})$ as a function of the number of \bmunu signal events is shown: the dotted curve shows the contour incorporating only the statistical uncertainty with all systematic nuisance parameters fixed at their best-fit value. The solid curve shows full likelihood contour including all systematic and statistical uncertainties. The orange data point and errors shows the determined best-fit value and the 1\,$\sigma$ (statistical + systematic) uncertainty.
 }
\label{fig:LL_scan}
\end{figure}

In Fig.~\ref{fig:categories_postfit} the muon momentum spectrum in the $B$ rest frame $p_\mu^B$ for the four signal categories is shown. The selected data events were used to maximize the likelihood Eq.~\ref{eq:likelihood}: in total $4 \times 22$ bins with $4 \times 132$ NPs parameterizing systematic uncertainties are determined. In Appendix~\ref{app:NPs} a full breakdown of the NP pulls is given. The recorded collision data are shown as data points and the fitted \bmunu signal and background components are displayed as colored histograms. The size of the systematic uncertainties is  shown on the histograms as a hatched band. We observe for the \bmunu branching fraction a value of
\begin{equation} \label{eq;meas_res}
 \bfRes \, ,
\end{equation}
with the first uncertainty denoting the statistical error and the second is from systematics. Fig.~\ref{fig:LL_scan} shows the profile likelihood ratio $\Lambda(\nu_{\rm sig})$ (cf. Eq.~\ref{eq:test_stat}). Assuming that all bins are described with approximately Gaussian uncertainty and including systematics with their full covariance, we calculate a $\chi^2$ value of 58.8 with 84 degrees of freedom using the predicted and observed bin values. The observed significance over the background-only hypothesis using the one-sided test statistics Eq.~\ref{eq:imp_test_stat} is $2.8$ standard deviations. This is in agreement with the median SM expectation of $2.4^{+0.8}_{-0.9}$ standard deviations, cf. Section~\ref{sec:stat}.

From the observed branching fraction we determine in combination with the $B$ meson decay constant $f_B$ a value for the CKM matrix element $\left| \Vub \right|$. Using  $f_B = 184 \pm 4 $ MeV~\cite{Aoki2017} we find
\begin{equation}
 \left| \Vub \right| = \left( 4.4^{+0.8}_{-0.9} \pm 0.4 \pm 0.1 \right) \times 10^{-3} \, ,
\end{equation}
where the first uncertainty is the statistical error, the second from systematics and the third from theory.
This value is compatible with both exclusive and inclusive measurements of $ \left| \Vub \right|$~\cite{pdg:2018}.

\begin{figure*}[t!]
  \vspace{2ex}
  \includegraphics[width=0.8\textwidth]{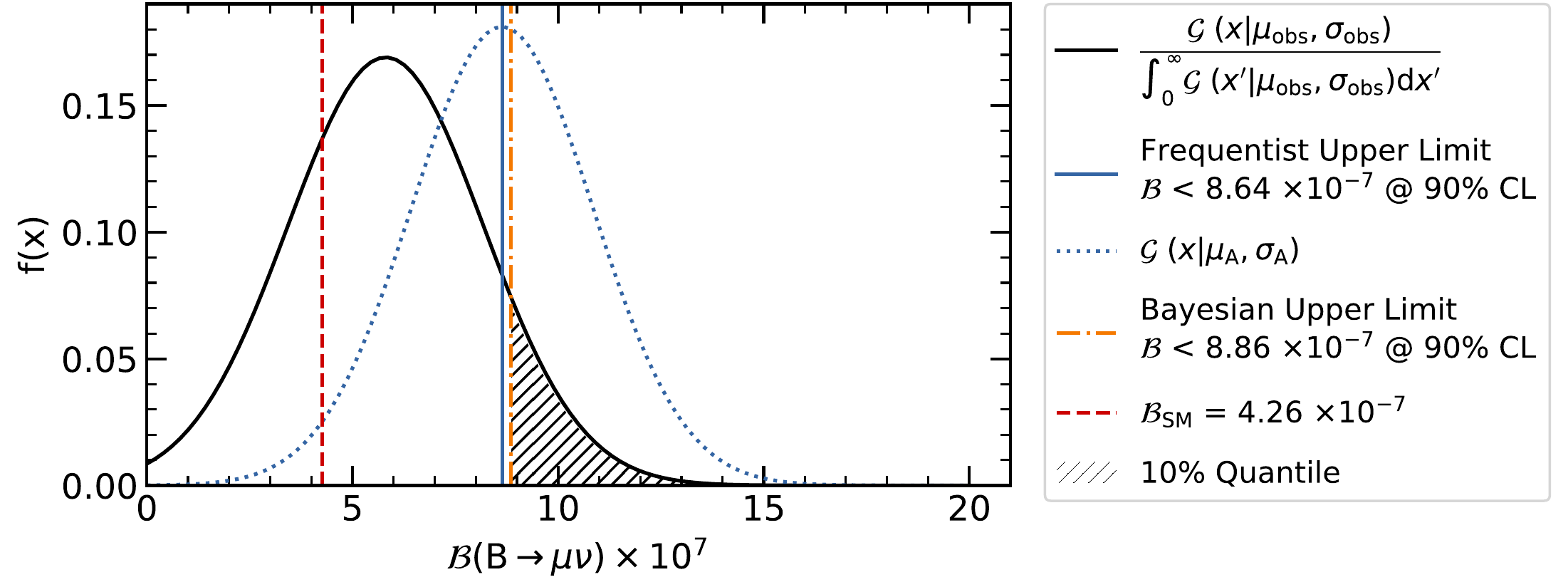}
\caption{
   The observed Bayesian (yellow dash-dotted) and Frequentist (blue) upper limits at 90\% CL are shown, along with the SM expectation of the \bmunu branching fraction and the Bayesian and Frequentist PDFs.
 }
\label{fig:limit}
\end{figure*}

Due to the low significance of the observed \bmunu signal, we calculate Bayesian and Frequentist upper limits of the branching fraction. We convert the likelihood into a Bayesian probability density function (PDF) using the procedure detailed in Section~\ref{sec:stat} and Eq.~\ref{eq:Bayesian_PDF}: Figure~\ref{fig:limit} shows the one-dimensional PDF, which was obtained using a flat prior in the partial branching fraction. The resulting Bayesian upper limit for \bmunu at 90\% confidence level (CL) is
\begin{equation}
  \mathcal{B}(\bmunu) < 8.9 \times 10^{-7} \, \text{at 90\% CL} \, .
\end{equation}
The Frequentist upper limit is determined using fits to ensembles of Asimov data sets with NPs shifted to the observed best fit values.  Figure~\ref{fig:limit} shows the corresponding Frequentist likelihood, for convenience also converted into a PDF (blue dotted line) and the resulting upper limit at 90\% CL is
\begin{equation}
  \mathcal{B}(\bmunu) < 8.6 \times 10^{-7} \, \text{at 90\% CL} \, .
\end{equation}

\begin{figure}[thb!]
	\includegraphics[width=0.45\textwidth]{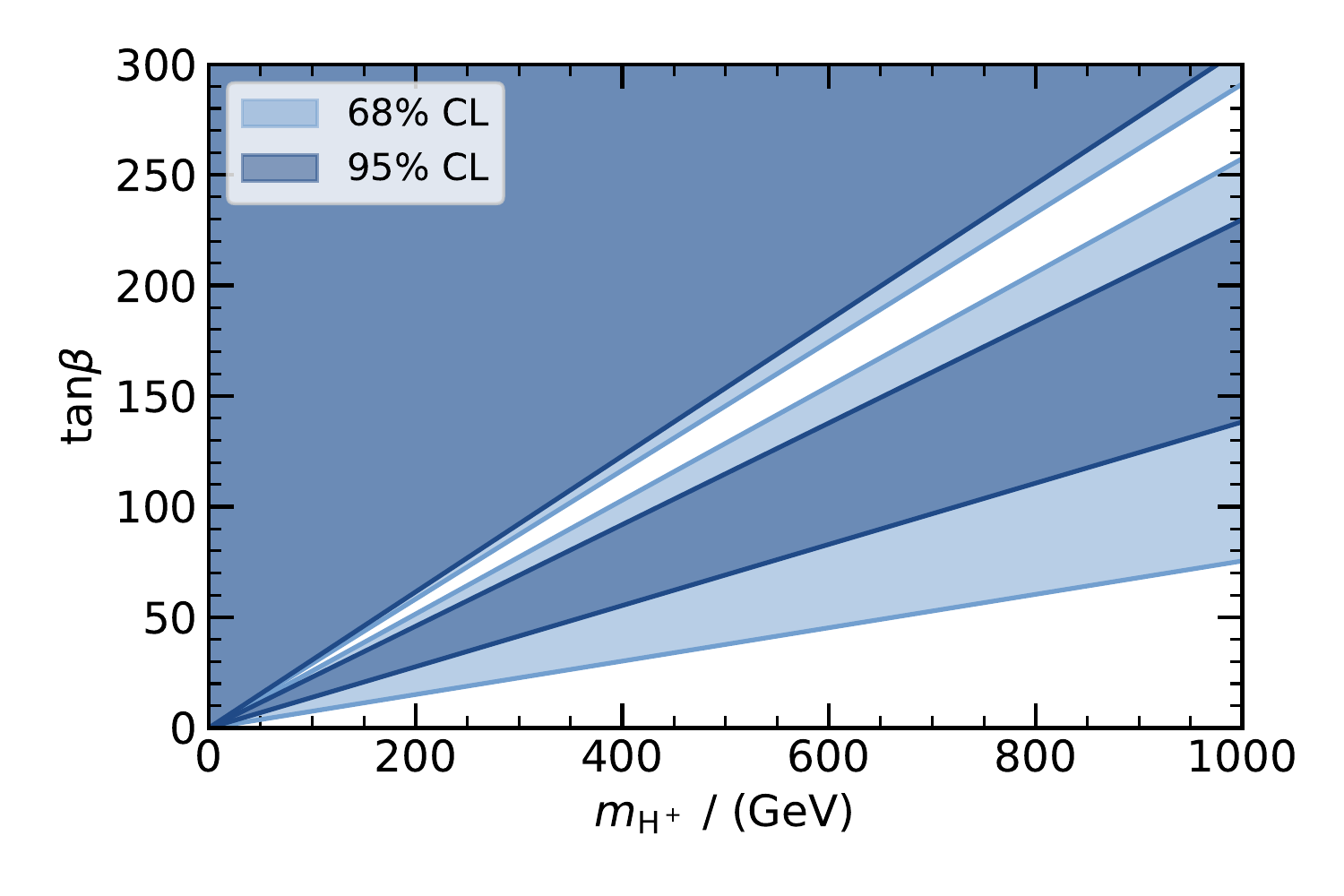}  \\
	\includegraphics[width=0.45\textwidth]{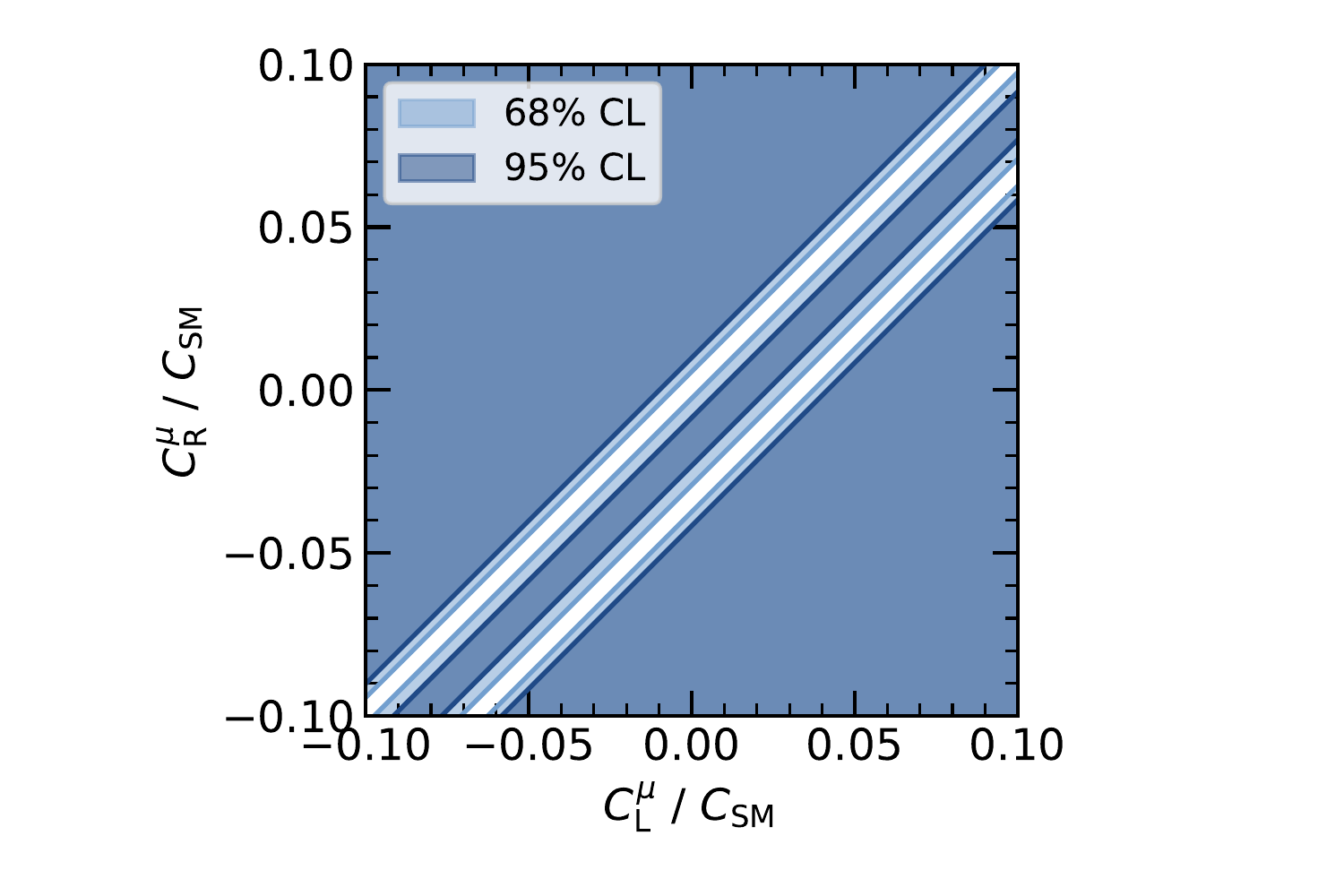}
	\caption{
		The 68\% and 95\% CL excluded model parameter space for the 2HDM type II ($\tan \beta$, $m_{H^+}$) and type III ($C_\mathrm{L}^{\Pmu}$, $C_\mathrm{R}^{\Pmu}$) are shown. The coefficients $C_\mathrm{L}^{\Pmu}$ and $C_\mathrm{R}^{\Pmu}$ are assumed to be real.
	}
	\label{fig:type2_type3}
\end{figure}

The observed branching fraction is used to constrain the allowed parameter space of the two-Higgs-doublet model (2HDM) of type II and type III. In these models the presence of charged Higgs bosons as a new mediator with specific couplings would modify the observed branching fraction, cf. Fig.~\ref{fig:btomunu_feynman}. The effect of the charged Higgs boson in the type II model is included in the expected \bmunu branching fraction by modifying Eq.~\ref{eq:BF_SM} according to Ref.~\cite{Hou:1992sy} to
\begin{equation}
\mathcal{B}( \bmunu) = \mathcal{B}^{\rm SM} \times \left( 1 - \frac{m_B^2 \, \tan^2 \beta}{m_{H^+}^2} \right)^2 \, ,
\end{equation}
with $\mathcal{B}^{\rm SM} $ denoting the SM branching fraction, $\tan \beta$ being the ratio of the vacuum expectation values of the two Higgs fields in the model, and $m_{H^+}$ the mass of the charged Higgs boson. The type III model further generalizes the couplings to \cite{Chen:2018hqy,Crivellin:2012ye}
\begin{equation}
\mathcal{B}( \bmunu) = \mathcal{B}^{\rm SM} \times \left| 1 + \frac{m_B^2}{m_b \, m_\mu} \left( \frac{C_\mathrm{R}^{\Pmu}}{C_\mathrm{SM}} - \frac{C_\mathrm{L}^{\Pmu}}{C_\mathrm{SM}} \right) \right|^2 \, ,
\end{equation}
with $m_b$ denoting the $b$ quark mass and the $C_{R/L}^{\Pmu}$ are the coefficients encoding the new physics contribution. Figure~\ref{fig:type2_type3} shows the allowed and excluded parameter regions at 68\% (light blue) and 95\% (dark blue) CL as calculated using the observed branching fraction Eq.~\ref{eq;meas_res} and by constructing a $\chi^2$ test. For the SM branching fraction prediction we use
\mbox{$\mathcal{B}^{\rm SM} = \left(4.3 \pm 0.8 \right) \times 10^{-7}$} calculated assuming an average value of $\left| \Vub \right| = \left( 3.94 \pm0.36) \right) \times 10^{-3}$ from Ref.~\cite{pdg:2018}. Due to the explicit lepton mass dependence in the type III model, the constructed bounds on $C_\mathrm{L/R}^{\Pmu}$ are more precise than any existing limits on $C_\mathrm{L/R}^{\Ptau}$ based on results from studying $\btaunu$ decays.

\begin{figure*}[thb!]
  \includegraphics[width=0.49\textwidth]{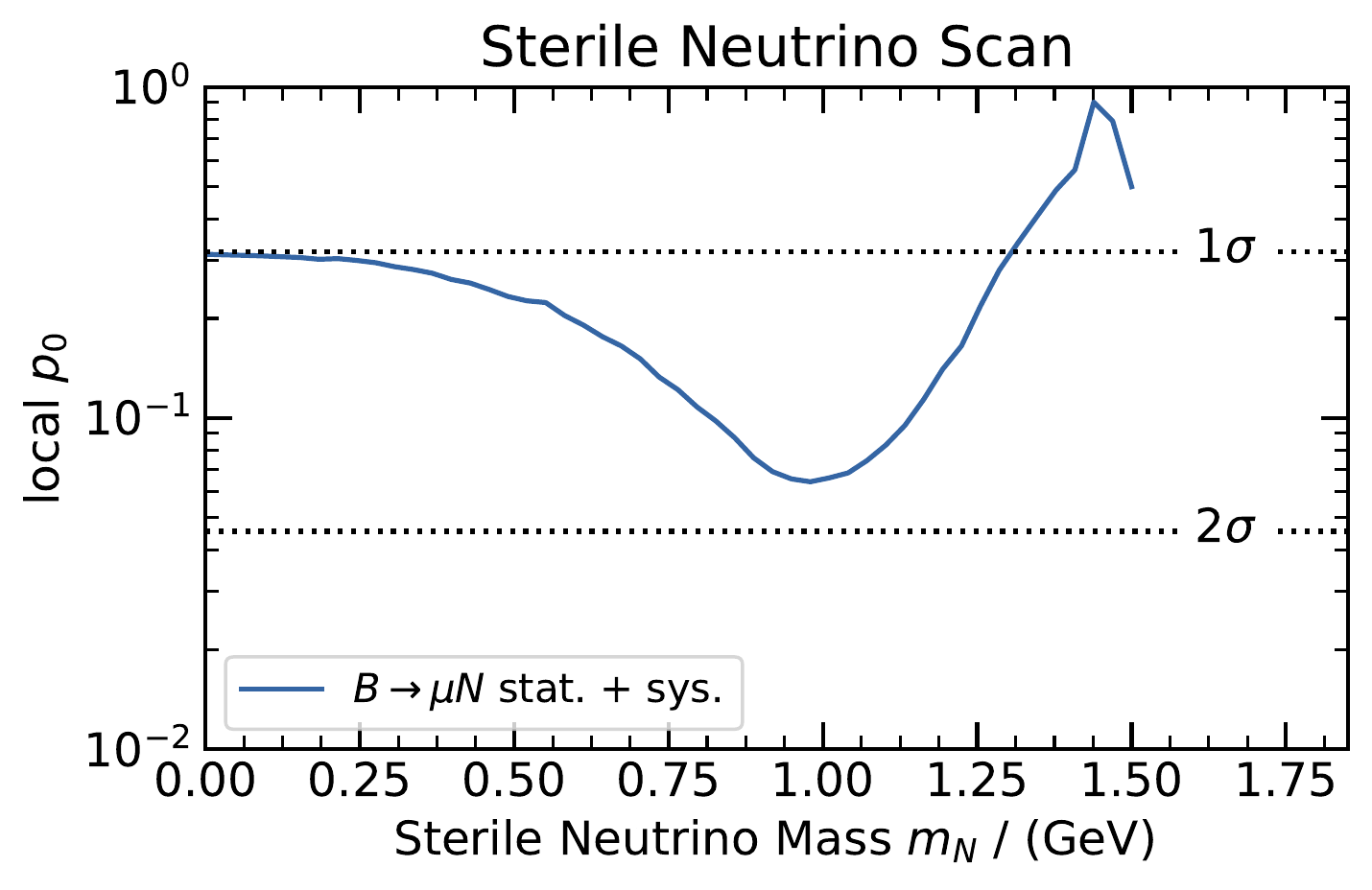}
  \includegraphics[width=0.49\textwidth]{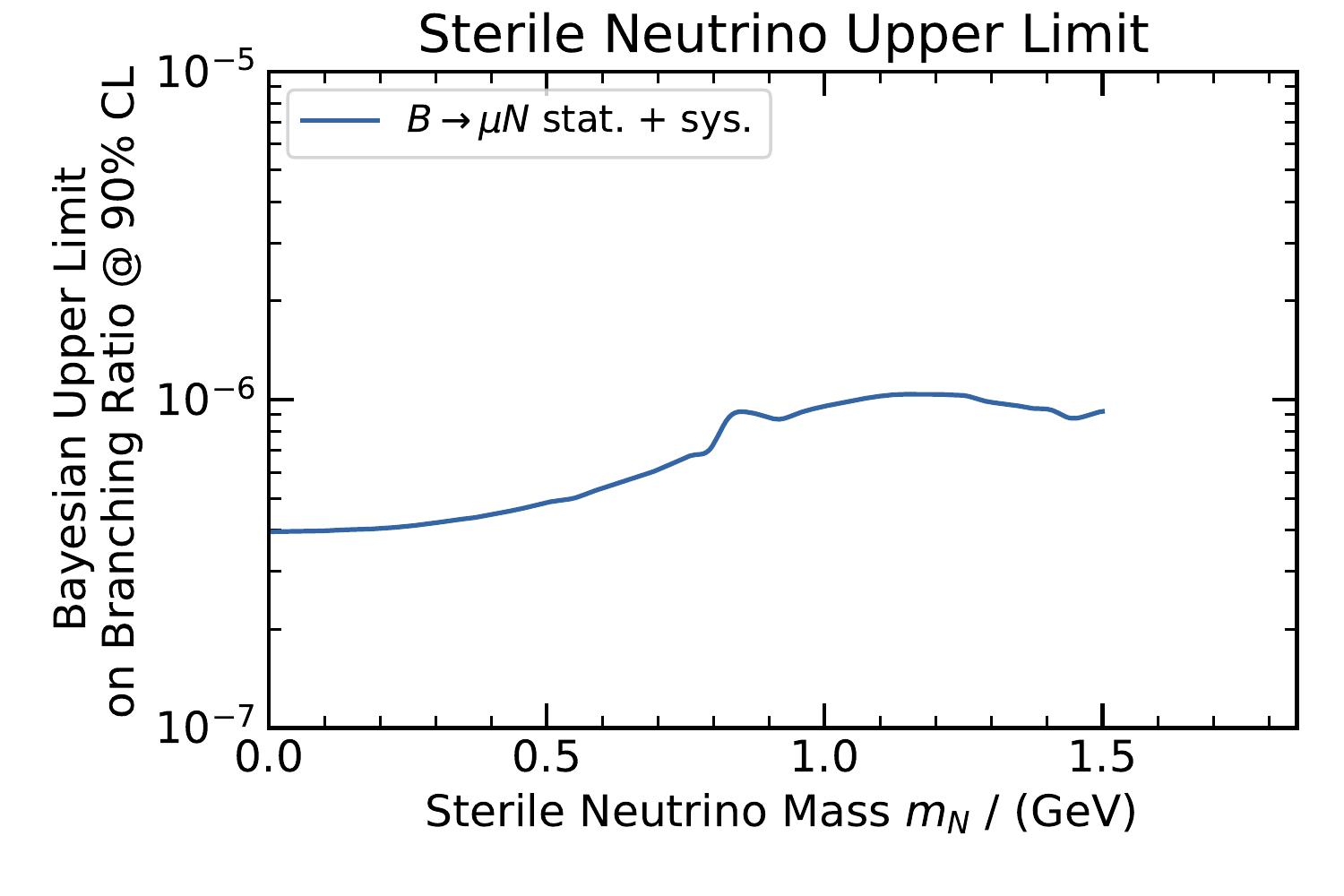} \\
  \includegraphics[width=0.49\textwidth]{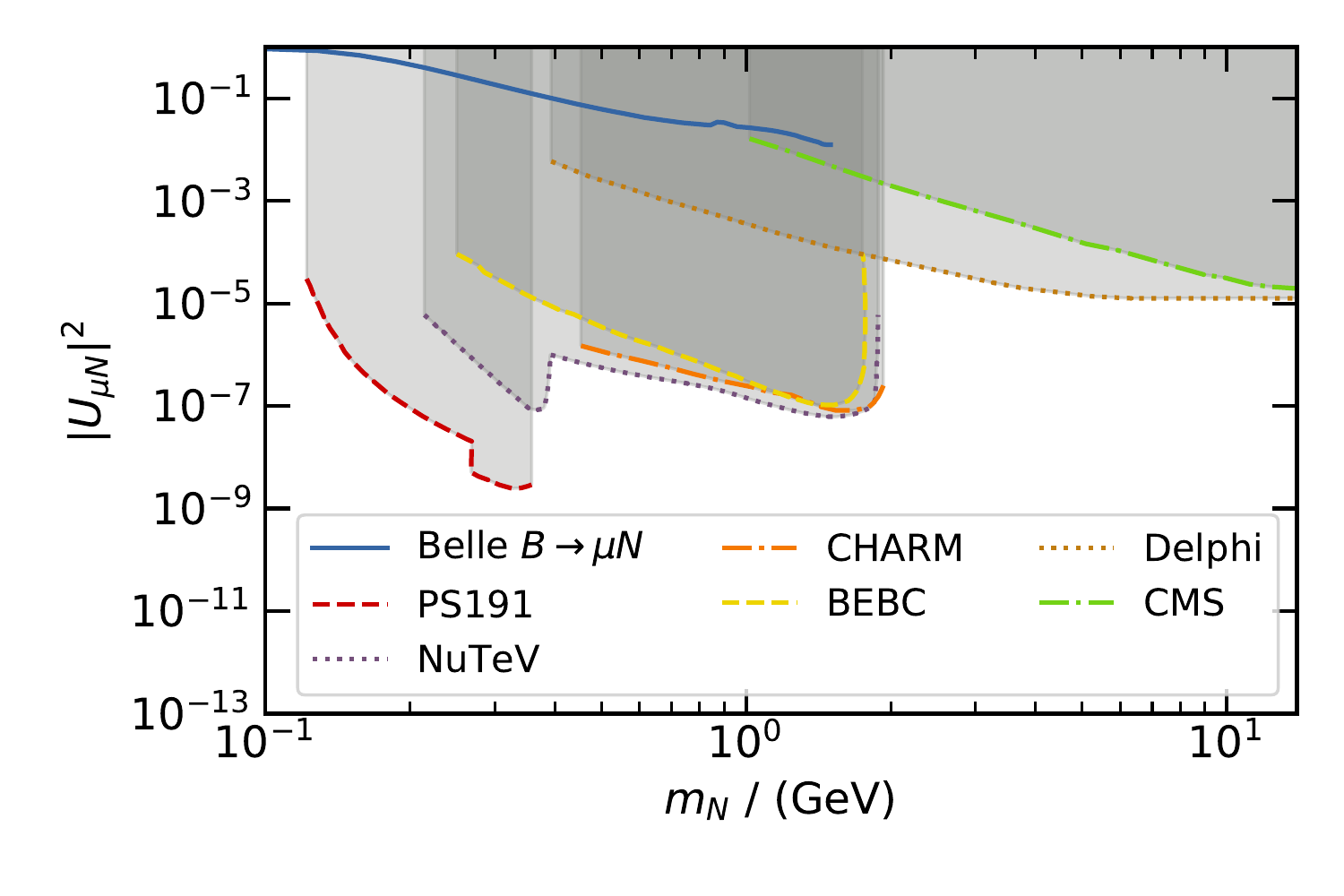}  
\caption{
    (top left) The observed local $p_0$ values for the sterile neutrino search \bmuN\ are shown with the SM process \bmunu included. If the SM process is accounted for, no significant excess is observed. The largest deviation from the background-only hypothesis is at $m_N = 1$ GeV. No correction for the look-elsewhere effect is included.
 (top right) The Bayesian upper limit on the branching fraction as calculated from the sterile neutrino signal yield. The $\bmunu$ process is fixed to its SM expectation.
 (bottom) The excluded area in the coupling-mass plane from this search in comparison to previous searches for sterile neutrinos.
 }
\label{fig:sterile_neutrino}
\end{figure*}

To search for sterile neutrinos in \bmuN\ we fix the \bmunu contribution to its SM value ($\mathcal{B}^{\rm SM}$) and search simultaneously in the four categories for an excess in the $p_\mu^B$ distributions. From the observed yields and our simulated predictions we calculate local $p_0$ values using the test statistic Eq.~\ref{eq:imp_test_stat}. The observed $p_0$ values are shown in Fig.~\ref{fig:sterile_neutrino} for sterile neutrino masses ranging from 0 - 1.5 GeV, and no significant excess over the background-only SM hypothesis is observed. The largest deviation is seen at a mass of $m_N = 1$ GeV with a significance of 1.8 $\sigma$. The result does not account for any corrections for the look-elsewhere effect.
We also calculate the Bayesian upper limit on the branching fraction from the extracted signal yield of the $\bmuN$ process with the \bmunu contribution fixed to its SM value. The upper limit as a function of the sterile neutrino mass is also shown in Fig.~\ref{fig:sterile_neutrino}.
To compare the upper limit from the \bmuN process to previous searches~\cite{PS191:1988fl, NuTeV:1999wq,CHARM:1986as,CHARM:2002de, CHARMII:1995sf, BEBC:1985sf,DELPHI:1997sf,CMS:2018mt} for sterile neutrinos we calculate the excluded values of the coupling $\left|U_{\mu N}\right|^2$ and the sterile neutrino mass $m_N$ using~\cite{DeanPrivateCom}
\begin{equation}
\begin{aligned}
\frac{\mathcal{B}(B^+\rightarrow \mu^+ N)}{\mathcal{B}(B^+ \rightarrow \mu^+ \Pnum)} &= \left|U_{\mu N}\right|^2 \frac{m_N^2 + m_{\mu}^2}{m_{\mu}^2} 
\frac{\sqrt{\lambda (r_{N\PB}, r_{\mu \PB})}}{\sqrt{\lambda(0, r_{\mu \PB})}}\\
&\times \frac{1 - (r_{N\PB}^2 - r_{\mu \PB}^2)^2 / (r_{N\PB}^2 + r_{\mu \PB}^2)}{1 - r_{\mu \PB}^2}\, ,
\end{aligned}
\end{equation}
with $r_{XY} = m_X / m_Y$ and the K\"all\'en function $\lambda(x,y) = (1 - (x-y)^2)(1 - (x+y)^2)$. The excluded values from this and the previous searches are shown in~Fig.~\ref{fig:sterile_neutrino}.

\section{Summary and Conclusions} \label{sec:conc}

In this paper results for the improved search of the \bmunu and \bmuN\ processes using the full Belle data set and an inclusive tag approach are shown. The measurement supersedes the previous result of ~Ref.~\cite{Sibidanov:2017vph} as it has a higher sensitivity and a more accurate modeling of the crucial semileptonic \bulnu\ background. The analysis is carried out in the approximate $B$ rest frame of the signal \bmunu decay, reconstructed from the remaining charged and neutral particles of the collision event. These are combined and calibrated to reconstruct the second $B$ meson produced in the collision. In combination with the known beam properties the four-momentum of the signal $B$ meson is then reconstructed and used to boost the reconstructed signal muon in the reference frame, where the signal $B$ meson is at rest. This results in a better signal resolution and improved sensitivity in contrast to carrying out the search in the c.m.\,frame of the colliding $e^+ \, e^-$-pair. The analysis is carried out in four analysis categories using the continuum suppression classifier and angular information of the $B$ meson and the muon. The branching fraction is determined using a binned maximum likelihood fit of the muon momentum spectrum. Shape and normalization uncertainties from the signal and background templates are directly incorporated into the likelihood. We report an observed branching fraction of
\begin{equation}
 \bfRes \, ,
\end{equation}
with a significance of 2.8 standard deviations over the background-only hypothesis. We also quote the corresponding 90\% upper limit using Bayesian and Frequentist approaches and use the observed branching fraction to set limits on type II and type III two-Higgs-doublet models. We find $\mathcal{B}(\bmunu) < 8.9 \times 10^{-7}$ and $\mathcal{B}(\bmunu) < 8.6 \times 10^{-7} \, \text{at 90\% CL}$ for the Bayesian and Frequentist upper limits, respectively. The type III constraints are the most precise determined to date. In addition, we use the reconstructed muon spectrum to search for the presence of a sterile neutrino created through the process of \bmuN and via a new mediator particle. No significant excess is observed for masses in the probed range of $m_N \in [0,1.5)$ GeV. The largest excess is seen at a sterile neutrino mass of 1 GeV with a local significance of 1.8 standard deviations.

\FloatBarrier

\clearpage

\acknowledgments


\vspace{-3ex}

We thank Marumi Kado and G\"unter Quast for discussions about one-sided test statistics and Ulrich Nierste and Dean Robinson for discussions about the sterile neutrino scenario. FB thanks UB for pulling through. 
We thank the KEKB group for the excellent operation of the
accelerator; the KEK cryogenics group for the efficient
operation of the solenoid; and the KEK computer group, and the Pacific Northwest National
Laboratory (PNNL) Environmental Molecular Sciences Laboratory (EMSL)
computing group for strong computing support; and the National
Institute of Informatics, and Science Information NETwork 5 (SINET5) for
valuable network support.  We acknowledge support from
the Ministry of Education, Culture, Sports, Science, and
Technology (MEXT) of Japan, the Japan Society for the
Promotion of Science (JSPS), and the Tau-Lepton Physics
Research Center of Nagoya University;
the Australian Research Council including grants
DP180102629, 
DP170102389, 
DP170102204, 
DP150103061, 
FT130100303; 
Austrian Science Fund (FWF);
the National Natural Science Foundation of China under Contracts
No.~11435013,  
No.~11475187,  
No.~11521505,  
No.~11575017,  
No.~11675166,  
No.~11705209;  
Key Research Program of Frontier Sciences, Chinese Academy of Sciences (CAS), Grant No.~QYZDJ-SSW-SLH011; 
the  CAS Center for Excellence in Particle Physics (CCEPP); 
the Shanghai Pujiang Program under Grant No.~18PJ1401000;  
the Ministry of Education, Youth and Sports of the Czech
Republic under Contract No.~LTT17020;
the Carl Zeiss Foundation, the Deutsche Forschungsgemeinschaft, the
Excellence Cluster Universe, and the VolkswagenStiftung;
the Department of Science and Technology of India;
the Istituto Nazionale di Fisica Nucleare of Italy;
National Research Foundation (NRF) of Korea Grants
No.~2015H1A2A1033649, No.~2016R1D1A1B01010135, No.~2016K1A3A7A09005
603, No.~2016R1D1A1B02012900, No.~2018R1A2B3003 643,
No.~2018R1A6A1A06024970, No.~2018R1D1 A1B07047294; Radiation Science Research Institute, Foreign Large-size Research Facility Application Supporting project, the Global Science Experimental Data Hub Center of the Korea Institute of Science and Technology Information and KREONET/GLORIAD;
the Polish Ministry of Science and Higher Education and
the National Science Center;
the Grant of the Russian Federation Government, Agreement No.~14.W03.31.0026; 
the Slovenian Research Agency;
Ikerbasque, Basque Foundation for Science, Spain;
the Swiss National Science Foundation;
the Ministry of Education and the Ministry of Science and Technology of Taiwan;
and the United States Department of Energy and the National Science Foundation.

\bibliographystyle{apsrev4-1}
\bibliography{paper.bib}

\begin{thebibliography}{55}
\bibitem [{cha()}]{chargeConjugateImplied}%
  \BibitemOpen
  \href@noop {} {}\bibinfo {note} {Charge conjugation is implied throughout
  this manuscript. Further we use natural units $c = \hbar = 1$
  throughout.}\BibitemShut {Stop}%
\end{thebibliography}

\vspace{-8ex}

\begin{appendix}
\vspace{5ex}
\section*{Appendix}
\section{Nuisance Parameter Pull Distributions}\label{app:NPs}
\begin{figure*}[tbh!]
  \includegraphics[width=0.9\textwidth]{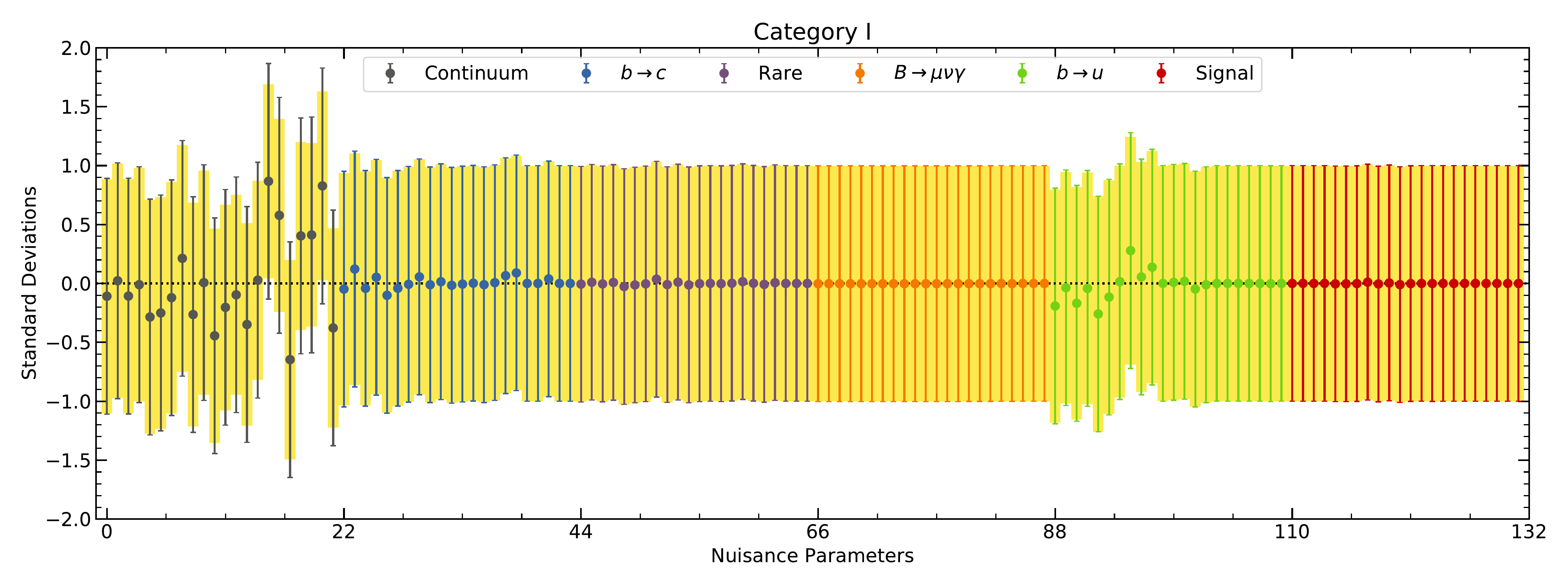} \\
  \vspace{1ex}
  \includegraphics[width=0.9\textwidth]{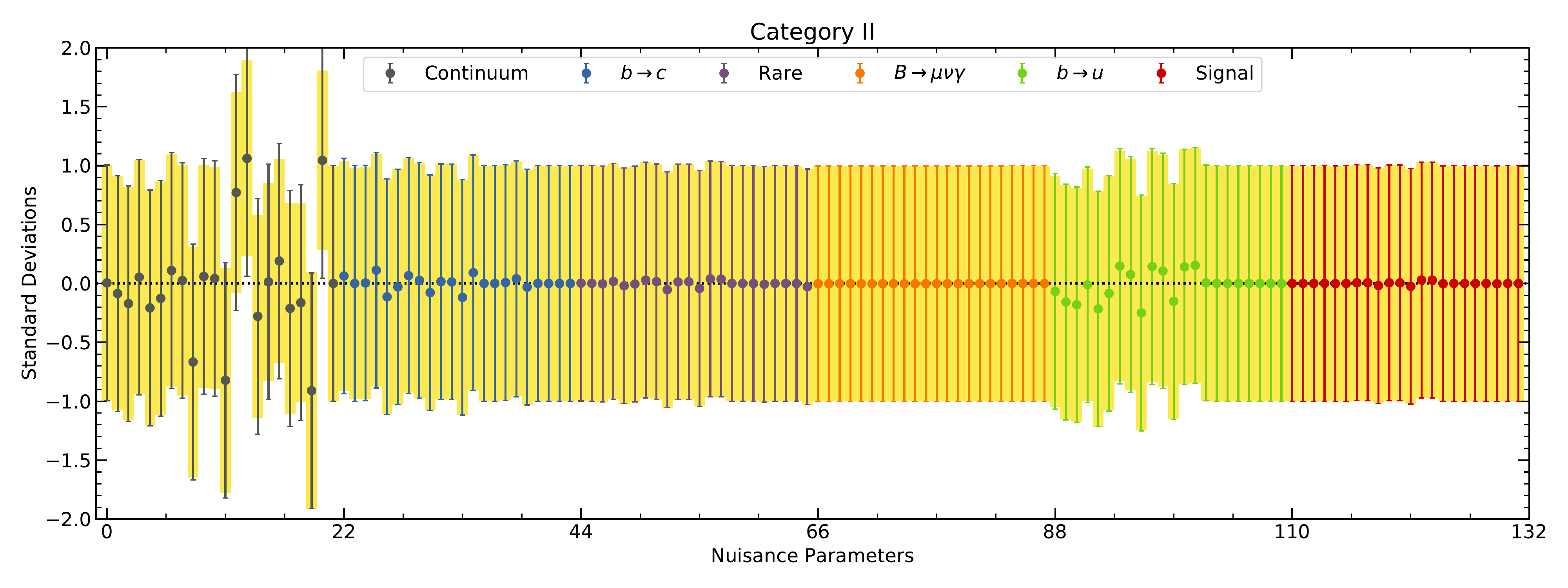} \\
  \vspace{1ex}
  \includegraphics[width=0.9\textwidth]{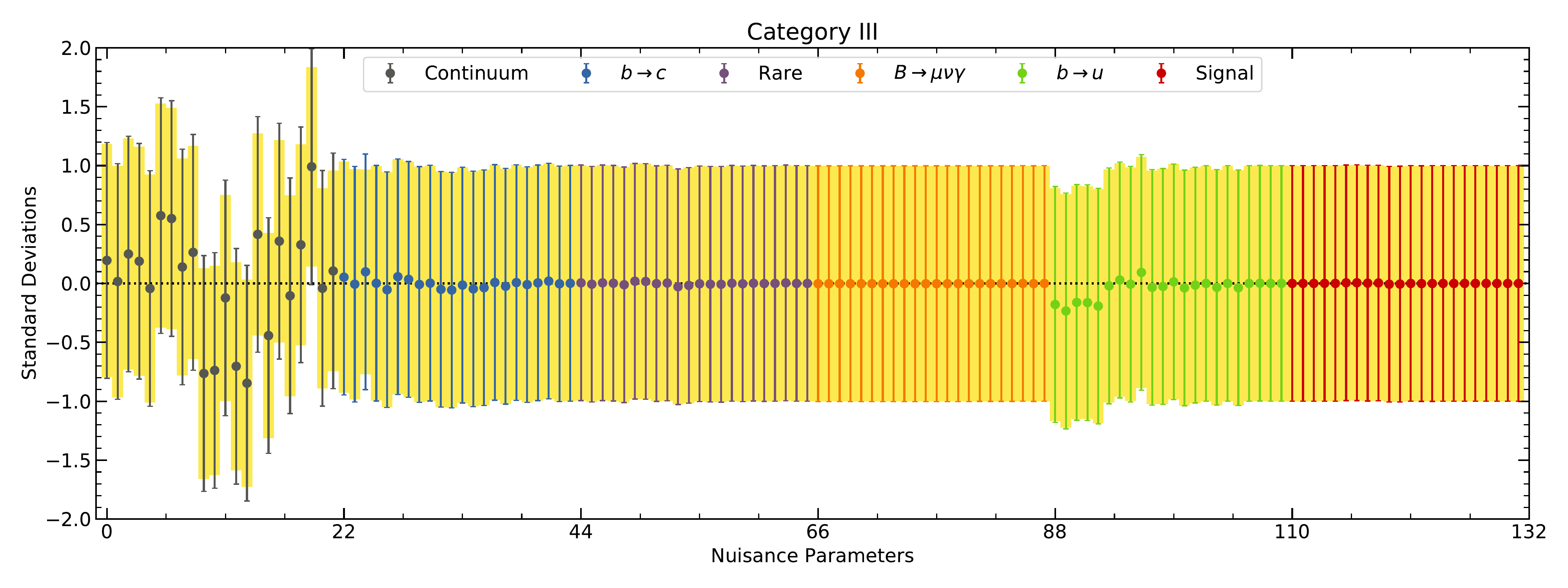} \\
  \vspace{1ex}
  \includegraphics[width=0.9\textwidth]{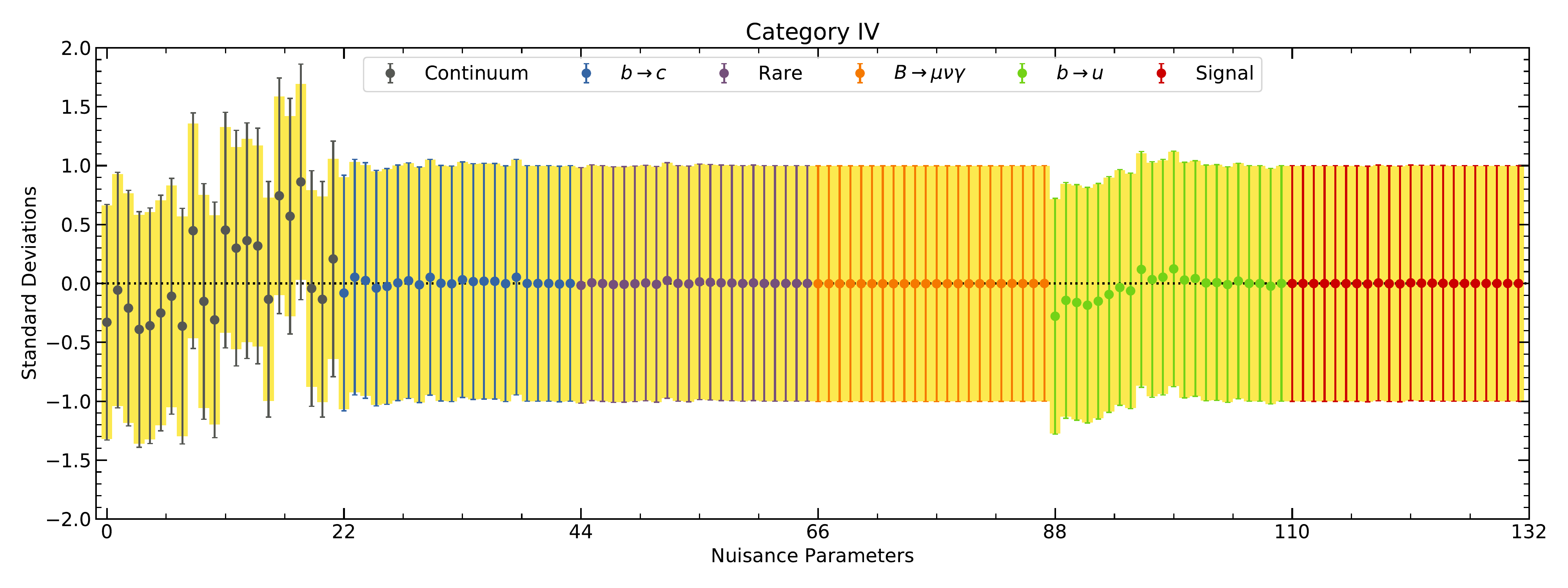}
\caption{
   The post-fit nuisance parameter distribution for each category is shown.
 }
\label{fig:NP_pulls}
\end{figure*}

The summary of the systematic nuisance parameters of the \bmunu fit is shown in Fig.~\ref{fig:NP_pulls}: pull distributions are displayed (defined as $(\theta - 0)/\sigma_{\theta}$) for each NP $\theta$ with uncertainty $\sigma_\theta$. In total 528 + 3 NPs were fitted, one for each fit template and bin. The same processes were correlated over the four categories and constraints were incorporated using multi-dimensional Gaussian PDFs. We observe mild pulls to adjust the \bulnu\ and continuum background shapes.

\section{\bulnu\ Hybrid Model Details}\label{app:Hybrid}

\begin{figure*}[tbh!]
  \includegraphics[width=0.49\textwidth]{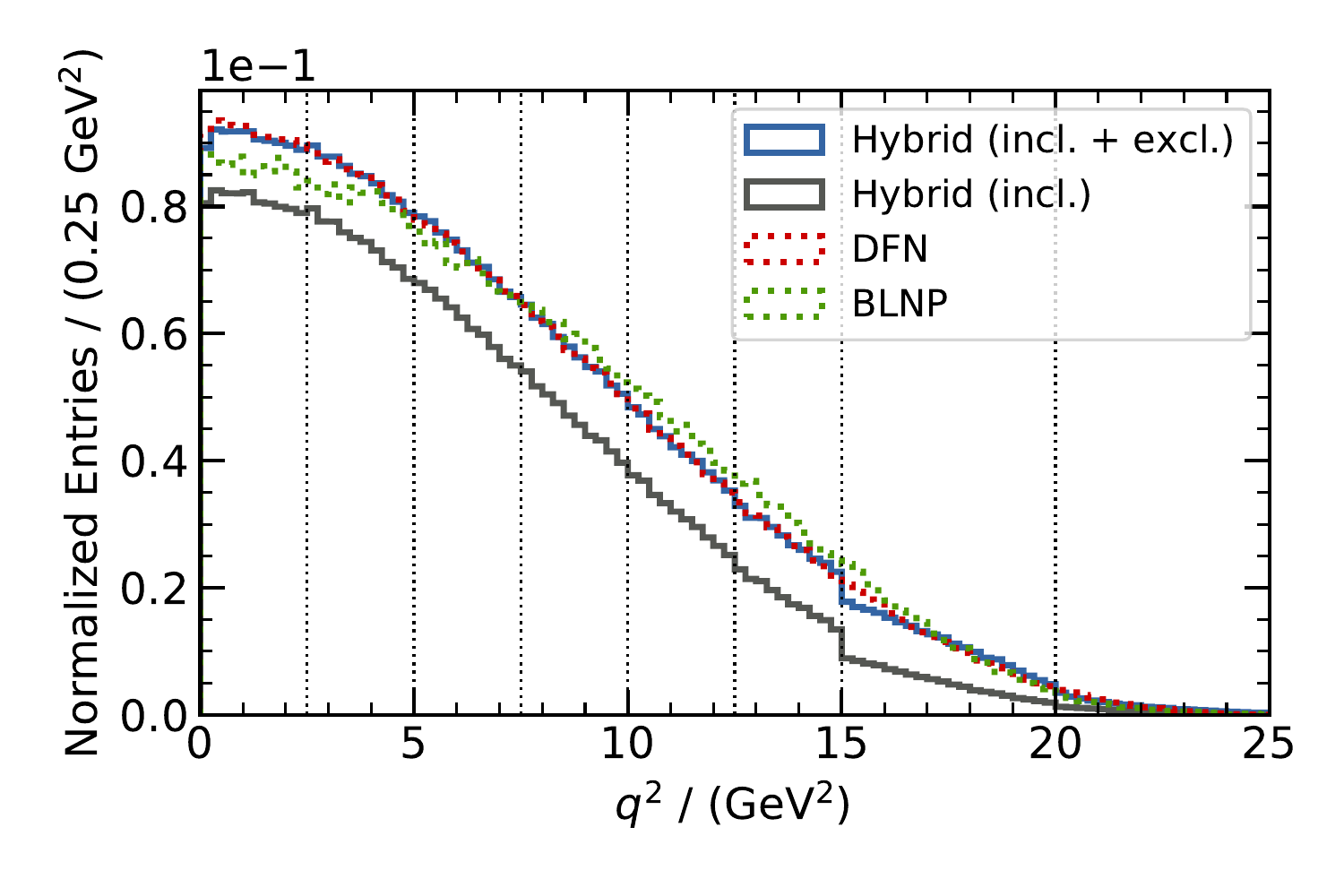}
  \includegraphics[width=0.49\textwidth]{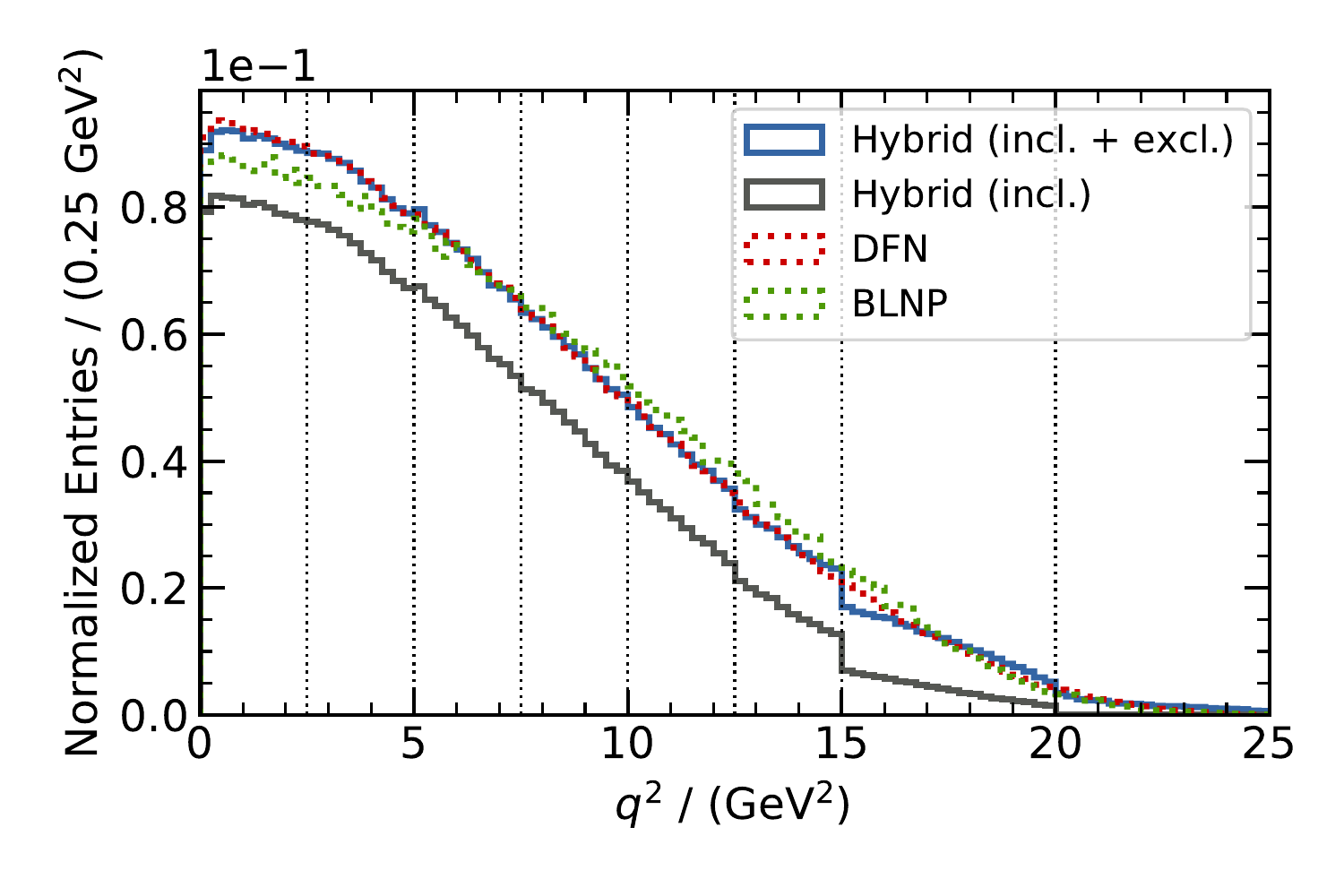} \\
  \includegraphics[width=0.49\textwidth]{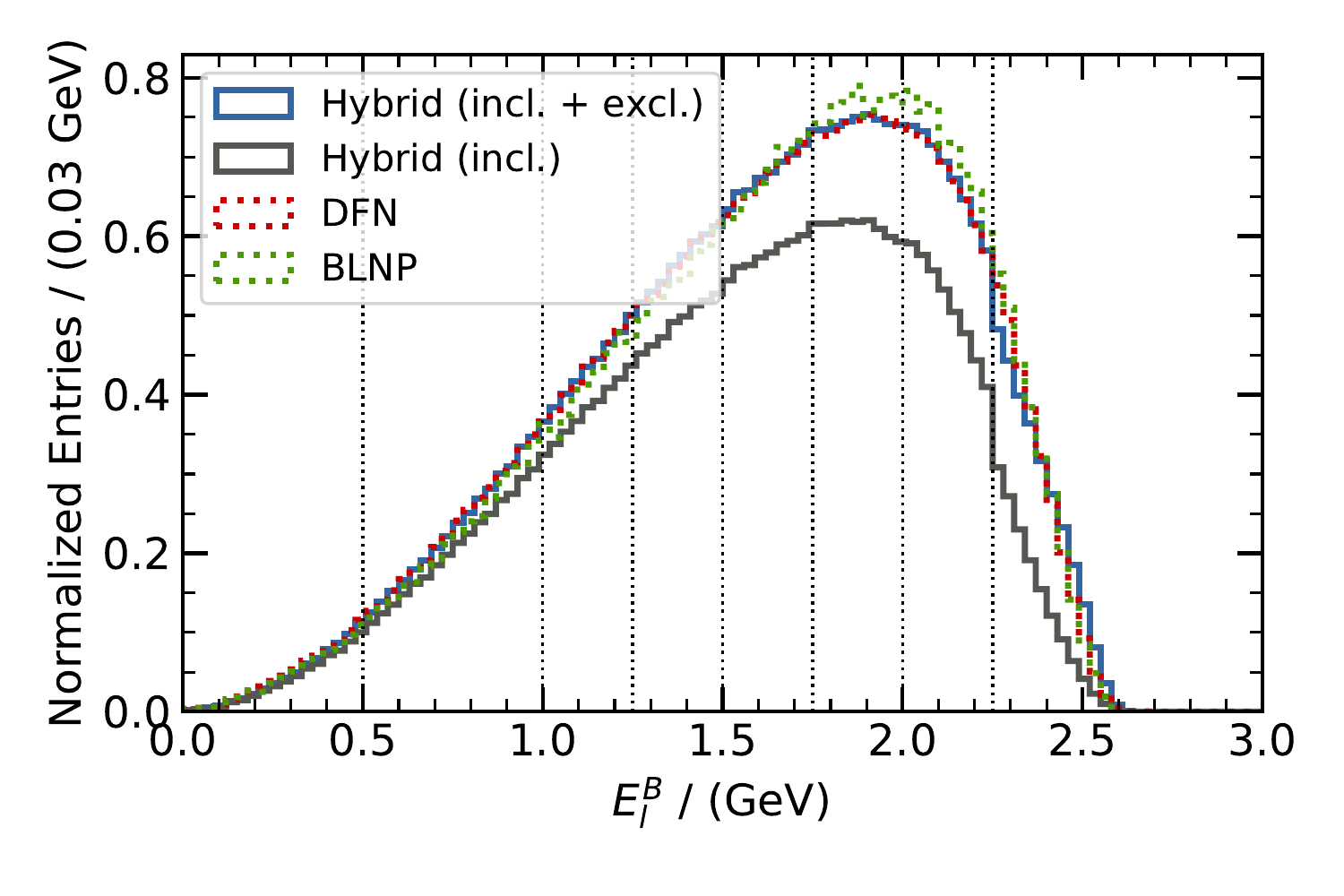}
  \includegraphics[width=0.49\textwidth]{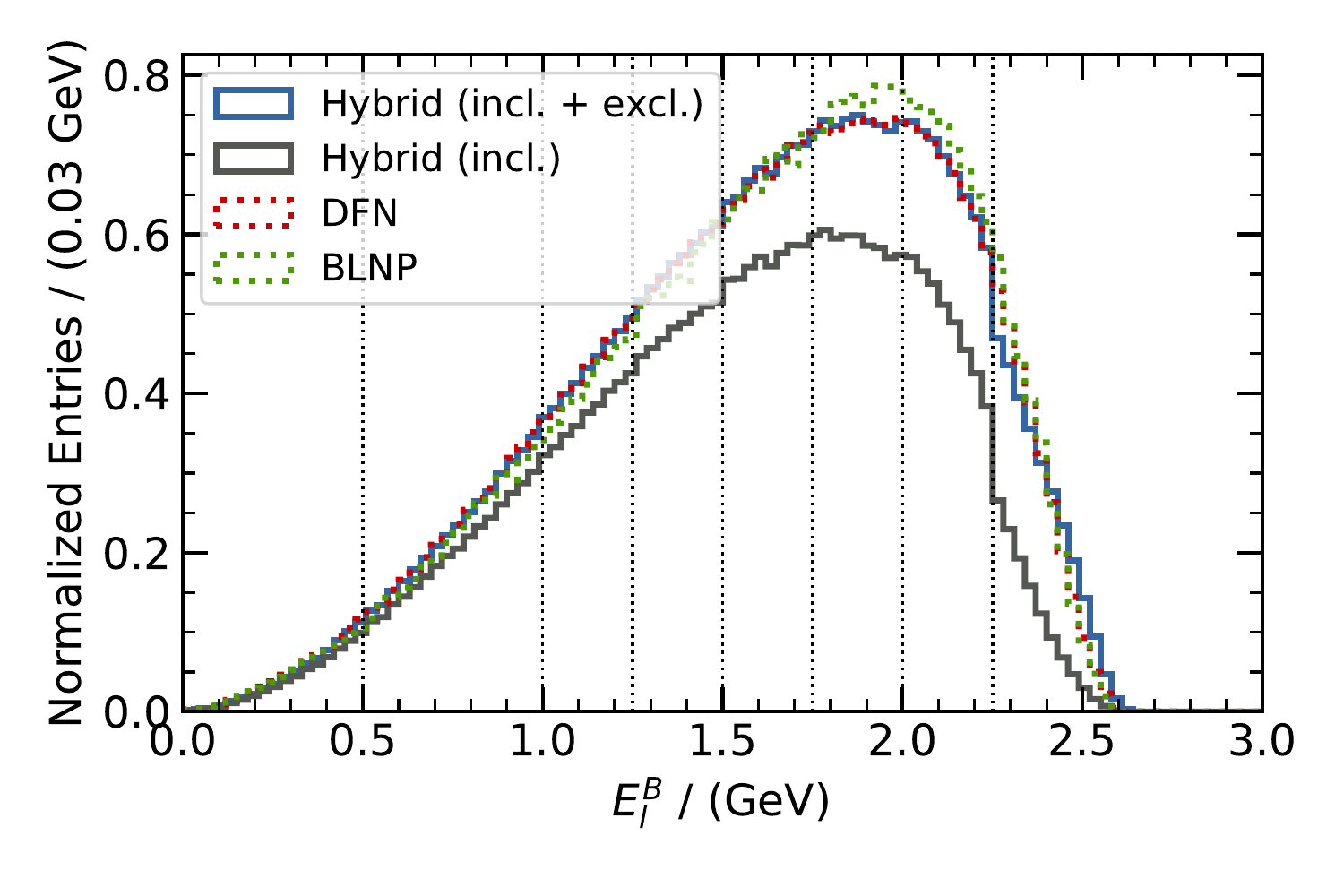} \\
  \includegraphics[width=0.49\textwidth]{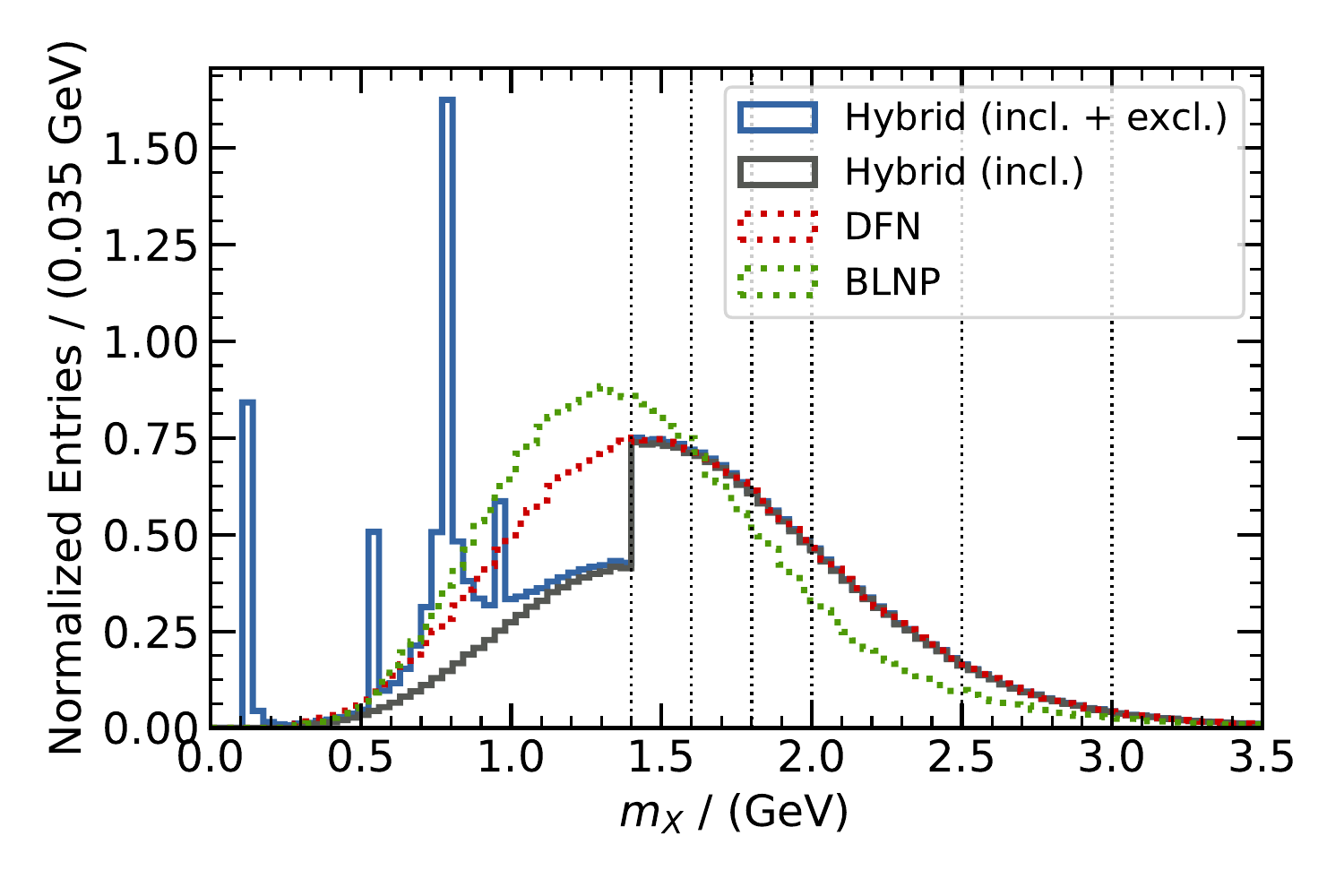}
  \includegraphics[width=0.49\textwidth]{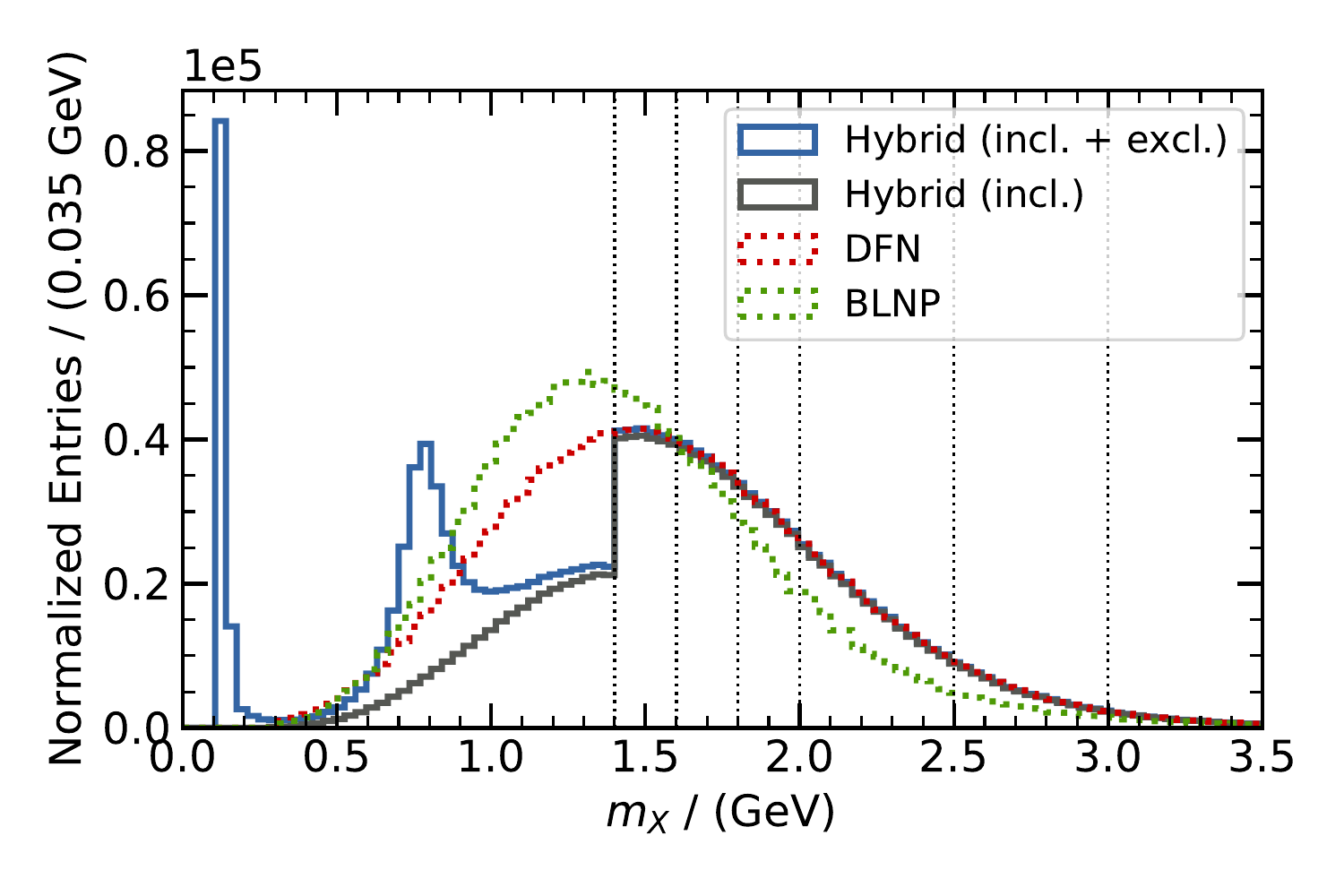} \\
\caption{
   The hybrid model predictions for semileptonic \bulnu decays for $B^+$ (left) and $B^0$ (right) as a function of $q^2$ ,$E_\ell^B$ and $m_X$ are shown. The dashed lines show the chosen hybrid binning.
 }
\label{fig:hybrid_details}
\end{figure*}

Figure~\ref{fig:hybrid_details} shows the \bulnu\ hybrid model: inclusive and exclusive decays are merged, such that for a given bin in the three dimensional space the total inclusive branching fraction is recovered. This is done by scaling down the inclusive prediction.

\section{Data vs.\ MC Reweighting}\label{app:d_vs_mc}
\begin{figure*}[tbh!]
  \includegraphics[width=0.49\textwidth]{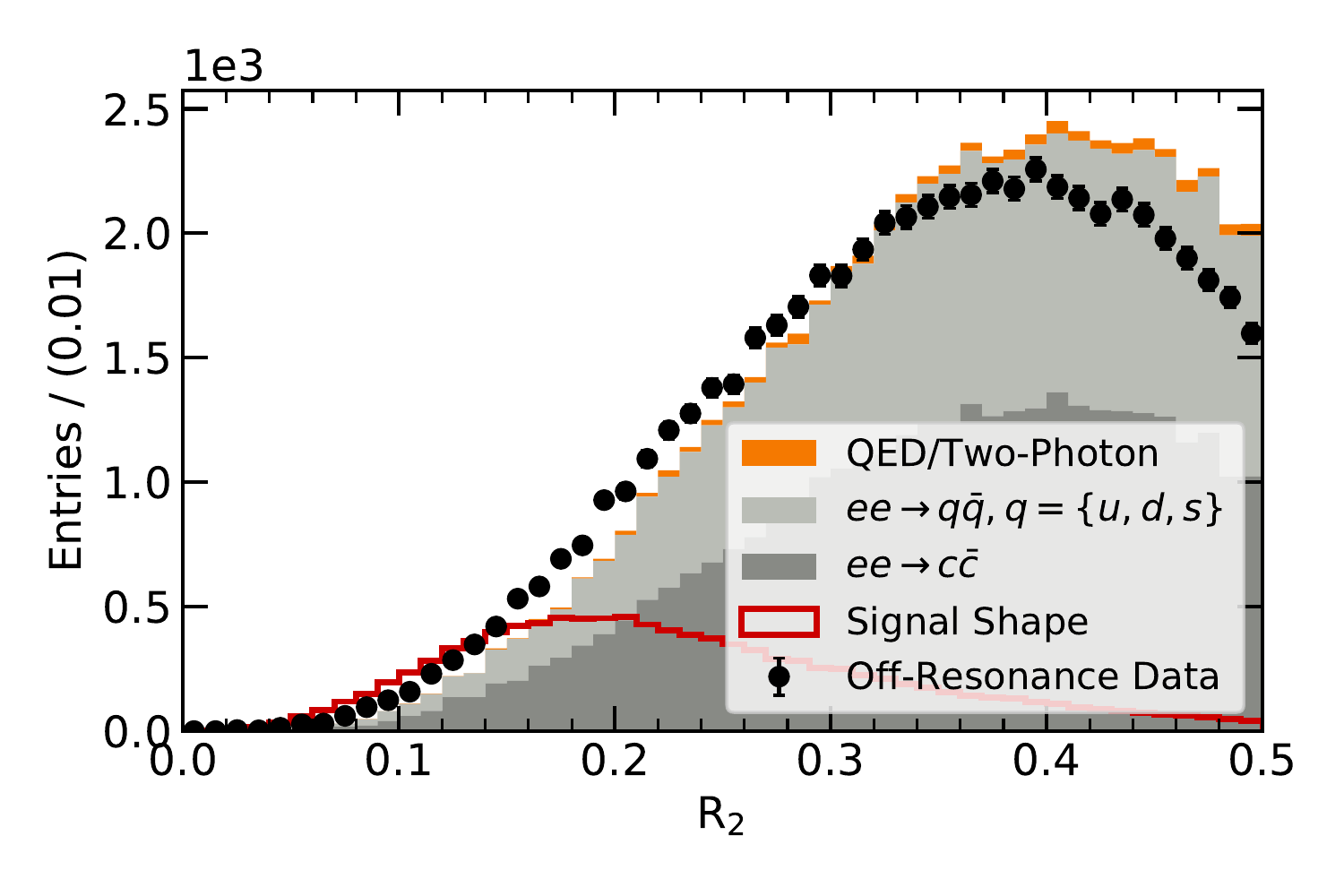}
  \includegraphics[width=0.49\textwidth]{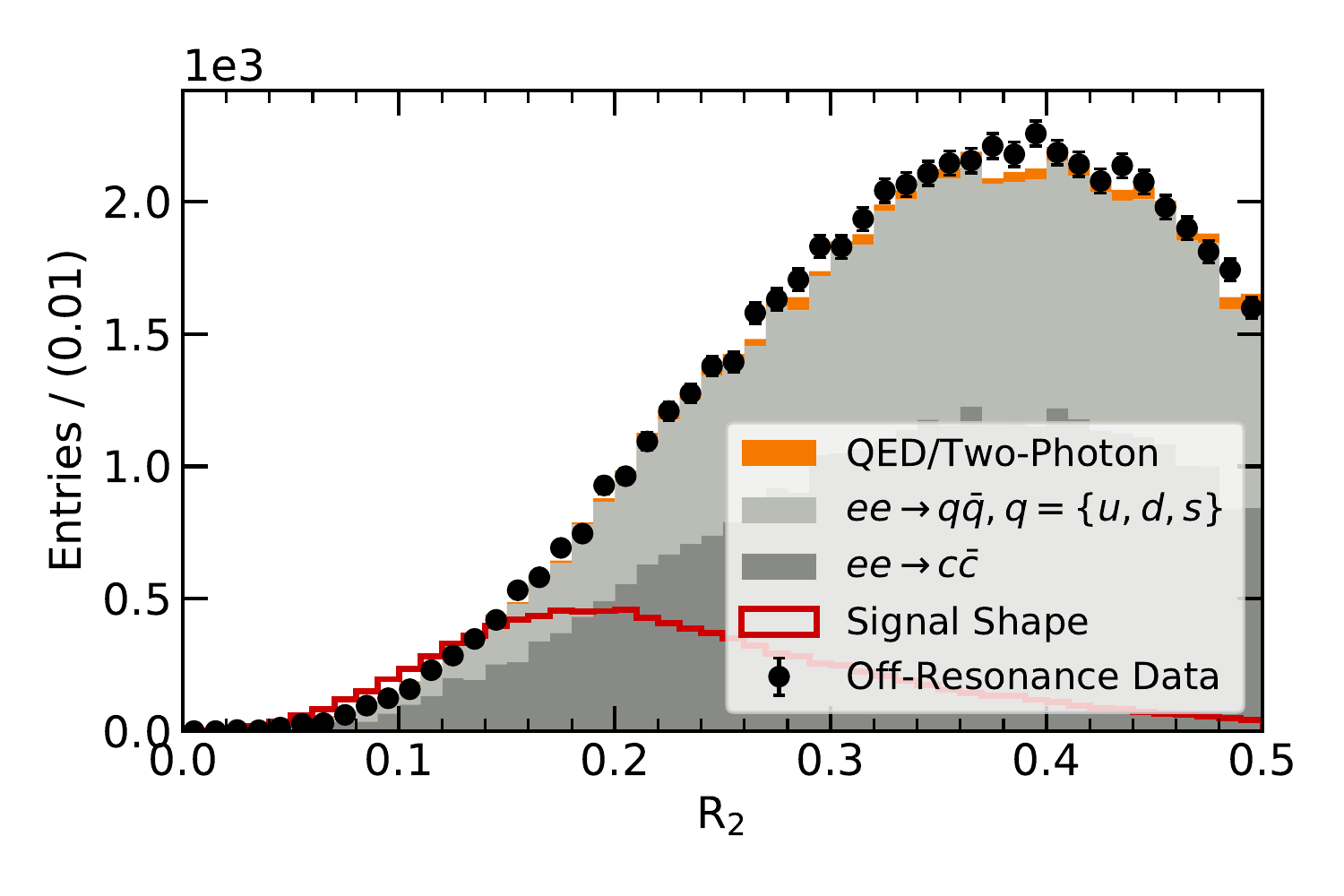} \\
  \includegraphics[width=0.49\textwidth]{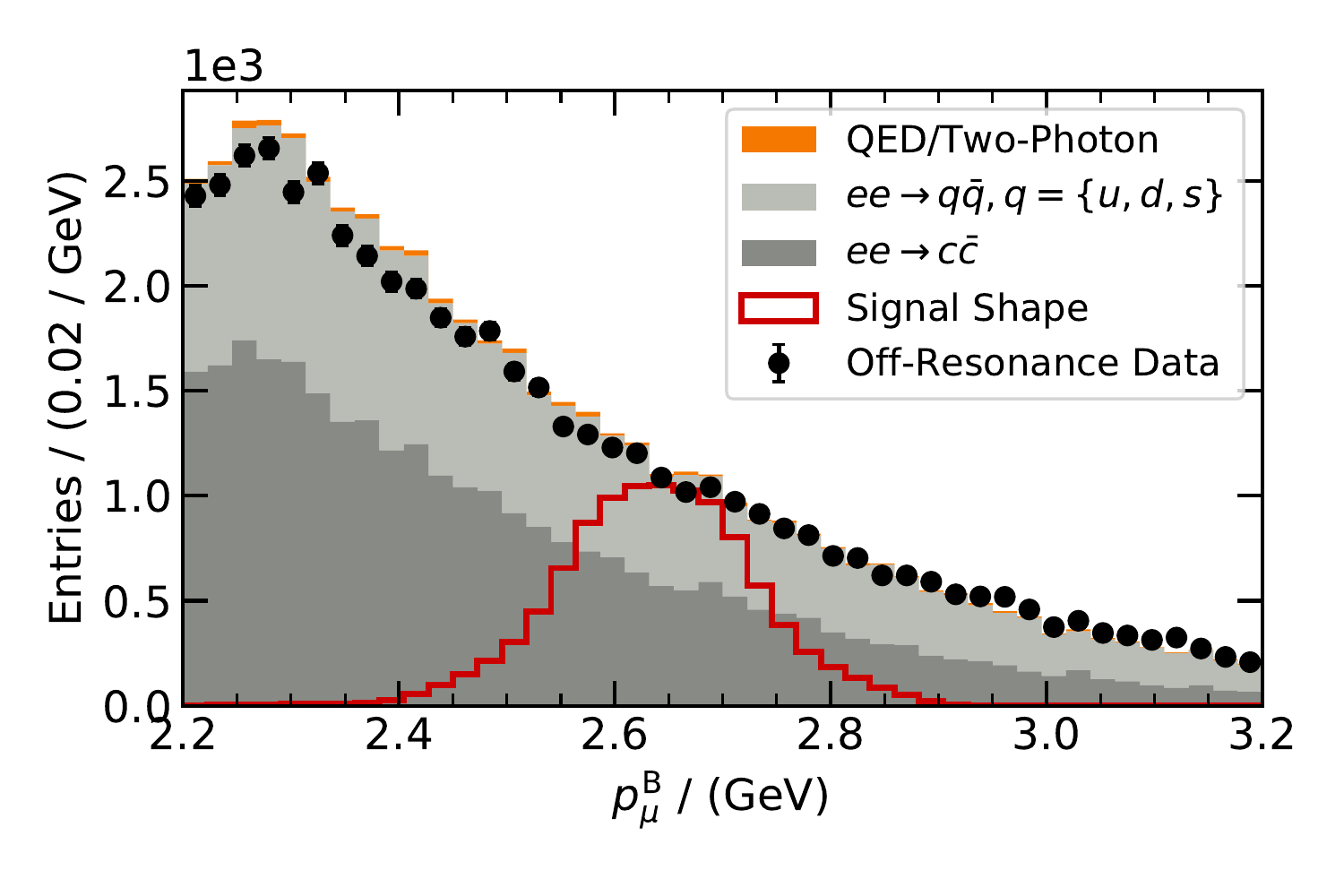}
  \includegraphics[width=0.49\textwidth]{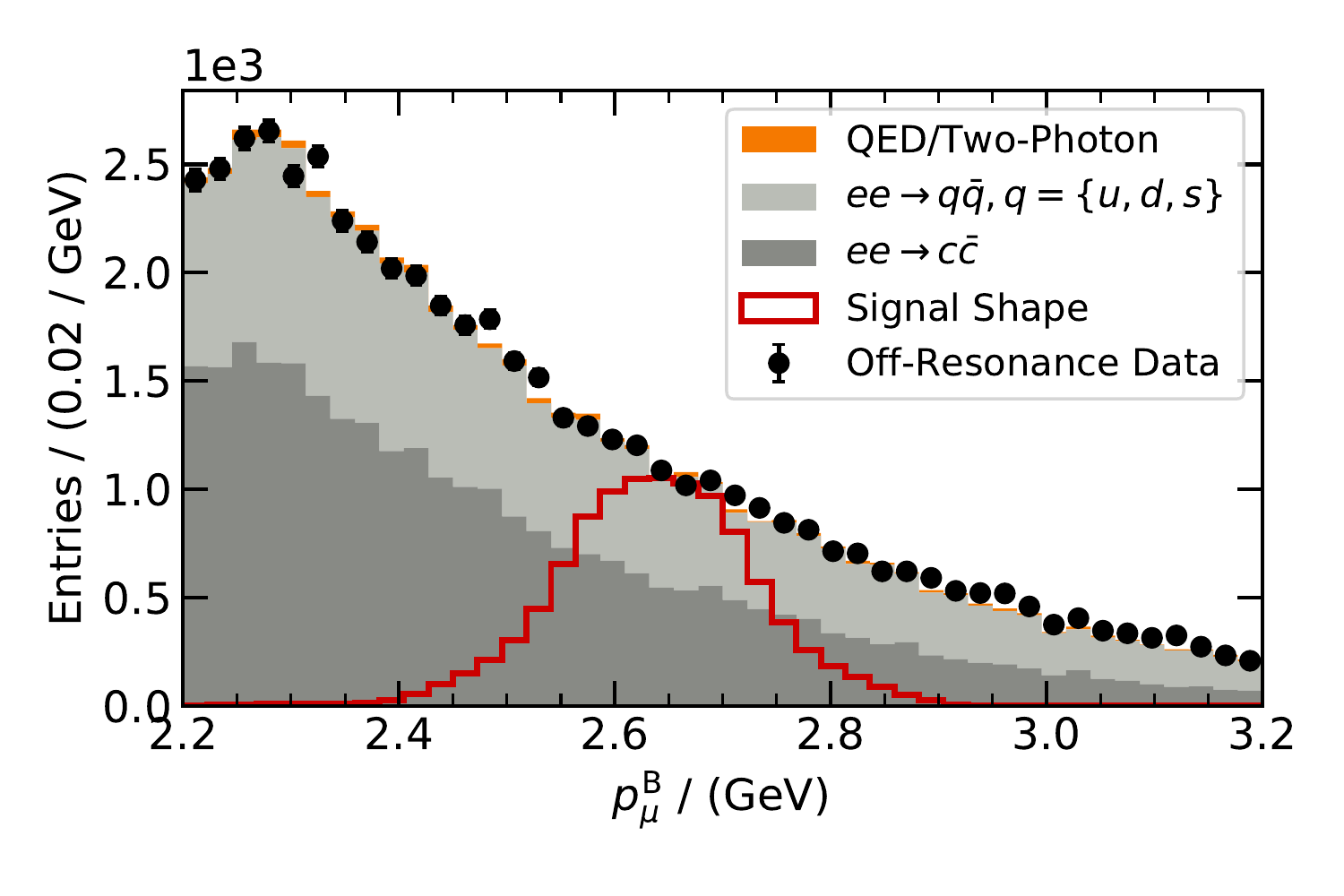} \\
  \includegraphics[width=0.49\textwidth]{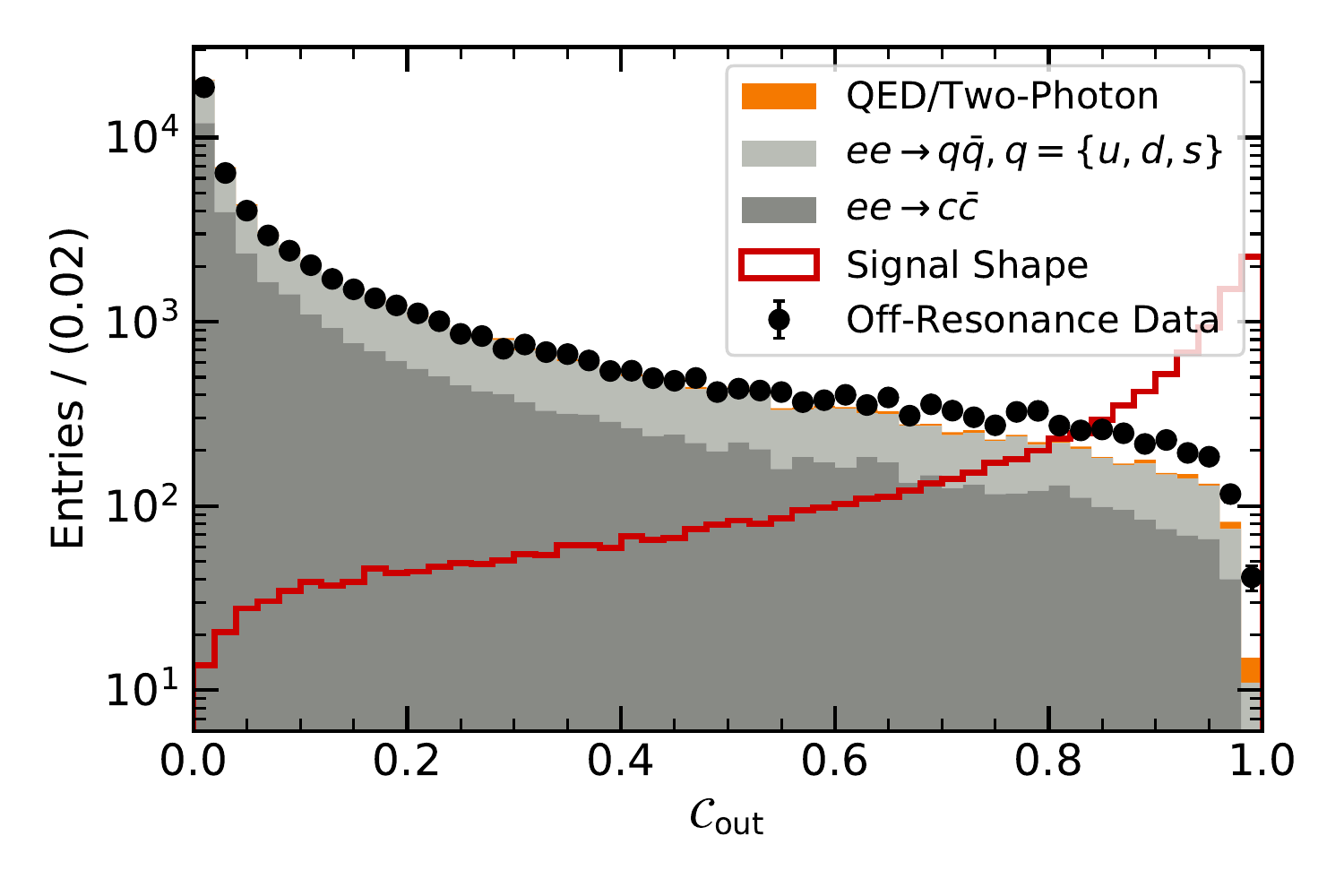}
  \includegraphics[width=0.49\textwidth]{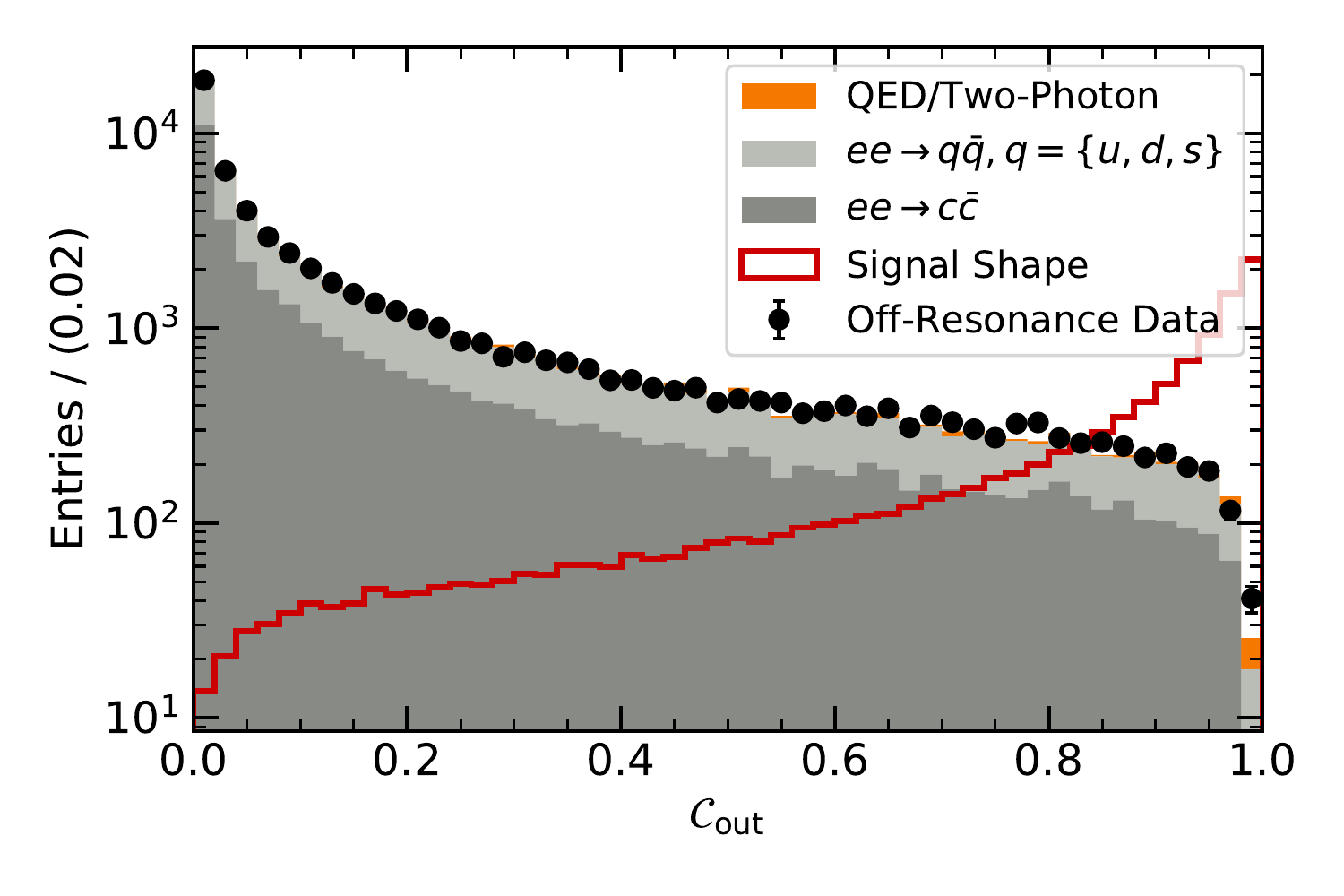} \\
\caption{
   A selection of variables used as input in the data vs.\ MC reweighting before (left) and after (right) the weights are applied. The simulated data is shown as histogram and the recorded off-resonance collision events as data points with uncertainties.
 }
\label{fig:d_vs_mc}
\end{figure*}

Figure~\ref{fig:d_vs_mc} shows the effect of the data vs.\ MC reweighting using the off-resonance collision events for three selected variables used in the training.

\end{appendix}

\end{document}